\documentclass[iop,twocolappendix,numberedappendix,apj]{emulateapj}
\usepackage{epsfig, natbib, graphicx, color, amsmath, cancel,icomma,bm}

\newcommand{\ptaken}{p_{\mathrm{taken}}}

\newcommand{\chisq}{\chi^2}

\newcommand{\lcol}{\lambda_{\mathrm{col}}}
\newcommand{\lgmr}{\lambda_{g-r}}
\newcommand{\lrmi}{\lambda_{r-i}}
\newcommand{\ltrue}{\lambda_{\mathrm{true}}}
\newcommand{\lobs}{\lambda_{\mathrm{obs}}}
\newcommand{\redmapper}{redMaPPer}

\newcommand{\bc}{\mathbf{c}}
\newcommand{\rnc}{\rho_{\nu}(\chisq)}
\newcommand{\zred}{z_{\mathrm{red}}}
\newcommand{\sigzred}{\sigma_{\zred}}
\newcommand{\zlambda}{z_\lambda}
\newcommand{\zlambdaraw}{z_{\lambda,\mathrm{raw}}}
\newcommand{\ztrue}{z_{\mathrm{true}}}
\newcommand{\sigint}{\sigma_{\mathrm{int}}}
\newcommand{\pmem}{p_{\mathrm{mem}}}
\newcommand{\refmag}{\widetilde{m_i}(z)}

\newcommand{\imag}{m_i}
\newcommand{\imagbar}{\bar{m}_i}
\newcommand{\cmod}{\left < c|z,\imag \right >}
\newcommand{\cmodbf}{\left < \mathbf{c}|z,\imag \right >}
\newcommand{\cbar}{\bar{c}(z)}

\newcommand{\sbar}{\bar{s}(z)}

\newcommand{\medcol}{\widetilde{c}(z)}
\newcommand{\medcoli}{\widetilde{c}(z_i)}
\newcommand{\madsigint}{\widetilde{\sigma}_{\mathrm{int}}(z)}

\newcommand{\madsig}{\widetilde{\sigma}(z)}

\newcommand{\cbarz}{\bar{c}_z}
\newcommand{\sbarz}{\bar{s}_z}

\newcommand{\corrmod}{\left < c_z|\ztrue,\imag \right >}
\newcommand{\zcg}{z_{\mathrm{CG}}}
\newcommand{\zphoto}{z_{\mathrm{photo}}}
\newcommand{\corrmodzl}{\left < c_{\ztrue}|\zlambda \right >}
\newcommand{\ucen}{u_{\mathrm{cen}}}
\newcommand{\usat}{u_{\mathrm{sat}}}
\newcommand{\ufg}{u_{\mathrm{fg}}}
\newcommand{\lsat}{\lambda_{\mathrm{sat}}}
\newcommand{\pcen}{p_{\mathrm{cen}}}
\newcommand{\Pcen}{P_{\mathrm{cen}}}
\newcommand{\Psat}{P_{\mathrm{sat}}}
\newcommand{\pfree}{p_{\mathrm{free}}}
\newcommand{\phicen}{\phi_{\mathrm{cen}}}
\newcommand{\Gcen}{G_{\mathrm{cen}}}
\newcommand{\fcen}{f_{\mathrm{cen}}}
\newcommand{\wcenbar}{\bar{w}_{\mathrm{cen}}}
\newcommand{\wsatbar}{\bar{w}_{\mathrm{sat}}}
\newcommand{\wfgbar}{\bar{w}_{\mathrm{fg}}}
\newcommand{\sigwcen}{\sigma_{w,\mathrm{cen}}}
\newcommand{\sigwsat}{\sigma_{w,\mathrm{sat}}}
\newcommand{\sigwfg}{\sigma_{w,\mathrm{fg}}}
\newcommand{\fsat}{f_{\mathrm{sat}}}
\newcommand{\phisat}{\phi_{\mathrm{sat}}}
\newcommand{\Gsat}{G_{\mathrm{sat}}}

\newcommand{\ffg}{f_{\mathrm{fg}}}
\newcommand{\barsigg}{\bar{\Sigma}_g}
\newcommand{\barsigz}{\bar{\Sigma}_{g,z}}
\newcommand{\mbarsat}{\bar{m}_{\mathrm{sat}}}
\newcommand{\covarint}{\bC_{\mathrm{int}}(z)}
\newcommand{\covarerr}{\bC_{\mathrm{err}}(z)}
\newcommand{\zin}{z_{\mathrm{in}}}
\newcommand{\zout}{z_{\mathrm{out}}}
\newcommand{\fmask}{f_{\mathrm{mask}}}

\newcommand{\photoz}{photo-$z$}
\newcommand{\photozs}{photo-$z$s}
\newcommand{\zinit}{z_{\mathrm{init}}}

\newcommand{\Mpc}{\mbox{Mpc}}
\newcommand{\hMpc}{h^{-1}\,\mathrm{Mpc}}

\newcommand{\avg}[1]{\left\langle #1 \right\rangle}

\newcommand{\lk}{{\cal{L}}}

\newcommand{\hatn}{\bm{\hat{n}}}

\newcommand{\bx}{\bm{x}}
\newcommand{\be}{\begin{equation}}
\newcommand{\ee}{\end{equation}}
\newcommand{\bea}{\begin{eqnarray}}
\newcommand{\eea}{\end{eqnarray}}

\newcommand{\bC}{\mathbf{C}}

\newcommand{\Rc}{R_c}

\newcommand{\kpc}{h^{-1}\,\mathrm{kpc}}

\newcommand{\erf}{\mbox{erf}}

\shortauthors{Rykoff et al.}
\shorttitle{\redmapper{} I: Algorithm and DR8 Catalog}

\begin{document}
\title{redMaPPer I: Algorithm and SDSS DR8 Catalog}
\author{E.~S.~Rykoff\altaffilmark{1}, E. Rozo\altaffilmark{1},
  M.~T.~Busha\altaffilmark{2,3},
  C.~E.~Cunha\altaffilmark{3}, A.~Finoguenov\altaffilmark{4},
  A.~Evrard\altaffilmark{5,6,7},
  J.~Hao\altaffilmark{8}, B.~P.~Koester\altaffilmark{5},
  A.~Leauthaud\altaffilmark{9}, B.~Nord\altaffilmark{8},
  M.~Pierre\altaffilmark{10}, R.~Reddick\altaffilmark{1,3},
  T.~Sadibekova\altaffilmark{10}, E.~S.~Sheldon\altaffilmark{11},
  R.~H.~Wechsler\altaffilmark{1,3}
}
\altaffiltext{1}{SLAC National Accelerator Laboratory, Menlo Park, CA 94025.}
\altaffiltext{2}{Institute for Theoretical Physics, University of
  Z\"urich, 8057 Z\"urich, Switzerland.}
\altaffiltext{3}{Kavli Institute for Particle Astrophysics and Cosmology,
  Stanford University, Palo Alto, CA 94305.}
\altaffiltext{4}{Department of Physics, University of Helsinki, FI-00014
  Helsinki, Finland.}
\altaffiltext{5}{Physics Department, University of Michigan, Ann Arbor, MI
  48109.}
\altaffiltext{6}{Astronomy Department, University of Michigan, Ann Arbor, MI
  48109.}
\altaffiltext{7}{Michigan Center for Theoretical Physics, Ann Arbor, MI 48109.}
\altaffiltext{8}{Center for Particle Astrophysics, Fermi National Accelerator
  Laboratory, Batavia, IL 60510.}
\altaffiltext{9}{Kavli Institute for the Physics and Mathematics of the
  Universe, University of Tokyo, Kashiwa 277-8583, Japan.}
\altaffiltext{10}{Service d'Astrophysique, CEA Saclay, F-91191 Gif sur Yvette
  Cedex, France.}
\altaffiltext{11}{Brookhaven National Laboratory, Upton, NY 11973.}

\begin{abstract}
  We describe \redmapper, a new red-sequence cluster finder
  specifically designed to make optimal use of ongoing and near-future
  large photometric surveys.  The algorithm has multiple attractive
  features: (1) It can iteratively self-train the red-sequence model
  based on a minimal spectroscopic training sample, an important feature
  for high redshift surveys;  (2) It can handle complex masks with
  varying depth;  (3) It produces cluster-appropriate random points to
  enable large-scale structure studies;  (4) All clusters are assigned
  a full redshift probability distribution $P(z)$;  (5) Similarly,
  clusters can have multiple candidate central galaxies, each with
  corresponding centering probabilities;  (6) The algorithm is
  parallel and numerically efficient: it can run a Dark Energy
  Survey-like catalog in $\sim$~500 CPU hours;  (7) The algorithm
  exhibits excellent photometric redshift performance, the richness
  estimates are tightly correlated with external mass proxies, and the
  completeness and purity of the corresponding catalogs is superb.  We
  apply the \redmapper{} algorithm to $\sim10,000\,\mathrm{deg}^{2}$
  of SDSS DR8 data, and present the resulting catalog of $\sim$~25,000
  clusters over the redshift range $z\in[0.08,0.55]$.  The
  \redmapper{} photometric redshifts are nearly Gaussian, with a
  scatter $\sigma_z \approx 0.006$ at $z\approx0.1$, increasing to
  $\sigma_z\approx 0.02$ at $z\approx 0.5$ due to increased
  photometric noise near the survey limit.  The median value for $|\Delta z|/(1+z)$
  for the full sample is $0.006$.  
  The incidence of
  projection effects is low ($\leq 5\%$).  Detailed performance
  comparisons of the \redmapper{} DR8 cluster catalog to X-ray and SZ
  catalogs are presented in a companion paper.
\end{abstract}


\section{Introduction}

Over the past several years, galaxy clusters have been recognized as powerful
cosmological
probes~\citep[e.g.,][]{henryetal09,vikhlininetal09b,mantzetal10a,rozoetal10a,clercetal12,bensonetal13,hasselfieldetal13}.
Galaxy clusters are one of the key probes of Dark Energy for ongoing
and upcoming photometric surveys such as the Dark Energy
Survey~\citep[DES:][]{des05}, Pan-STARRS \citep{panstarrs02}, the Hyper-Suprime
Camera (HSC)\footnote{http://www.naoj.org/Projects/HSC/HSCProject.html}, and the
Large Synoptic Survey Telescope \citep[LSST:][]{lsst12}.

Because galaxies are obviously clustered on the sky, rich galaxy clusters have
been detected as far back as the 1800's~\citep{biviano00}, with the first large
catalogs generated from galaxy overdensities on photographic plates created 50 years
ago~\citep[e.g.,][]{abell58,zwickyetal68,aco89}.  More recently, the advent of
multi-band data has led to a proliferation of optical cluster finding
algorithms.  These algorithms use various techniques for measuring clustering
in angular position plus color/redshift space, ranging from simple matched-filters
to more complicated Voronoi tesselations.  These cluster finders can be divided
into roughly two classes, those based on photometric
redshifts~\citep[e.g.,][]{kepneretal99,vanbreukelenetal09,milkeraitisetal10,durretetal11b,spdpg11,soaressantosetal11,wenetal12},
and those utilizing the cluster
red-sequence~\citep[e.g.,][]{annisetal99,gladderyee00,koesteretal07a,gladdersetal07,galetal09,thanjavuretal09,hmkrr10,mgb12}.
However, relatively few of these optical catalogs have been utilized for
cosmological parameter constraints~\citep[e.g.,][]{rwkme07a,rozoetal10a,manaetal13}. 

Given the above landscape, it is a fair question to ask whether the world
really needs yet another photometric cluster finding algorithm.  As we describe
below, we believe that the answer to this question is yes.  In
particular, there are a variety of important features that any reasonable
optical cluster finder must have in order to properly exploit the photometric
data that will become available with ongoing or near-future photometric surveys
such as the DES or LSST.

What must we require of current photometric cluster finders?  The key features
are as follows.
\begin{enumerate}
  \item{The algorithm must be able to smoothly detect galaxy clusters in a
    consistent way across a braod redshift range.  This can be a challenge for
    photometric redshift (``\photoz'') and red-sequence based algorithms alike.
    For photometric redshift based algorithms, one must be cautious because
    biases and scatter in reported \photozs{} increase at fainter magnitudes
    where spectroscopic training and validation samples can be highly
    incomplete.  For red-sequence based cluster finders, one must confront the
    fact that the $4000\,\mathrm{\AA}$ break characteristic of the
    early-type galaxy spectra moves across filters.  While $g-r$ is an ideal
    color for cluster detection at low redshift, one should rely primarily on
    $r-i$ at itermediate redshifts, and $i-z$ at higher redshifts (and we note
    that this will also affect \photoz-based finders).  By $z \approx 1$,
    near-infrared (NIR) photometry is required.  Being able to smoothly
    transition from one color to the next --- or better yet, to always use all
    available photometric data --- is paramount.}
  \item{To the extent possible, the algorithm should self-train to the
    available data.  For instance, algorithms reliant on \emph{a priori}
    parameterizations of the red sequence could easily result in systematic
    biases if the \emph{a priori} parameterization differs from reality.  Note
    that this also impacts \photoz-based cluster finders, since there can be
    unknown and difficult to calibrate biases in the photometric redshifts of
    cluster galaxies.}
  \item{The algorithm should be numerically efficient, capable of running on
    extremely arge data sets within reasonable timeframes with modest
    computational resources.}
  \item{The algorithm must be able to properly account for complex survey
    masks, including varying depth.}
  \item{The algorithm must allow the construction of proper cluster-random
    points that adequately characterize the effective survey volume for cluster
    detection in order to enable large scale structure studies.  In particular,
    it is worth emphasizing that because galaxy clusters are extended objects
    on the sky, the galaxy mask used to construct the cluster catalog is
    \emph{not} the appropriate mask characterizing the angular and redshift
    selection of galaxy clusters for any particular cluster finder.}
  \item{The algorithm should produce a full $P(z)$ distribution for every
    cluster.  Similarly, given that the center of a galaxy cluster can be
    observationally uncertain, there should be a corresponding centering
    distribution in the plane of the sky $P(\hatn)$.  Note that if one adopts
    the prior that a galaxy resides at the center of a galaxy cluster, then the
    probability $P(\hatn)$ collapses to the probability that any given cluster
    galaxy is the correct cluster center.  Our expectation is that just as
    $P(z)$ allows one to adequately recover the redshift distribution of galaxy
    clusters in a statistical sense, so too will centering probabilities for
    cluster galaxies allow one to statistically recover the angular
    distribution of clusters in the sky, a point that is of critical importance
    for large-scale structure studies.}
  \item{In order to maximize the cosmological utility of the derived cluster
    samples, the richness estimators should be fully optimized for the purpose
    of minimizing the scatter in the richness--mass relation.}
\end{enumerate}

The {\bf red}-sequence {\bf Ma}tched-filter {\bf P}robabilistic {\bf
  Per}colation (\redmapper) cluster finding algorithm is our solution to the
above list of must-haves.  Concerning the last point in particular, over the
past several years we have empirically explored what works and what does not
work in estimating cluster richness~\citep[][henceforth
  R12]{rrkmh09,rrknw11,rkrae12}.  For instance, we have demonstrated that
estimating membership probabilities for every galaxy is very effective, while
using hard color cuts to derive cluster membership can lead to large biases.
We have fully optimized the optical detection radii, as well as the luminosity
cuts employed when counting galaxies.  We have also investigated whether total
galaxy counts or total cluster luminosity is a better mass proxy, and whether
or not trying to add blue galaxies into richness estimates results in
improvements.  All of these lessons have gone into the creation of the
\redmapper{} cluster finder.

There, is however, one feature of \redmapper\ that represents more of a
personal bias as opposed to an empirically driven choice, namely, the fact that
\redmapper\ is a red-sequence cluster finder.  Indeed, operationally,
\redmapper\ can be easily adapted to work in \photoz-space rather than working
directly in color-space.  However, we are wary of reliance on photometric
redshifts, which become increasingly difficult to characterize for faint
galaxies due to a lack of spectroscopic training and validation samples.
Furthermore, cluster galaxies are a very particular population, and \photoz{}
estimates tailored for clusters should be derived separately from the total
galaxy population.  Though there have been some studies comparing different
cluster finders \citep[e.g.,][Paper II]{gotoetal02, bahcalletal03b,
  lopesetal04}, we have not seen any conclusive evidence for \photoz-based
algorithms outperforming red-sequence methods or vice-versa.  Here, we have
opted to rely on a red-sequence method when developing \redmapper.  Of course,
at high enough redshift, as the red-sequence begins to disappear, it is obvious
that photometric redshift methods must necessarily perform
better~\citep[e.g.,][]{ebgss08,bsvsg11}.  We do not, however, expect this to be
a problem for near-future large photometric surveys.  Note that while it is
true that \redmapper{} also relies on spectroscopic training samples, an
important advantage of our novel red-sequence modeling alogirhtm is that we do
not require a locally representative training sample: our spectroscopic
training galaxies can be the brightest cluster galaxies at all redshifts, with
no degradation in the performance of our photometric redshift estimates.

The \redmapper{} algorithm is designed to handle an arbitrary photometric galaxy
catalog, with an arbitrary number of photometric bands ($\geq 3$), and will
perform well provided the photometric bands span the $4000\,\mathrm{\AA}$ break
over the redshift range of interest.  It is thus well suited to current surveys
such as the Sloan Digital Sky Survey~\citep[SDSS:][]{yorketal00} for low and
moderate redshift clusters ($z\in[0.05,0.55]$), as well as upcoming surveys
such as DES for low and high redshift clusters ($z<1$).  As a case study, in
this paper we present the \redmapper{} catalog as run on
$10,400\,\mathrm{deg}^2$ of photometric data from the Eighth Data
Release~\citep[DR8:][]{dr8} of the SDSS.  We will make the full DR8
\redmapper{} catalog available after this paper is accepted for publication.

The layout of this paper is as follows.  In Section~\ref{sec:data}, we describe
the SDSS data used for this work, followed in Section~\ref{sec:outline} with an
overall outline of the \redmapper{} cluster finder.  In
Section~\ref{sec:lambdachi} we describe the multi-color richness estimator
$\lambda$, which is an update of the single-color richness estimator used in
R12.  In Section~\ref{sec:holes} we describe our strategy for dealing with
stellar masks and regions of limited depth in the survey.  In
Section~\ref{sec:calibration}, we describe the self-training of the red
sequence parametrization used to detect clusters, as well as measure their
photometric redshifts, which is described in Section~\ref{sec:photoz}.  In
Section~\ref{sec:centering} we describe our new probabilistic cluster centering
algorithm.  Finally, in Section~\ref{sec:clusterfinder} we put all these pieces
together into the \redmapper{} cluster finder.  The resulting SDSS DR8
\redmapper{} cluster catalog is described in Section~\ref{sec:catalog}.  Next,
in Section~\ref{sec:purcomp}, we describe a new, more accurate method of using
the survey data to estimate the purity and completeness of the cluster catalog,
and in Section~\ref{sec:clustermasks} we describe how these methods can be
applied to generate a cluster detection mask over the full sky.  A summary is
presented in Section~\ref{sec:summary}.  In the appendices we present several
systematic checks, including Appendix~\ref{app:ntrain} which contains an
estimate of the minimum number of training spectra required for an accurate
red-sequence calibration.  A full detailed comparison of \redmapper{} and other
large photometric survey catalogs to X-ray cluster catalogs is presented in a
companion paper~\citep[Paper II:][]{rozorykoff14}.  When necessary, distances
are estimated assuming a flat $\Lambda$CDM model with $\Omega_m=0.27$, and
$h=1.0\ \Mpc$, i.e., all quoted distances are in $\hMpc$.


\section{Data}
\label{sec:data}

As discussed above, the \redmapper{} algorithm is designed to handle an
arbitrary photometric galaxy catalog, with an arbitrary number of photometric
bands ($\geq 3$).  Of course, the quality of the output depends on the quality
of the photometry.  As a case study, in this paper we run \redmapper{} on SDSS
DR8 data, due to its large area and uniform coverage.


\subsection{SDSS DR8 Photometry}

The input galaxy catalog for this work is derived from SDSS DR8
data~\citep{dr8}.  This data release includes more than $14,000\,\mathrm{deg}^2$
of drift-scan imaging in the Northern and Southern Galactic caps.  The survey
edge used is the same as that used for Baryon Acoustic Oscillation Survey
(BOSS) target selection~\citep{dsaaa13}, which reduces the total area to
$\approx 10,500\ \deg^2$ with high-quality observations and a well-defined
contiguous footprint.  Similarly, bad field and bright star masks are based
on those used for BOSS.

The BOSS bright star mask is based on the Tycho catalog~\citep{hfmuc00}.
However, this catalog is incomplete at the bright end.  Cross-matching Tycho to
the Yale Bright Star Catalog~\citep{hj91}, covering 9000 of the brightest stars
in the sky (mostly visible to the naked eye), we have found an extra 70 stars
--- including very bright stars such as Arcturus and Regulus --- that obviously
impacted galaxy photometry and detection.  We have also found that very large,
bright galaxies such as M33 cause significant problems for photometry in the
area, including many spurious sources marked as galaxies.  To handle these
issues, we have visually inspected and masked obviously bad regions around 63
objects brighter than $V<10$ from the New General
Catalogue~\citep[NGC:][]{sinnott88} that are in the DR8 footprint, as well as
the bright stars mentioned above.  In total, an additional $36\,\mathrm{deg}^2$
($\sim0.3\%$ of the total area) is removed by our combined bright star and
galaxy mask.  After accounting for all the masked regions, the input galaxy
catalog covers $10,400\,\mathrm{deg}^2$.

As discussed in R12, the careful selection of a clean input galaxy catalog is
required for proper cluster finding and richness estimation.  Our input catalog
cuts are similar to those from \citet{scmbw12} used for BOSS
target selection, with some modifications.  First, we select galaxies as
classified by the default SDSS star/galaxy separator.  We further limit our
catalog to $i<21.0$, approximately the $10\sigma$ limiting magnitude for the
survey.\footnote{Although we note that the limiting magnitude is not precisely constant
  over the survey.}  We then filter all
objects with any of the following flags set in the $g$, $r$, or $i$ bands: {\tt
  SATUR\_CENTER}, {\tt BRIGHT}, {\tt TOO\_MANY\_PEAKS}, and ({\tt NOT BLENDED}
{\tt OR} {\tt NODEBLEND}).  Unlike the BOSS target selection, we have chosen to
\emph{keep} objects flagged with {\tt SATURATED}, {\tt NOTCHECKED}, and {\tt
  PEAKCENTER}. 

Particular care has to be made in avoiding over-aggressive flag cuts because of
the way that the SDSS \emph{photo} pipeline handles dense regions such as
cluster cores.  In these cases, the central galaxy and many satellites may be
originally blended into one object and then deblended.  However, if there is a
problem with one part of the parent object --- such as a cosmic ray hit that is
not properly interpolated --- then this bad flag is propagated to \emph{all}
the children.  We have found that removing objects marked with {\tt SATURATED},
{\tt NOTCHECKED}, and {\tt PEAKCENTER} often mask out cluster centers, while
truly saturated objects such as improperly classified stars are also rejected
via the {\tt SATUR\_CENTER} flag cut.  Overall, by including these objects we
increase the number of galaxies in the input catalog by less than $2\%$, and
our tests have shown no significant effect on the richness measurements except
for a few clusters for which the cores were inadvertently masked out when
galaxies with the above flags were removed. In total, there are 56.5 million
galaxies in the input catalog.

In this work, we use {\tt CMODEL\_MAG} as our total magnitude in the $i$ band,
and {\tt MODEL\_MAG} for $u$, $g$, $r$, $i$, and $z$ when computing colors.  We
limit our input catalog to galaxies that have $\imag < 21.0$, approximately the
$10\sigma$ limit for galaxy detection such that the characteristic magnitude
error for our faintest objects is $\sim0.1$.  The DR8 \"{u}bercalibration
procedure yields magnitude uniformity on the order of 1\% in $griz$ and 2\% in
$u$.  The resulting color scatter introduced by the photometry is significantly
narrower than the width of the cluster red sequence.  All magnitudes and colors
are corrected for Galactic extinction using the dust maps and reddening law of
\citet[SFD:][]{sfd98}.


\subsection{Spectroscopic Catalog}
 
Although our cluster finder uses only photometric data, we require
spectroscopic data to calibrate the red sequence and to validate our
photometric redshifts.  For this purpose we use the SDSS DR9 spectroscopic
catalog~\citep{dr9}.  This spectroscopic catalog has over 1.3 million galaxy
spectra, including over 500,000 Luminous Red Galaxies (LRGs) at $z\sim0.5$ from
the CMASS sample.  As detailed below, we only use $\approx 20\%$ of the
available data in our training, and use the remaining data set to validate our
photometric redshifts.


\section{Outline of the Cluster Finder}
\label{sec:outline}

The \redmapper{} algorithm finds optical clusters via the red sequence
technique.  More specifically, it is built around the optimized richness
estimator $\lambda$ developed in R09 and R12.  The algorithm is divided into
two stages: a calibration stage, where we empirically calibrate the properties
of the red sequence as a function of redshift; and a cluster-finding stage,
where we utilize our calibrated model to identify galaxy clusters and measure
their richness.  The
algorithm is iterative.  First, an initial rough color calibration is used to
identify clusters.  These clusters are then used to better calibrate the
red sequence, which enables a new cluster finding run \citep[see also][for
a similar approach within the context of cluster finding with spectroscopic
data sets]{blackburnekochanek12}.  These two
calibration/cluster finding stages are iterated several times before a final
cluster finding run is made.

The calibration itself is also an iterative procedure described in detail in
Section~\ref{sec:calibration}.  We start with a set of ``training clusters''
that have a red central galaxy with a spectroscopic redshift to calibrate the red
sequence model.  As we show in Appendix~\ref{app:ntrain}, our minimal training
requirements for unbiased cluster richness and photometric redshift estimation
are $\approx 40$ clusters per redshift bin of width $\pm 0.025$.  In the case
of SDSS DR8, the spectroscopic availability greatly surpasses this requirement
by many orders of magnitude.  However, for upcoming surveys such as DES probing
much higher redshifts, this will no longer be the case, and we have developed
\redmapper{} with these limitations in mind.

For the present work on DR8, we construct our sample of training clusters from
red spectroscopic galaxies.  These red galaxies are used as ``seeds'' to look
for significant overdensities of galaxies of the same color.  The significant
overdensities thus become our training clusters that are used to fit a full
linear red-sequence model (including zero-point, tilt, and scatter) to the
sample of all high-probability cluster members with a luminosity $L\geq
0.2L_*$.  This luminosity threshold is optimal for richness measurements (R12).
In this way we effectively transfer the ``seed'' spectroscopic redshift to all
high probability cluster members, which enables a much more accurate
measurement of the red sequence.  This is especially true for fainter
magnitudes where there is very limited spectroscopic coverage.  Note that
because the algorithm utilizes all colors simultaneously, the ``scatter'' is
characterized not by a single number but by a covariance matrix.

Given a red-sequence model, the cluster finding proceeds as follows (see
Section~\ref{sec:clusterfinder} for details).  First, we consider all
photometric galaxies as candidate cluster centers (thereby assuming the center
of a cluster is located at a galaxy position).  We use our red-sequence model
to calculate a photometric redshift for each galaxy ($\zred$; see
Section~\ref{sec:zred}), and evaluate the goodness of fit of our red galaxy
template.  Galaxies that are not a reasonable fit to the red galaxy template at
any redshift are not considered as possible central galaxies for the purpose of
cluster ranking.  We note that as long as a cluster has at least \emph{one}
galaxy that is a reasonable fit to the template, that cluster will be
considered in the first step of the cluster-finding stage.  We then use the
$\zred$ value of the candidate central galaxy as an initial guess for the cluster
redshift.  Based on this redshift, we compute the cluster richness $\lambda$
and its corresponding likelihood using a multi-color generalization of the
method of R12 (see Section~\ref{sec:lambdachi}).  When a significant number of
red-sequence galaxies ($\geq 3$) is detected within a $500\,\kpc$ aperture, we
re-estimate the cluster redshift by performing a simultaneous fit of all the
high probability cluster members to the red sequence model.  This procedure is
iterated until convergence is achieved between member selection and cluster
photometric redshift ($\zlambda$; see Section~\ref{sec:zlambda}).  The
resulting list of candidate cluster centers is then rank-ordered according to
likelihood.  Starting with the highest ranked cluster we measure the richness
and membership probabilities.  These probabilities are then used to mask out
member galaxies for lower-ranked clusters in a process we term
``percolation'' (see Section~\ref{sec:cf:percolation}).  In this way we prevent double-counting of galaxy clusters.


\section{Richness Estimator $\lambda$}
\label{sec:lambdachi}

The \redmapper{} richness estimator, $\lambda$, is a
multi-color extension of the richness estimator developed 
in R09 and R12, which we now denote
$\lcol$ to indicate that it is a single-color richness.  Here we review how
we calculate $\lambda$ and highlight the differences relative to R12.

Let $\bm{x}$ be a vector describing the observable properties of a galaxy
(e.g., multiple galaxy colors, $i$-band magnitude, and position).  We model the
projected distribution within and around clusters as a sum $S(\bx)=\lambda
u(\bx|\lambda)+b(\bx)$ where $\lambda$ is the number of cluster galaxies,
$u(\bx|\lambda)$ is the density profile of the cluster normalized to unity, and
$b(\bx)$ is the density of background (i.e., non-member) galaxies.  The
probability that a galaxy found near a cluster is actually a cluster member is
simply
\begin{equation}
\pmem = p(\bx) = \frac{\lambda u(\bx|\lambda)}{\lambda u(\bx|\lambda)+b(\bx)}.
\label{eqn:pmem}
\end{equation}
We note that in Section \ref{sec:cf:percolation}, the definition of the membership
probability will be modified to allow for proper percolation of the cluster finder.
This modification will only impact clusters that are close to each other along
the line of sight and at comparable redshifts.
Regardless, the total number of cluster galaxies $\lambda$ must satisfy the
constraint equation
\begin{equation}
\lambda  = \sum p(\bx|\lambda) = \sum_{R<\Rc(\lambda)} \frac{\lambda u(\bx|\lambda)}{\lambda u(\bx|\lambda)+b(\bx)}.
\label{eqn:lambdadef}
\end{equation}
The corresponding statistical uncertainty in $\lambda$ is given by
\begin{equation}
\mbox{Var}(\lambda) = \sum p(\bx|\lambda)\left[ 1-p(\bx|\lambda) \right].
\label{eqn:error}
\end{equation}
In principle, these sums should extend over all galaxies.  In practice, one
needs to define a cutoff radius $R_c$ and a luminosity cut
$L_{\mathrm{cut}}$.  In R12 and \citet{rrknw11} we showed that the scatter in the
mass--richness relation is expected to be minimized when $L_{\mathrm{cut}} = 0.2L_*$,
while the optical radial cut scales with richness via  
\begin{equation}
  \Rc(\lambda) = R_0(\lambda/100.0)^\beta.
  \label{eqn:radius}
\end{equation}
where $R_0 =1.0\,\hMpc$ and $\beta=0.2$.  We adopt these parameters
in \redmapper.

To determine the cluster richness of a galaxy cluster, note that $\lambda$ is
the only unknown in Eqns.~\ref{eqn:lambdadef} and \ref{eqn:radius}.  Therefore,
we can numerically solve Eqn.~\ref{eqn:lambdadef} for $\lambda$ using a
zero-finding algorithm.  The solution to Eqn.~\ref{eqn:lambdadef} defines
$\lambda$, and naturally produces a cluster radius estimate $\Rc$ via
Eqn.~\ref{eqn:radius}.  We emphasize this cluster radius is not a proxy for any
sort of standard overdensity radius such as $R_{500c}$ or $R_{200m}$.\footnote{
$R_{500c}$ ($R_{200m}$) is the radius enclosing an overdensity of 500 (200) times the critical (mean) 
density of the Universe.}

As in R12, we consider three observable properties of galaxies for our filter
function $u(\bx)$: $R$, the projected distance from the cluster center;
$\imag$, the galaxy $i$-band magnitude, and a color variable.  Ideally, our
color variable would be the full color vector (e.g., $\bc=\{u-g,g-r, r-i,i-z\}$ in
the case of SDSS data).  However, practical considerations forced us to reduce
this information to a single $\chi^2$ value which gives the goodness of fit of
our red-sequence template.  In doing so, we effectively compress the
information contained in the multi-dimensional color vector into a single
number that measures the ``distance'' in color space between the galaxy of
interest and our red sequence model.  This is described in more detail below.

We adopt a separable filter function
\begin{equation}
  u(\bx) = [2\pi R \Sigma(R)]\phi(\imag)\rnc,
\end{equation}
where $\Sigma(R)$ is the two dimensional cluster galaxy density profile,
$\phi(m)$ is the cluster luminosity function (expressed in apparent
magnitudes), and $\rnc$ is the $\chi^2$ distribution with $\nu$ degrees of
freedom.  The pre-factor $2\pi R$ in front of $\Sigma(R)$ accounts for the fact
that given $\Sigma(R)$, the radial probability density distribution is $2\pi R
\Sigma(R)$.  We summarize below the filters used in \redmapper.


\subsection{The $\chisq$ Filter}
\label{sec:chi2filter}

Assume we have a multicolor red sequence model for which we have $\left
<\bc|z,m_i \right >$, the mean color of the red sequence galaxies for any given
redshift $z$ and $i$-band magnitude $m_i$.  Furthermore, assume we have a corresponding
covariance matrix $\bC_{\mathrm{int}}(z)$ to describe the intrinsic scatter and
correlations of galaxy colors about the mean.

When comparing a given galaxy with color vector $\bc$ to the model color, with
the assumption of Gaussian errors the distribution of galaxies will be
represented by a $\chisq$ distribution:
\begin{equation}
  \chisq(z) = \left ( \bc - \cmodbf \right ) \left (
  \bC_{\mathrm{int}}(z) + \bC_{\mathrm{err}} \right )^{-1} \left ( \bc - \cmodbf \right )
  \label{eqn:chi2}
\end{equation}
where $\bc$ is the color vector of the galaxy under consideration, $m_i$ is
the observed galaxy magnitude, $\cmodbf$ 
is the model color, and $\bC_{\mathrm{int}}(z)$ is the corresponding
covariance matrix, which itself depends on redshift.  The matrix
$\bC_{\mathrm{err}}$ describes the photometric error of the galaxy under
consideration.

For red sequence cluster members, $\chi^2$ will be distributed according
to the $\chi^2$ distribution with $\nu$ degrees of freedom,
\begin{equation}
  \rnc = \frac{(\chisq)^{(\nu/2 - 1)} e^{-\chisq/2}}{2^{\nu/2}\Gamma(\nu/2)},
  \label{eqn:rnc}
\end{equation}
where $\nu$ is the number of colors employed when estimating $\chisq$.  Note
that for $\nu=1$ the $\chisq$ filter does not reduce to the single color filter
of R12.  This is because our distance measurement $\chi^2$ does not
distinguish between galaxies that are too red from galaxies that are too blue,
so there is some loss of information when moving from color-space to $\chi^2$.
While a full $\nu$-dimensional Gaussian color filter
would work better than our $\chi^2$ filter --- and would exactly reduce
to the single color $\lcol$ from R12 when $\nu=1$ ---   the problem
of background estimation for such a filter is much more difficult.  In
particular, in the case of DR8, it requires
one to estimate the galaxy density in a five dimensional space: $\{\imag,u-g,g-r,r-i,i-z\}$.  We found these background estimates to be very noisy,
so we compressed the color information to a single variable $\chisq$.
In this way, at any given redshift the background depends only on 
$\imag$ and $\chisq$.


\subsection{The Radial and Luminosity Filters}
\label{sec:nfwandlumfilter}

For the radial filter, we follow R09 and R12 and adopt a projected NFW
profile~\citep{navarroetal95}, which is a good description of the dark matter
profile in N-body simulations.  In addition, it has been found to be a good
description of the radial distribution of cluster
galaxies~\citep{linmohr04,hmwas05,popessoetal07}.  In R12 it was shown that in
order to minimize the scatter in the mass--richness relation the NFW
filter works as well or better than other possible radial profiles.  Therefore,
we refer readers to Section 3.1 of R12 for details on the form of the radial
filter.

For the luminosity filter, we similarly follow R09 and R12 and adopt a
Schechter function~\citep[e.g.,][]{hswk09}, written as
\begin{equation}
\phi(\imag) \propto 10^{-0.4(\imag-m_*)(\alpha+1)}\exp\left(-10^{-0.4(\imag-m_*)}\right).
\label{eqn:lumfilter}
\end{equation}
In an update from R12, we have set $\alpha = 1.0$ independent of redshift,
which provides a better description of the data.  The
characteristic magnitude, $m_*$, is the same as used in R12, calculated for a
k-corrected passively evolving stellar population~\citep{kmawe07a}.  In the
redshift range $0.05<z<0.7$, appropriate for DR8, $m_*(z)$ is
well approximated ($\delta < 0.02\,\mathrm{mag}$) by the following polynomials:
%
\begin{equation*}
  \label{eqn:mstar}
  m_*(z) =
  \begin{cases}
    22.44 + 3.36\ln(z) + 0.273\ln(z)^2 & \text{if $z \leq 0.5$,}
    \\
    \,\,- 0.0618\ln(z)^3  - 0.0227\ln(z)^4 &
    \\
    22.94 + 3.08\ln(z) - 11.22\ln(z)^2  & \text{if $z > 0.5$.}
    \\
    \,\,- 27.11\ln(z)^3 - 18.02\ln(z)^4
  \end{cases}
\end{equation*}
For each cluster, $m_*$ is taken at the appropriate redshift and the luminosity
filter is normalized to unity at the appropriate magnitude cutoff.
As with R12,
this is taken to be $0.2L_*$, or $m_* + 1.75\,\mathrm{mag}$. Although in the
current version of \redmapper\ both $\alpha$ and $m_*$ are fixed
as described above, in future releases we will replace these parameters
with those directly measured from calibration clusters.  We emphasize, however,
that modest changes to the shape of the luminosity filter result in insignificant changes
to the recovered richness.  Of course, changes to the magnitude limit above which one
counts galaxies has an obvious systematic impact on the richness as one moves up and
down the luminosity function, although we have found these modest shifts no not
signifcantly impact the mass--richness scatter~(see R12).


\subsection{Background Estimation}
\label{sec:background}

As in R12, we assume that the background density\footnote{We use the term
  ``background'' to mean all unassociated galaxies, including objects behind
  and in front of the cluster.} is uniform, such that
$b(\bx|z) = 2\pi R \barsigg(\imag,\chisq|z)$ where $\barsigg(\imag,\chisq|z)$
is the galaxy density as a function of galaxy $i$-band magnitude and $\chisq$,
where $\chisq$ is evaluated using the red sequence model at redshift $z$.  In
this way, the \emph{effective} background for every cluster is different and depends
on the cluster redshift.  Note that since clusters reside in overdense regions,
this mean background is likely an underestimate of the true background,
but the incurred bias is small \citep[see R12 and][]{rrknw11}, and is irrelevant
for cosmological purposes, as it is completely degenerate with the amplitude
of the richness--mass relation.  Variability of the background (which is expected)
is more problematic, and can lead to catastrophic projections.  These are
expected to be small \citep{rrknw11}, and will be addressed in a future paper (Rozo et al., 
in preparation).

To calculate the mean galaxy density, we first calculate the $\chisq$ value for
all galaxies in a grid of redshifts with spacing 0.02.  For computation
purposes, we only calculate the $\chisq$ for galaxies that are brighter than
$0.1L_*$ at a given redshift bin.  This implies that the magnitude range being
sampled is different for each redshift bin.  At each redshift we bin the full galaxy
catalog in $\chisq$ and magnitude using a cloud-in-cells (CIC)
algorithm~\citep[e.g.,][]{hockney81}, and divide by the survey area.  For our
cells, we use $\chisq \in [0,20]$ with a bin size of 0.5, and $i \in
[12,m_\mathrm{lim}]$ with a bin size of 0.2.  The $\chisq<20$ cut can be
justified by the fact that the sum total of cluster membership probability in
the \redmapper{} catalog for galaxies with $\chisq\in[15,20]$ is only $0.7\%$.
That is, our $\chisq<20$ cut impacts our results at well below the $1\%$ level.
The final galaxy number density is normalized by the width of each color and
magnitude bin.  To evaluate the background at an arbitrary redshift, we
linearly interpolate between the backgrounds computed along our redshift grid.
As noted in R12, because the background is measured per square degree, the
average number of background galaxies as a function of $\chisq$, magnitude, and
redshift is automatically accounted for as the angular size of the clusters
changes with redshift.


\section{Handling Masked Regions and Limited Depth}
\label{sec:holes}

In an ideal world, our survey would have uniform depth, be deep enough
to reach $0.2L_*$ at all redshifts of interest, and there would be no missing 
and/or masked regions, e.g., due to bright stars.  Most previous optical cluster
finders make this simple assumption\footnote{An exception is
  3DMF~\citep{milkeraitisetal10}, for which they calculate the fractional area
  masked for each cluster}. Here, we describe how we can properly
correct for these effects.  Our approach is conceptually straightforward.  Given a
  cluster model and a geometric and magnitude mask, we can effectively
  calculate the fraction of cluster galaxies that we expect to be masked out.
  This correction factor is then applied to the ``raw'' richness to compensate
  for the masked region. In practice, this correction can be self-consistently incorporated
  into the richness estimation, as described below. 
   Our method is simple to implement with any geometric
  mask, including those that describe variations in depth.  However, we do not
  take into account masks that contain one or more missing bands.


\subsection{The Correction Term}
\label{sec:corrterm}

Looking back at Eqn.~\ref{eqn:lambdadef}, we have:
\begin{equation}
  1 = \sum_i \frac{u(\bx_i)}{\lambda u(\bx_i) + b(\bx_i)},
\end{equation}
where $\bx_i$ describes the radial position, color (via $\chisq$), and
luminosity (via $\imag$) of each galaxy.  This formulation works if
we can see all galaxies, but in reality we cannot.  Let us then pixelize all
observable space $\bx$ into infinitesimal pixels, and let $N_i$ be the number
of galaxies in pixel $i$.  Most pixels have $N_i=0$, but a few have $N_i=1$.
Thus, the sum over all galaxies can be re-written in terms of a sum over all
pixels via:
\begin{equation}
  1 = \sum_i N_i \frac{u(\bx_i)}{\lambda u(\bx_i)+b(\bx_i)}.
\end{equation}
In the case of masking, we can only observe the galaxies that are inside the
mask, so we can split this sum into:
\begin{equation}
  1 = \sum_{\mathrm{in}} N_i \frac{u(\bx_i)}{\lambda u(\bx_i) + b(\bx_i)} +
  \sum_{\mathrm{out}}N_i \frac{u(\bx_i)}{\lambda u(\bx_i) + b(\bx_i)}.
\end{equation}
The ``in'' term is the raw $\lambda$ that we usually compute, and can be
replaced by the standard sum over all observed galaxies.  The ``out'' term is
now a correction to the standard expression, call it $C$,
\begin{equation}
  C = \sum_{\mathrm{out}}N_i \frac{u(\bx_i)}{\lambda u(\bx_i) + b(\bx_i)}.
\end{equation}
In terms of $C$, Eqn.~\ref{eqn:lambdadef} can be rewritten as
\begin{equation}
  1 - C = \sum_{\mathrm{gals}} \frac{u(\bx)}{\lambda u(\bx) + b(\bx)}.
\end{equation}

Now, while $C$ is unknown (we cannot see the masked region), we can
compute its expected value for a cluster of richness $\lambda$.
Using the fact that
\begin{equation}
  \left < N_i \right > = [\lambda u(\bx_i) + b(\bx_i)]\Delta \bx_i,
\end{equation}
we see that the expectation value of $C$ is given by 
\begin{equation}
  \left < C  |\lambda \right > = \int_{\mathrm{out}} d\bx\ u(\bx).
  \label{eqn:meanc}
\end{equation}
In the above equation, we have made explicit the fact that $C$ depends on
$\lambda$, both via the cutoff radius used in the sum over galaxies, and because
the radial filter depends on $\lambda$.  Thus, in the presence of missing data,
our richness estimate is given by the solution to
\begin{equation}
  1 - \avg{C|\lambda} = \sum_{\mathrm{gals}} \frac{u(\bx)}{\lambda u(\bx) + b(\bx)}.
  \label{eq:Clambda}
\end{equation}

Note, however, that because $C$ is a unknown, there must also be additional
measurement error associated with this unknown correction.
To calculate the variance of $C$, we note that $\mathrm{Var}(N_i) = \left < N_i
\right >$.  To compute $\mathrm{Cov}(N_i,N_j) = \left <N_i N_j \right > - \left
< N_i \right > \left <N_j \right >$, we first compute $\left <N_i N_j \right
>$.  For infinitesimal pixels, $N_i N_j = 1$ implies that either both pixels
$i$ and $j$ contain cluster galaxies, or that one pixel contains a cluster
galaxy while the other pixel contains a background galaxy, or that both pixels
contain a background galaxy.  Consequently,
\begin{equation}
  P(N_i N_j = 1) = \Delta\bx^2 [ \lambda(\lambda-1)u_i u_j + \lambda(u_i b_j +
    u_j b_i) + b_i b_j].
\end{equation}
Since $\left < N_i N_j \right > = P(N_i N_j = 1)$, subtracting off $\left < N_i
\right > \left <N_j \right >$ we arrive at
\begin{equation}
  \mathrm{Cov}(N_i,N_j) = -\Delta \bx^2 u_i u_j
\end{equation}
for $i \neq j$.  Putting it all together, we arrive at
\begin{eqnarray}
\mathrm{Var}(C) & = & \sum_{i} \avg{N_i} p_i - \sum_{i}\sum_{j\neq i} \mathrm{Cov}(N_i,N_j) p_i p_j \\
	& = &  \int_{\mathrm{out}} d\bx\ u(\bx)p(\bx) - \left[ \int_{\mathrm{out}} d\bx\ u(\bx) p(\bx) \right]^2.
\label{eqn:varc}	
\end{eqnarray}

To propagate the error
in $C$ into the error in $\lambda$, we set
\begin{equation}
  \sigma_\lambda = \frac{d\lambda}{dC}\bigg\vert_{\lambda(\left < C \right >)}
  \sigma_C,
  \label{eqn:siglambda}
\end{equation}
where $\sigma_C = [\mathrm{Var}(C)]^{1/2}$.  
The derivative of $\lambda$ with respect to $C$ is evaluated numerically about the
expectation value of $\avg{C|\lambda}$, where $\lambda$ is the solution to Eqn.~\ref{eq:Clambda}.

For future reference, it will be useful to define the ``scale factor''
\be
S(z) = \frac{1}{1-\avg{C|z}} \label{eqn:sz}
\ee
for the case in which the only source of masking is due to limited depth.  With this definition,
a cluster with richness $\lambda$ has a total of $\lambda/S(z)$ galaxies above the
limiting magnitude of the survey.

In addition, it is useful to calculate the fraction of the effective cluster area that is masked
solely by geometrical factors such as bright stars, bad fields, and survey
edges.  This is complementary to $S(z)$ defined above in that it contains all
the masking \emph{except} the magnitude limit.  The cluster mask fraction is then
\begin{equation}
  \fmask = \frac{\int_{\mathrm{out}}dx\ u(x)}{\int dx\ u(x)}.
\label{eqn:maskfrac}
\end{equation}
This quantity is very useful because clusters that are strongly affected by edges
are more likely to be catastrophically miscentered, and to have poor
richness estimates.  Consequently, when defining our cluster catalog we will
apply a cut in $\fmask$.


\subsection{Evaluating the Mask Correction}
\label{sec:correval}

Evaluation of the mask correction and its associated error on $\lambda$ can be
difficult.  Fortunately, this problem is well suited to Monte Carlo
integration.  First, define the selection function ${\cal S}(\bx)$, so that
${\cal S}(\bx)=1$ if the galaxy is in a region where it is detected, and ${\cal S}(\bx)=0$ if
the position $\bx$ is masked out.  Then, for any function $f(x)$,
\begin{equation}
  \int_{\mathrm{out}} d\bx\ f(\bx) = \int_{\mathrm{cluster}}
      [1-{\cal S}(\bx)]f(\bx)d\bx,
      \label{eqn:intf}
\end{equation}
where the integral on the right hand side is over the full cluster region,
i.e., $R \in [0,R(\lambda)]$, $L \geq L_{\mathrm{min}}$, and over all colors.

Applying this to Eqn.~\ref{eqn:meanc} we find
\begin{equation}
  \left < C \right > = \int_{\mathrm{cluster}} [1-{\cal S}(\bx)] u(\bx) d\bx.
\end{equation}
Since $u(\bx)$ is the probability distribution for $\bx$, we can evaluate the above integral using
Monte Carlo integration by randomly sampling $N$ sets of model galaxy parameters from the $u(\bx)$ filter function,
and then computing the sample mean of the function $1-{\cal S}(\bx)$.  That is, $\avg{C}$ is simply
the fraction of random draws that fall in the masked region.
Similarly, we can evaluate the integrals defining $\mathrm{Var}(C)$ via
\begin{equation}
  \begin{split}
  \mathrm{Var}(C) = \frac{1}{\lambda}\frac{1}{N}\sum_{\mathrm{out}} p(\bx_i) - 
    \left [ \frac{1}{\lambda}\frac{1}{N}\sum_{\mathrm{out}} p(\bx_i) \right ]^2,
  \end{split}
\end{equation}
where $N$ is again the total number of random draws.  For simplicity, we have
replaced the $1-{\cal S}(\bx)$ terms with a summation
over all galaxies that are outside the detectable region due to the mask, as
these are the only galaxies that contribute to the summand.

One of the slowest part of this process is drawing random
realizations of $\bx_i$.  However, these random draws do not need to be
independent from cluster to cluster.  In practice, we generate a template
distribution of 5000 galaxies, and scale the radius and magnitude 
to the appropriate values for each galaxy cluster.  We find that this number of
galaxies gives accurate results for the recovered richnesses and richness
errors, except for galaxy clusters that are largely masked out.  Consequently,
our final cluster selection criteria includes the requirement that the cluster
mask fraction, $\fmask$, is less than 20\%.  In Appendix~\ref{sec:corrtest} we
demonstrate with DR8 data that given our filter function this formalism
accurately corrects for masked regions and limited depth.


\section{Calibration of the Red Sequence}
\label{sec:calibration}

\subsection{Outline}

Suppose we have a complete sample of red galaxies with spectroscopic redshifts
down to the limiting magnitude of the survey.  One can then directly fit a
red-sequence model to these galaxies to calibrate the color as a function of
redshift for these galaxies.  The question then becomes: how does one get a
sample of red spectroscopic galaxies?  Note that in order to calibrate the
\emph{tilt} of the red sequence, it is important to include a significant
number of less luminous galaxies, which are more difficult to come by.

Our solution is to simply use the cluster members themselves.  If we know the
spectroscopic redshift of a cluster, then all the cluster members can be tagged
with the spectroscopic redshift of their central galaxy, leveraging one
spectroscopic redshift into many.  Of course, from photometric data one can
only identify \emph{likely} cluster members, so the fit of the red-sequence
model must account for contamination by non-cluster members.  

In order to fit the red-sequence model, all that is required is a sample of
galaxy clusters with known spectroscopic redshifts.  As discussed in
Section~\ref{sec:outline}, we only require a limited set of these training
clusters.  These training clusters can be derived from external X-ray and SZ
catalogs, or from spectroscopic follow-up of likely centrals in dense regions
by running \redmapper{} with an ad-hoc red-sequence model.

In the specific case of SDSS DR8, we construct this training cluster sample
based on existing spectroscopic galaxy samples.  Each spectroscopic galaxy is
used as a ``seed'' around which we look for galaxy clusters by identifying
nearby overdensities of galaxies with the same color as the seed galaxy.  The
details of these steps are described below.  We emphasize that our calibration
is performed using only $2,000\,\mathrm{deg}^2$ of SDSS data.  As we show in
Appendix~\ref{app:ntrain}, we show that while we have the full wealth of SDSS
spectroscopic data at our disposal, an equivalent red-sequence model may be
derived from only 400 central galaxies from $z\in[0.05,0.6]$.


\subsection{Selecting Seed Galaxies and the Initial Color Model}
\label{sec:seltrain}

The initial calibration of the red sequence relies on spectroscopic ``seed''
galaxies.  This may simply be a set of training clusters with
spectroscopy (see Appendix~\ref{app:ntrain}), or in the case of DR8, a broad
spectroscopic sample that contains a sufficient number of red galaxies in galaxy
clusters.  For SDSS spectroscopy, the first step is to identify the subsample
of spectroscopic galaxies that are red.  This is achieved by using a single
color that samples the $4000\mathrm{\AA}$ break for early type galaxies.  With
SDSS data, we use $g-r$ for $z<0.35$, and $r-i$ for $z>0.35$.  Because we wish
to have a relatively clean selection of red-sequence galaxies as our seeds, we
approach the problem of selecting these galaxies in several steps.  We
emphasize that some of these steps are only necessary for cutting the full list
of SDSS spectra to an appropriate red galaxy sample.

\emph{Step 1: Perform an approximate red galaxy selection.} To make this
selection, we bin the galaxies in redshift bins of width $\pm 0.025$.  We then
use the error-corrected Gaussian Mixture Method~\citep{hkmrr09} to estimate the
mean and intrinsic width $\sigint(z)$ of the red-sequence galaxies.  Those
galaxies within $2\sigma$ of the expected mean for red galaxies are considered
approximately red.

\emph{Step 2: Use this approximate red galaxy selection to measure the mean
  color of the red galaxies as a function of redshift.}  Given our approximate red galaxy
sample, we refine our initial estimate of the mean color--redshift relation by
minimizing the function
\begin{equation}
  s = \sum_i |c_i - \medcoli|
  \label{eqn:median}
\end{equation}
where $\medcol$ is our model for the color as a function of redshift.
The function $\medcol$ is defined via spline interpolation, and
the value of the spline nodes are the parameters with respect to which
$s$ is minimized.  The spline nodes
are placed on a redshift grid with spacing of 0.1.  
Our use of $\medcol$ indicates that this is our
early calibration model, which is distinct from the full model color
$\avg{c|z}$ that is derived at the end of the calibration procedure. In defining
the function $s$, we rely on the sum of absolute values rather than the sum of
the squares to make the resulting minimization more robust to gross outliers.
The function $s$ is minimized using the downhill-simplex method of
\citet{nelder65} as implemented in the IDL \texttt{AMOEBA} function.

\emph{Step 3: Use the mean color--redshift relation to estimate the width of the
  color--redshift relation.}  We can now improve upon our initial estimate of
the width of the color--redshift relation by minimizing the function
\begin{equation}
  s = \sum_i \left | |c_i - \medcoli| - MAD \right |,
  \label{eqn:madsig}
\end{equation}
where $\madsigint = 1.4826\times MAD$, where $MAD$ is the median absolute
deviation of the sample about the median, and the factor of $1.4826$ relates
the $MAD$ to the standard deviation for a Gaussian distribution.  The value
$MAD$ is again defined via spline interpolation, with the free parameters being
the values of the function at the nodes.

\emph{Step 4: Generate a final sample of seed galaxies.}  Finally, with the
full red spectroscopic galaxy model in hand ($\medcol$, $\madsigint$), we can
cleanly select our seed sample.  We select all galaxies within $2\madsigint$ of
the model color at the spectroscopic redshift of the galaxy, using $g-r$ at
$z<0.35$ and $r-i$ at $z>0.35$.

\begin{figure}
  \begin{center}
    \scalebox{1.2}{\plotone{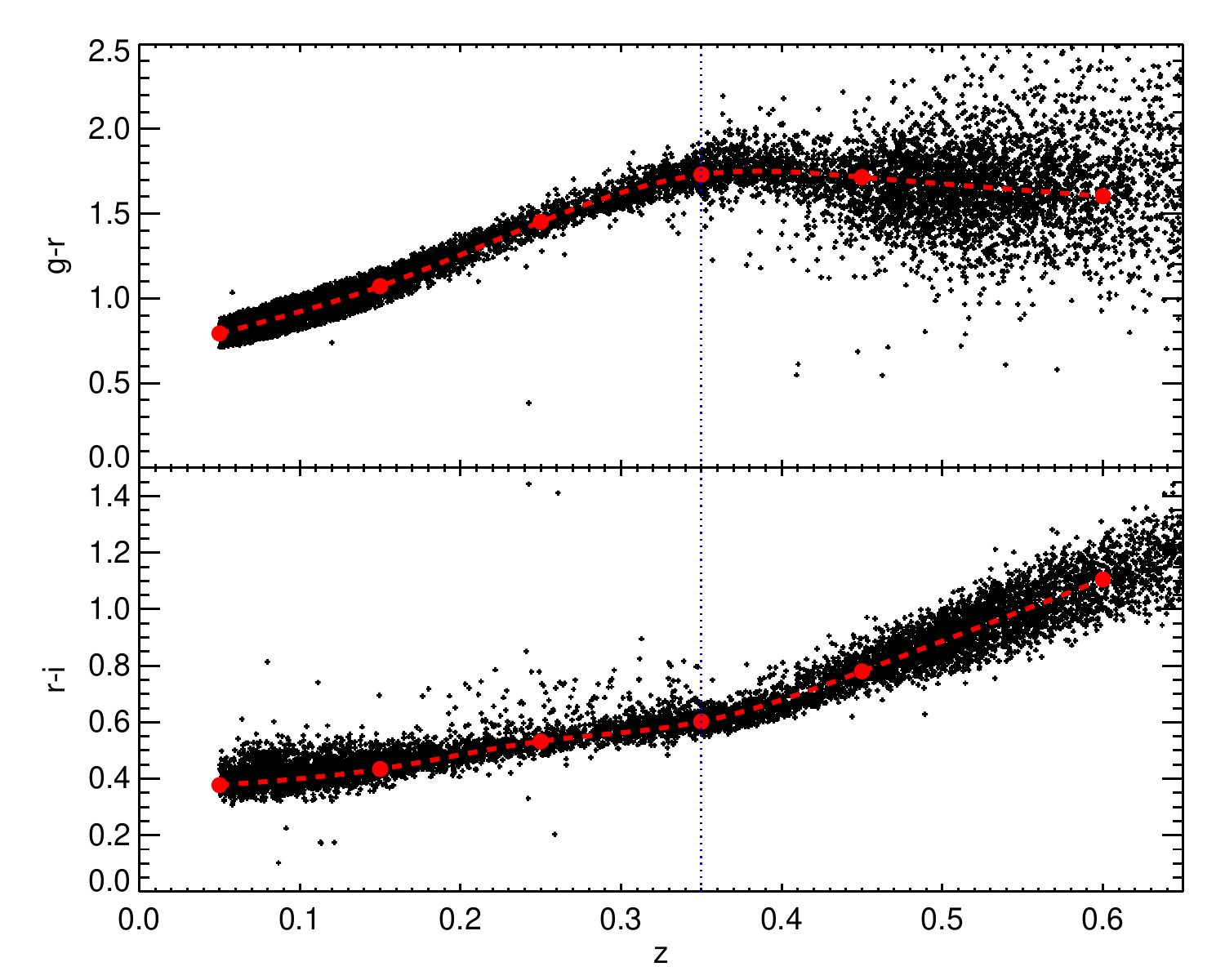}}
    \caption{Sample of red spectroscopic galaxies selected for training in 2000
      $\mathrm{deg}^2$ of DR8.  Top panel shows $g-r$ color and bottom panel
      $r-i$ color.  The red galaxy selection is done in $g-r$ ($r-i$) at $z<0.35$
      ($z>0.35$), selecting all galaxies within $2\madsigint$ of the
      spectroscopic redshift of the galaxy.  Note that this selection leaves a
      small number of outliers in the complementary color, as well as a small
      number of large outliers with anomalously large photometric errors.
      }
    \label{fig:redgals}
  \end{center}
\end{figure}

In Figure~\ref{fig:redgals} we show the final seed spectroscopic galaxy selection
for the $g-r$ and $r-i$ colors.  The large red points show the median colors at
the node positions, and the dashed red lines show the cubic spline
interpolation.  Note that the single-color selection leaves a small number of
outliers in the complementary color.  In addition to the seed galaxies, we will
make use of our red spectroscopic galaxy color model in the following section.

\subsection{Single Color Member Selection}

Having selected our seed galaxies and calibrated a rough initial
color--redshift relation, we now proceed to find likely cluster members around each of
our seed galaxies.  For this first iteration we rely on single-color based
membership.  Specifically, in R12 we demonstrated that for moderately rich
($\lambda \gtrsim 20$) clusters, one can reliably estimate the red sequence
directly from the data as follows.  First, we select all galaxies within a
color window around the seed galaxy.  Next, we fit for the amplitude and tilt
of the red-sequence of that galaxy cluster directly from the galaxy data.
However, in
extending this algorithm to high redshift, we found that large photometric
errors can introduce an unacceptable amount of noise in
the initial color-box selection of galaxies.  Therefore, rather than drawing a
color box around the color of the \emph{central galaxy} for the initial fit, we draw the color box around
the \emph{model color} $\medcol$ calibrated in the previous section.  In detail, we:

\begin{enumerate}
  \item{Take a red galaxy of known spectroscopic redshift (the ``seed'').}
  \item{Select all galaxies within $500\,\kpc$ of the spectroscopic galaxy,
    as well as $2\sigma$ of the model color determined in
    Section~\ref{sec:seltrain}.}  For the model color, we use
    $g-r$ at $z\leq 0.35$ and $r-i$ at $z\geq 0.35$.  The width $\sigma$
    of the color box is set to $0.05$ and $0.03$ respectively, which we expect
    to be the approximate red sequence width (e.g., R12).
  \item{Fit the red sequence (slope and intercept) of these galaxies.}
  \item{Measure the single-color $\lcol$ using the method of R12 and a
    fixed aperture of $500\,\kpc$.}
  \item{For all overdensities with $\lcol > 10$, take the galaxies with
    non-zero $\pmem$ and assign them the spectroscopic redshift of the initial
    seed galaxy.  In practice, we limit our analysis to those galaxies with $\pmem>0.7$.}
\end{enumerate}

At this point, we have leveraged the spectroscopic seed galaxies to generate a
set of red galaxies as faint as $0.2L_*$ over the redshift range of interest.
Although not all of these galaxies are true cluster members, we have an
estimate $\pmem$ of the probability that each such galaxy is indeed a red
sequence cluster member, as in Eqn.~\ref{eqn:pmem}.  Consequently, we can model
the contamination of non-red-sequence galaxies in our sample, as shown below.

\begin{figure}
  \begin{center}
    \scalebox{1.2}{\plotone{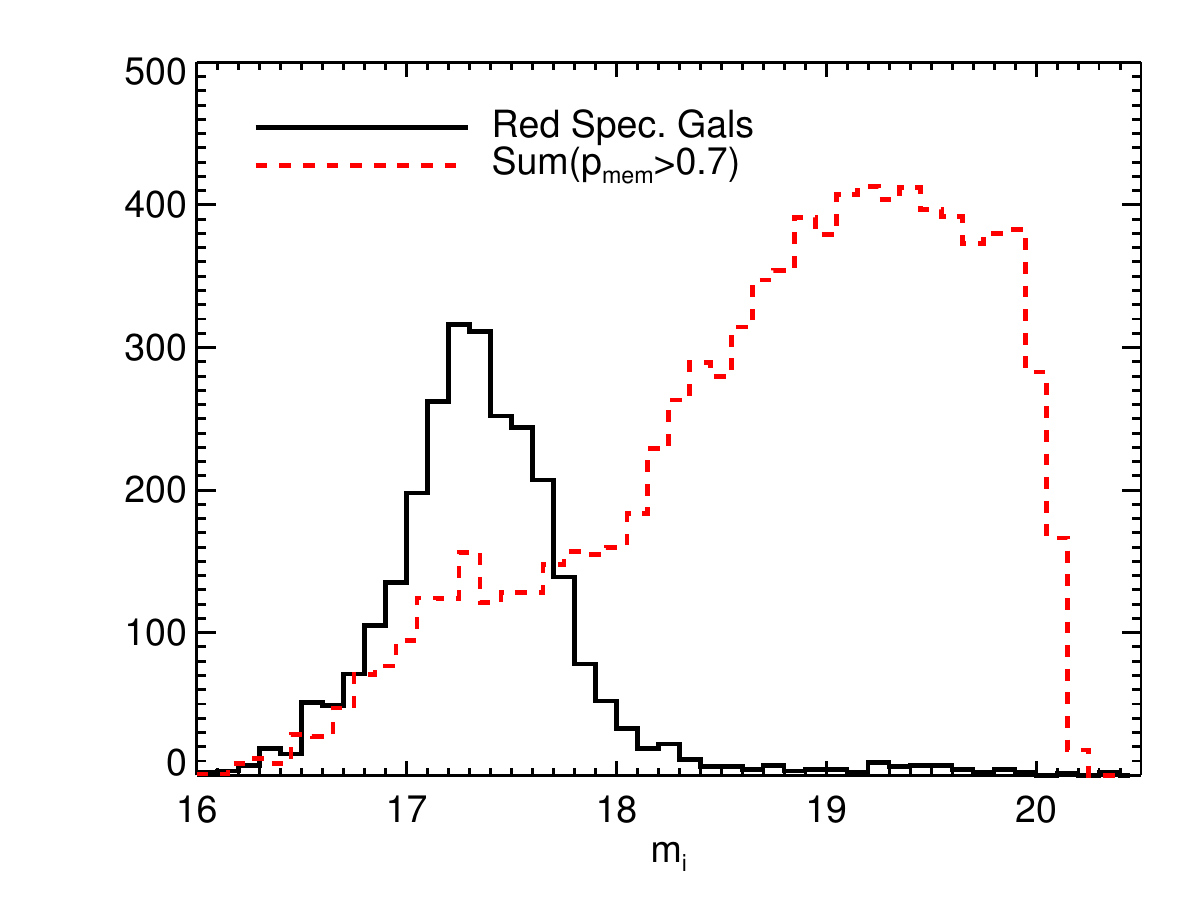}}
    \caption{Magnitude distribution of the red spectroscopic seed galaxies (black histogram)
    and photometrically selected cluster galaxies with $\pmem>0.7$ (red dashed histogram) in the redshift
    slice $z\in[0.24,0.26]$.  The ``cluster member'' histogram weights each galaxy
    by its membership probability.  The gain in the training sample of faint red-sequence
    galaxies through our photometric selection is of critical importance for an accurate
    calibration of the red-sequence as a function of redshift, particularly at the faint end.
    }
    \label{fig:cal_gain}
  \end{center}
\end{figure}

We emphasize that it is essential that we leverage our spectroscopic redshifts
to fainter magnitudes to properly model the red sequence.  In the case of DR8,
our initial seed galaxy sample is comprised of $42,000$ galaxies associated
with $\lambda>5$ clusters, almost all of which are preferentially bright.  By
contrast, our final calibration sample (see Section~\ref{sec:iterating}) is
comprised of over $600,000$ red-sequence galaxies that extend to much fainter
magnitudes.  This is illustrated in Figure~\ref{fig:cal_gain}.  The magnitude
distribution of our seed galaxies in the redshift slice $z\in[0.24,0.26]$
(solid black histogram) is contrasted with the membership-weighted magnitude
distribution of our final photometrically selected training sample (red dashed
histogram).  We see that the gain in the effective number of red sequence
training galaxies is enormous, allowing for an accurate calibration of the
red-sequence (amplitude, tilt, and scatter) as a function of redshift.  We 
have explicitly verified that modest changes to the cuts applied in this section
do not impact our final calibration of the red-sequence resulting
from the subsequent analysis described in the next section.

\subsection{Modeling The Red Sequence}
\label{sec:modrs}

\begin{figure}
  \begin{center}
    \scalebox{1.2}{\plotone{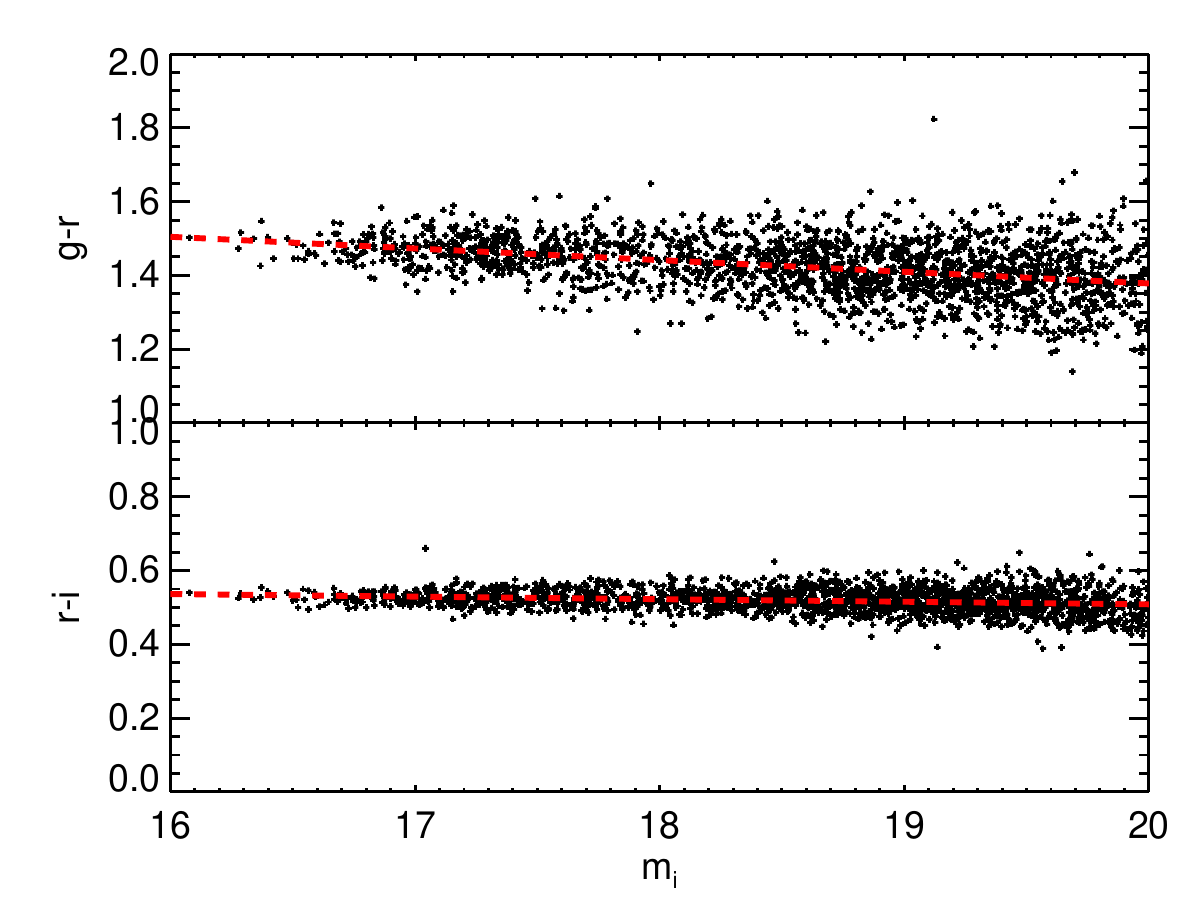}}
    \caption{Composite red sequence at $0.25<z<0.26$ for
      color-selected galaxies with $\pmem>0.9$.  A linear model (red
      dashed line) with roughly constant intrinsic scatter is a good
      representation of the red sequence in both $g-r$ and $r-i$.  We note that
    the $\pmem>0.9$ cut is employed for illustration purposes only.}
    \label{fig:comprs}
  \end{center}
\end{figure}

Given a list of galaxies with multidimensional color ($\mathbf{c}$), redshift
($z$, taken to be the cluster redshift), and membership probability ($\pmem$), we can now 
proceed to calibrate the full red sequence model.  Our model is well motivated by observations of
galaxy clusters, in that the red sequence at any given redshift in a given
color can be described by a simple linear relation between color and $i$-band
magnitude $m_i$ with intrinsic scatter $\sigint$.  For example,
Figure~\ref{fig:comprs} shows the composite red sequence at $z=0.25$ for both
$g-r$ and $r-i$ colors, for all galaxies selected in the final calibration
iteration with $\pmem>0.9$.

Our red-sequence model we is defined in terms of smoothly evolving functions of redshift
characterizing the amplitude and slope of the mean color--redshift relation, and
the corresponding covariance matrices.
We have
opted to use a cubic spline interpolation to parameterize these functions.
Given the large number of colors (four for SDSS), and broad redshift range, our
model necessarily contains a large number of free parameters.  For instance, 
in our SDSS DR8 implementation, we required a total of 118 parameters
to fully characterize the red-sequence model.   In principle, we
would like to fit the full red sequence model simultaneously.  However, to make
the problem more tractable we fit the red sequence parameters governing the
mean relation and the diagonal elements of the covariance matrix one color at a
time.  Once these terms are in place, we fit the off-diagonal terms of the
covariance matrix.  We are also cautious that our model does not have too many
free parameters given the training data such that over-fitting becomes
possible.  As shown in Appendix~\ref{app:ntrain}, this can be a problem in the
case of very sparse training data.

An additional complication comes from the fact that our selection of red
sequence galaxies is not entirely clean.  Our fit of the red sequence
must take into account the background density of non-member galaxies, as described below.
In addition, we also have to contend with blue cluster galaxies that are not taken
into account by a global background term.  These blue galaxies will tend to
have two effects.  First, as the blue fraction increases at lower luminosities,
they will tend to steepen the apparent red sequence tilt.  Second, the blend of
red and blue galaxies will tend to broaden the apparent intrinsic width of the
red sequence.

In order to deal with both of these effects of blue cluster galaxies, we have
taken a pragmatic approach.  When fitting the red sequence for a given
color, we first perform a sharp color cut to concentrate on the core of the red galaxy
distribution.  Naively, this cut would introduce biases in the recovered red-sequence
model, leading to under-estimates of the scatter.  
We avoid this difficulty by explicitly modeling such a color cut into our
likelihood function.  All that remains is to specify the color cut.  Here, we apply
a color cut of $1.5\sigma$ about the median color of the high probability
member galaxies, where $\sigma$ is the median absolute deviation of the 
color about the median.

\subsubsection{Measuring the Model Mean and Color Scatter}

As noted above, we begin by measuring the model color $\avg{c|m_i,z}$
as a function of galaxy magnitude $m_i$ and cluster redshift $z$ for each color,
one at a time.  The first step in this process is to define the pivot point $\refmag$ used
to calibrate the amplitude and tilt of the mean red sequence relation at redshift $z$.
We write
\begin{equation}
  \cmod = \cbar + \sbar [\imag - \refmag],
  \label{eqn:cmod}
\end{equation}

We wish to select a pivot point that is characteristic of most cluster members.  To do
so, starting from our full members list, we apply a $\pmem>0.7$ cut.
Using this sub-sample, we minimize the cost function $E$ where
\begin{equation}
E = \sum |m_i - \refmag|
\end{equation}
where $\refmag$ is defined via spline interpolation, and the model parameters are the value
of $\refmag$ at the nodes.  

Having defined our pivot point as a function of redshift, we turn to calibrating the amplitude
and slope of the mean relation, i.e., $\cbar$ and $\sbar$ in Eqn.~\ref{eqn:cmod}.
As a first step, we do a rough estimate of the amplitude and scatter, which
we will use to isolate the core of the color distribution of member galaxies.
These rough estimates for the amplitude and scatter are denoted $\medcol$
and $\madsig$, and are obtained  by selecting galaxies with $\pmem>0.7$, and then fitting
for these functions as was done in Section~\ref{sec:seltrain}.  Specifically, the
functions are spline interpolated, with model parameters being the value of these
functions at the nodes. The best fit parameters are found by 
minimizing Eqn.~\ref{eqn:median}, and $\madsig$
is defined by minimizing Eqn.~\ref{eqn:madsig}.
The primary difference between these new color estimates and scatter
relative to those derived in Section~\ref{sec:seltrain} is that these
parameters are now appropriate to the full red sequence rather than simply the
(brightest) spectroscopic galaxies.

We now turn to measuring the actual model parameters defining the amplitude
$\cbar$, slope $\sbar$, and scatter $C_{ii}^{\mathrm{int}}(z)$.  As before, we
use a cubic spline interpolation to parameterize these smoothly evolving
functions of redshift.  For DR8, we have chosen to use a node spacing of $0.05$
for $\cbar$, $0.1$ for $\sbar$, and $0.15$ for $C_{jj}^{\mathrm{int}}(z)$.  We
have found that a relatively tight spacing is required for $\cbar$, as this
function can change relatively rapidly at filter transitions.  Fortunately,
$\cbar$ is the most robust parameter, and thus is amenable to smaller node
spacings.  The slope and scatter are not expected to vary as rapidly, and are
also noisier to estimate, so we have chosen wider node spacings.  Overall, the
calibration is not very sensitive to the node spacings chosen provided there
are sufficient calibration galaxies (though see Appendix~\ref{app:ntrain}).

Starting from the photometrically selected galaxy training set from the previous section,
we first apply a color cut $|c - \medcol| < 1.5\madsig$, which ensures that the
red-sequence parameters are based on the core of the red galaxy distribution, 
and are therefore less likely to be biased by blue galaxies.  In our model,
the probability that a red-sequence cluster galaxy has a color $c$ is given by
a truncated Gaussian distribution,
\begin{equation}
  G(c) = \frac{\frac{1}{\sqrt{2\pi}\sigma}e^{-(c - \cmod)^2/2\sigma^2}}
  {\erf\left ( \frac{1.5\madsig}{\sqrt{2}\sigma} \right )},
\end{equation}
where the expectation value $\avg{c|m_i,z}$ is defined in terms of our model
functions $\cbar$ and $\sbar$ as per Eqn.~\ref{eqn:cmod}, and the scatter $\sigma$
is the sum in quadrature of the intrinsic scatter and the photometric error of the galaxy,
\begin{equation}
  \sigma = \sqrt{\sigma^2 + \sigint^2(z)},
\end{equation}
where $\sigint(z) = \sqrt{C_{jj}^{\mathrm{int}}}$ is the intrinsic scatter
of the red sequence.  The  `erf' term in the denominator accounts for the
fact that $G(c)$ is truncated at $\medcol \pm \madsig$, under the approximation
$\cbar = \medcol$.  This approximation is only used in the overall normalization
of the distribution.

The total probability distribution for all of our calibration galaxies must account for the fact
that some of our galaxies are in fact background galaxies, so the full color distribution 
is given by
\begin{equation}
  P(c) = \pmem G(c) + (1-\pmem)b(c,\imag),
  \label{eqn:rspdf}
\end{equation}
where $b(c,\imag)$ is the distribution in color and magnitude of galaxies about random points.
The shape of the background function is obtained by
binning all galaxies in color and magnitude bins and using a CIC algorithm as
in Section~\ref{sec:background}.

In the end, our task is to calculate the set of $\cbar$, $\sbar$, and
$C_{jj}^{\mathrm{int}}(z)$ values at the given cubic spline nodes that
maximizes the total likelihood given by
\begin{equation}
  \ln\lk = \sum_i\ln P_i.
  \label{eqn:rslike}
\end{equation}
As above, we accomplish this maximization by making use of the downhill-simplex
method.  The maximum likelihood point defines the model functions $\cbar$, $\sbar$,
and $C_{jj}^{\mathrm{int}}(z)$.
We emphasize that the likelihood is explicitly truncated as the data is, so
that the recovered scatter is unbiased relative to the full population of
cluster member galaxies, as we have confirmed with simple mock red sequences
and blue clouds.

\begin{figure}
  \begin{center}
    \hspace{-0.2in} \scalebox{1.2}{\plotone{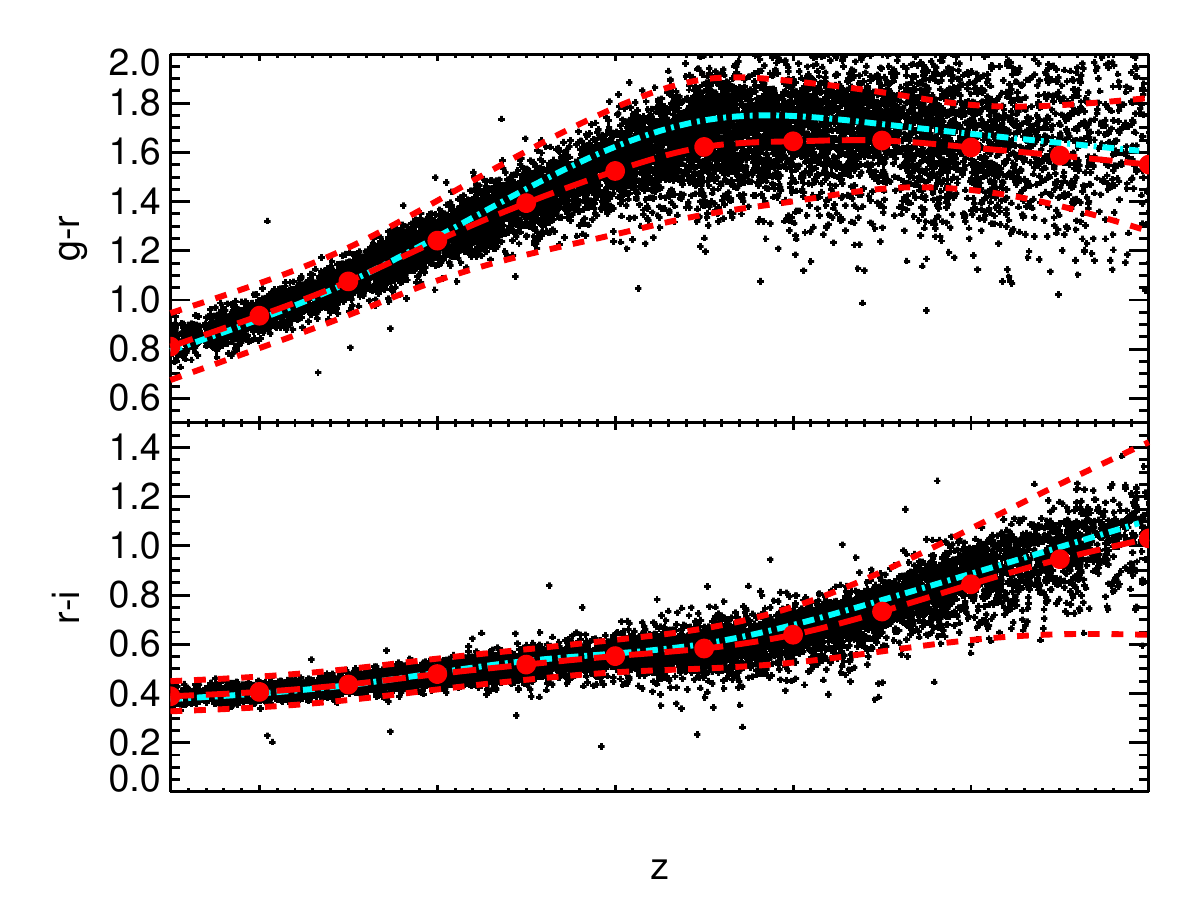}}
    \caption{Color as a function of redshift for the sample of red-sequence
      galaxies with $\pmem>0.9$.  The red points indicate the $\cbar$ values at
      the spline node positions, and the long-dashed lines are the spline
      interpolation.  The short-dashed red lines indicate the $3\sigint$ range.
      Note that the colors in the figure are not corrected for red sequence
      tilt.  The cyan dash-dotted line shows the color model for the bright
      spectroscopic sample from Figure~\ref{fig:redgals}, which tend to be
      brighter and redder than the full population.  We caution that the
      intrinsic width of the red sequence can be wider than the $\pmem>0.9$
      sub-population of galaxies in this illustration suggests, since high
      probability membership requires the galaxy to fall close to the expected
      average color.  Conversely, the larger number of outliers in $g-r$ above
      reflects the fact that the photometric errors in $g-r$ at high redshift
      are larger than the intrinsic width of the red sequence.}
    \label{fig:cz_iter3}
  \end{center}
\end{figure}

In Figure~\ref{fig:cz_iter3} we show the color evolution of red sequence galaxies
with $\pmem>0.9$ for the $g-r$ and $r-i$ colors in DR8.  The red points indicate
the $\cbar$ values at the spline node positions, and the long-dashed lines are
the smooth interpolation.  The short-dashed lines indicate the $3\sigint$
range.  Note that the colors in the figure are not corrected for red sequence
tilt.  We caution that the intrinsic width of the red sequence can be wider than 
naively indicated by the $\pmem>0.9$ galaxies, since high probability galaxies
must reside closer to the average red-sequence model.

\subsubsection{Measuring $C_{jk}^{\mathrm{int}}(z)$}

With the intercept and slope of the red sequence in hand, as well as the
diagonal elements of the covariance matrix, we now estimate the off-diagonal
elements of the covariance matrix, $C_{jk}^{\mathrm{int}}(z)$.  Once again, we
use a cubic spline interpolation, with the same $0.15$ node spacing as used for
$C_{jj}^{\mathrm{int}}(z)$.

In order to make the calculation tractable, to constrain the off-diagonal elements
of the covariance matrix we consider the problem piecewise, tackling two colors
at a time.  Each individual piece of the covariance matrix constrained in this
way will be positive-definite and thus a valid covariance matrix.
Unfortunately, due to noise in the estimation of the parameters, this method
does not guarantee that the total covariance matrix,
$\covarint$, will also be positive-definite.

To ensure that $\covarint$ is positive-definite, we constrain the parameters
for pairs of colors in a specific priority order, ensuring that the best
constrained colors have precedence.  In the case of DR8 data in the redshift
range $z\in[0.05,0.6]$, these are $g-r$ and $r-i$.   Then, at each step
in the downhill-simplex estimation described below we do not allow any terms
in $C_{jk}^{\mathrm{int}}(z)$ that result in a minimum eigenvalue in the
\emph{total} covariance matrix $\covarint$ that is less than $0.01^2$.  In this
way, the first color pair to be constrained ($g-r$,$r-i$) is essentially
free, while the final (and noisiest) color pair to be constrained ($u-g$,$i-z$)
will not result in a non-invertable covariance matrix $\covarint$. 

To perform the pairwise constraints on the off-diagonal elements, let us
consider the residuals in two colors $x_j$ and $x_k$.  We start with Eqn.~\ref{eqn:cmod},
\begin{equation}
  x = c - \avg{c|\imag,z} = (\bar{c}(z) + \bar{s}(z) [\imag - \refmag]).
\end{equation}
The probability distribution function is again a Gaussian, though this time we
explicitly leave the covariance matrix in the equation:
\begin{equation}
  G(\mathbf{x}) =
  \frac{1}{\sqrt{2\pi}|\bC|^{1/2}}\exp \left [
    -\frac{1}{2}\mathbf{x}\bC^{-1}\mathbf{x} \right ],
\end{equation}
where $\mathbf{x} = \{x_j,x_k\}$ is the vector of residuals, and the total
covariance matrix $\bC$ is
\begin{equation}
  \bC = \covarint + \covarerr.
\end{equation}
Here $\covarint$ and $\covarerr$ are the
covariance matrices characterizing the intrinsic scatter and photometric error
respectively.  The intrinsic scatter is simply
\begin{equation}
  \bC_{\mathrm{int}} = \left (
  \begin{array}{cc}
    \sigma_{\mathrm{int},j}^2 & r\sigma_{\mathrm{int},j}\sigma_{\mathrm{int},k}\\
    r\sigma_{\mathrm{int},j}\sigma_{\mathrm{int},k} & \sigma_{\mathrm{int},k}^2\\
  \end{array} \right ),
\end{equation}
where $\sigma_j$ and $\sigma_k$ are known from the previous section, and $r$ is
the only unknown.  The covariance matrix $\covarerr$ is derived from
the photometric error in each band.  Given two colors $c_j =
m_\alpha-m_\beta$ and $c_k = m_\gamma - m_\delta$, the covariance matrix
characterizing the photometric error is given by
\begin{equation}
  \covarerr = \left ( \begin{array}{cc}
    \sigma_\alpha^2 + \sigma_\beta^2 & \eta \\
    \eta & \sigma_\gamma^2 + \sigma_\delta^2\\
  \end{array} \right ),
\end{equation}
and
\begin{equation}
  \eta =
  \begin{cases}
    -\sigma_\beta^2 & \text{if $\gamma=\beta$}\\
    0 & \text{otherwise}.
  \end{cases}
\end{equation}
Here, we are assuming that neighboring colors are of the form $c_{\alpha\beta}$
and $c_{\gamma\delta}$, i.e., that the ``shared'' magnitude is
$m_\beta=m_\gamma$.  The covariance between photometric errors
arises precisely because, for example, the neighboring colors $g-r$ and $r-i$
are both derived from the same $r$-band magnitude.

The color distribution function of the full galaxy population is again given by
Eqn.~\ref{eqn:rspdf}, noting that now the background term $b(c_j,c_k,m_i)$ is
given by a three dimensional binning in two colors and $i$-band magnitude.  In
addition, we implement a prior on $r$ with $0$ mean and width $0.45$ for each
of the nodes.  We find that this prior reduces the noise in the parameter
constraints, which is especially important at high redshift where the
photometric errors dominate and the covariance matrix is largely unconstrained.
At the same time, this prior allows high correlations ($r\sim0.9$) if strongly
favored by the data.  Our total likelihood is now given by:
\begin{equation}
  \ln\lk = \sum_i\ln P_i - \sum_n\frac{(r_n/0.45)^2}{2},
\end{equation}
where $\sum_n$ is a sum over all the nodes, and $r_n$ is the correlation
coefficient at that node.  That is, the prior is placed at each of the nodes.
Maximization of the likelihood function defines the final values for the
correlation coefficients that characterize the intrinsic scatter covariance matrix.


\subsection{Iterating The Red-Sequence Model}
\label{sec:iterating}

We emphasize that the estimation of the red sequence parameters in the previous
section depends on the membership probabilities ($\pmem$) of the red sequence
galaxies.  Of course, the membership probabilities themselves depend on the red
sequence model.  In order to obtain a red sequence model that is consistent
with the membership probabilities, we take an iterative approach.

After we calibrate the red sequence parameters based on single color membership
probabilities, we run the cluster finder on the training data, as described in
Section~\ref{sec:clusterfinder}.  During these calibration runs we restrict
ourselves to finding clusters associated with our seed galaxies so that we can
affirmatively associate a spectroscopic redshift with each cluster.  We note
that our cluster finder starts with the photometric redshift estimate ($\zred$)
of each cluster galaxy, so spectroscopic galaxies whose colors are incompatible
with the red-sequence at the spectroscopic redshift never result in galaxy
clusters.  Thus, our training sample at this point results in robust clusters
with spectroscopic redshifts. Further, failures in the photoz of the galaxies
for red-sequence galaxies are rare (see Figure~\ref{fig:zred}), so any such
failures simply slightly reduce the sample of training clusters, without
otherwise adversely affecting our training sample.  The resulting cluster
catalog includes cluster member lists and new membership probability estimates
$\pmem$ based on the full color model.  With these in hand we can re-estimate
the red sequence model as described in Section~\ref{sec:modrs}.

\begin{figure}
  \label{fig:csz}
  \begin{center}
    \scalebox{1.0}{\plotone{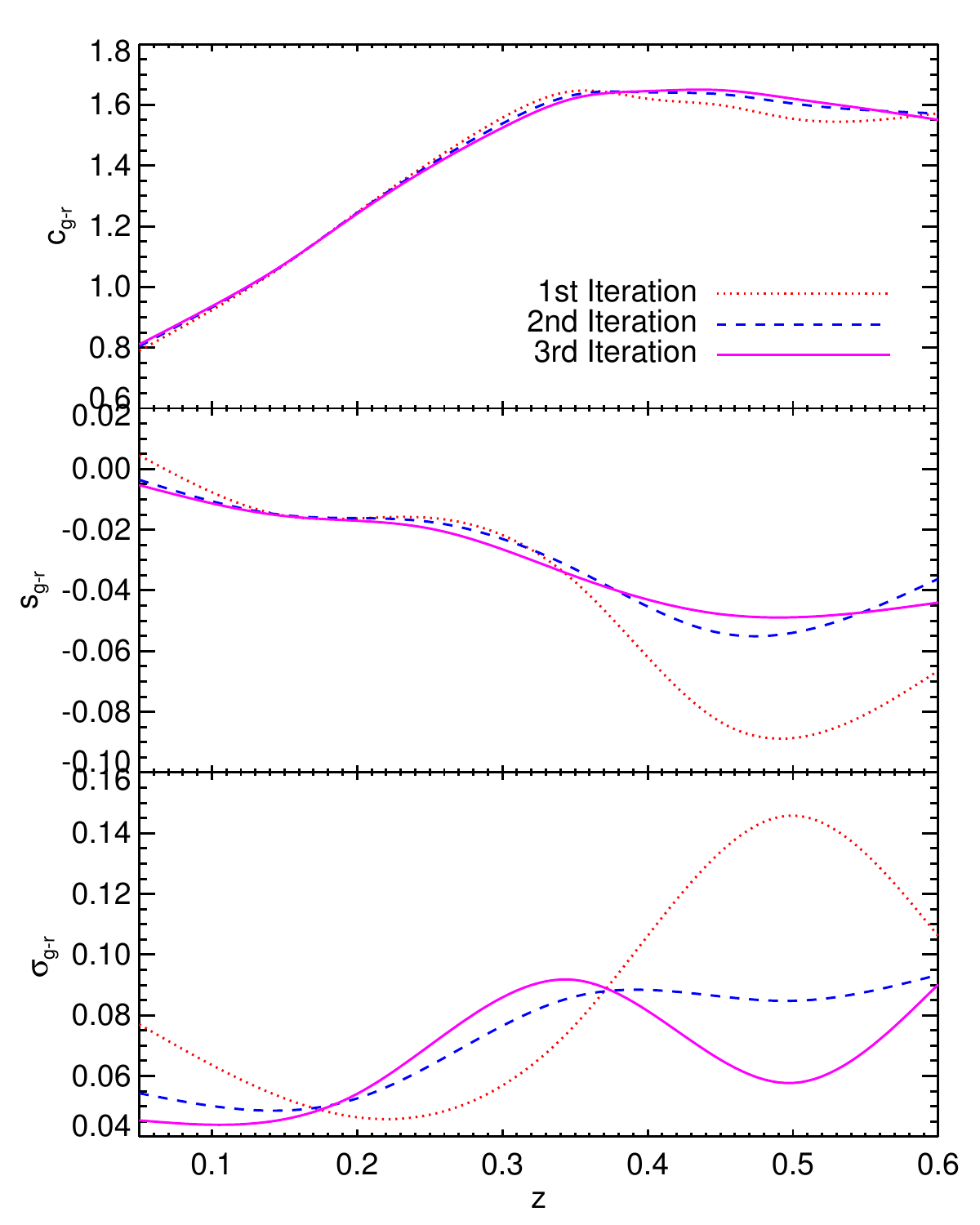}}
    \caption{\emph{Top:} Average color, $\cbar$ for $g-r$, at the pivot magnitude
      $\refmag$, for the first (red dotted line), second (blue dashed line),
      and third (magenta solid line) iterations of the calibration as a
      function of redshift. \emph{Middle:} As for top, with red sequence slope
      $\sbar$.  \emph{Bottom:} As for top, with intrinsic scatter
      $\sigma_j^{\mathrm{int}} = \sqrt{C_{jj}^{\mathrm{int}}}$.}      
  \end{center}
\end{figure}

As we iterate, the largest shifts in the model occur between the first and
second iteration, reflecting the shift from estimating membership probabilities
based on a single color, and estimating membership probabilities with the full
multi-color data.  Figure~\ref{fig:csz} shows the red sequence parameters
($\cbar$, $\sbar$, and $\sigma_j^{\mathrm{int}}$) for each of the first 3 iterations of
our red sequence calibration.  For illustration, only the figures for the
$g-r$ color are shown.  The color at the reference magnitude $\refmag$ and
slopes characterizing the average color of red sequence galaxies converge
quickly, and is generally well measured, except for $u-g$ at high redshift,
where the large photometric errors in $u$ make our model estimates noisy. The
scatter model, on the other hand, converges slowly, particularly at high
redshift, where the intrinsic scatter is often sub-dominant to photometric
errors.  As we now show, however, by the third iteration our model is well
converged.

We define convergence of the red sequence model in terms of the relevant
quantity for our purposes, i.e., the cluster richness $\lambda$.  That is, we
require that cluster richness estimates be insensitive to further iterations.
To this end, we have run the calibration through ten iterations.  Given the red
sequence model for each of these ten iterations, we estimate the photometric
redshift and cluster richness of a standard set of galaxy clusters while fixing
the central galaxy of these systems.  Let then $\lambda_i$ and $z_i$ denote the
richness and redshift estimates from iteration $i$.  We bin the clusters in
narrow redshift slices ($\pm 0.01$), and we calculate: 1- the median ratio
$\lambda_i/\lambda_3$, and 2- the median offset
$(\lambda_i-\lambda_3)/\sigma_3$, where $\sigma_3$ is the error estimate in the
richness as estimated from iteration $3$.

\begin{figure}
  \begin{center}
    \hspace{-0.15in} \scalebox{1.2}{\plotone{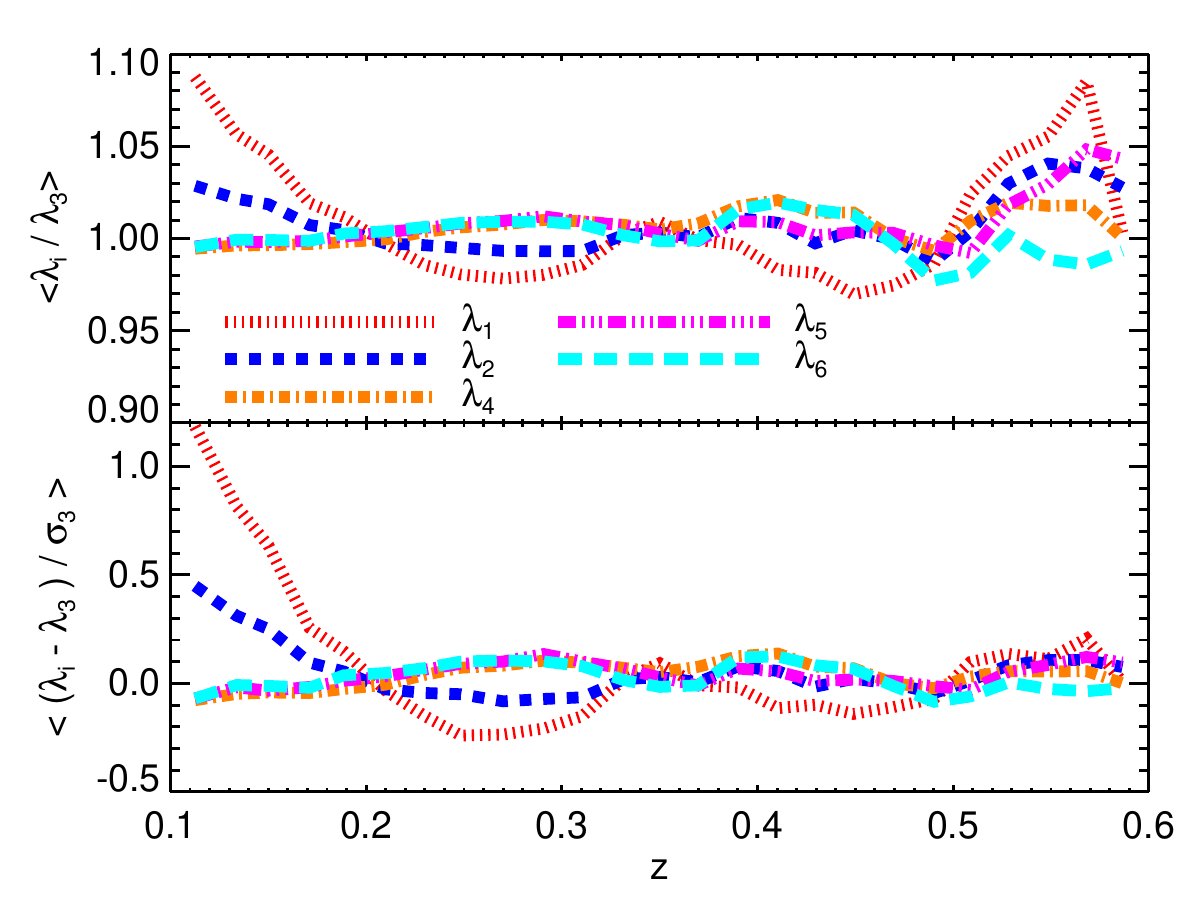}}
    \caption{\emph{Top:} Average richness bias as a function of redshift for
      the first six iterations of the red-sequence model for the DR8 training
      region, as compared to $\lambda_3$, the richness computed in the third
      iteration.  Even for the first iteration, the bias is $<10\%$ at all
      redshifts.  After the third iteration, the biases are always $<1\%$ at
      low redshift and $<5\%$ at high redshift. \emph{Bottom:} Error normalized
      average deviation relative to the baseline.  After the third iteration
      the bias is always $<0.1\sigma$.  }
    \label{fig:iterbias}
  \end{center}
\end{figure}

In Figure~\ref{fig:iterbias} we show the results of these iteration checks for
the first 6 iterations in the DR8 training region.  Even for the first
iteration, for which $\pmem$ was estimated using a single color, the bias is
always $<10\%$ (though at the lowest redshift that shift is $\sim1\sigma$).
However, after the third iteration, the biases are always $<1\%$ at low
redshift and $<5\%$ at high redshift.  The bottom panel shows that after the
third iteration the biases are $<0.1\sigma$.  Thus, we rely on the output of
our third iteration for our final cluster catalog.

\section{Photometric Redshift Estimation}
\label{sec:photoz}

At the end of our calibration we have a complete red sequence model as a function of redshift.
Note, however, that in order to estimate the richness of a photometric cluster we need to know
the cluster redshift.  If we have some initial, reasonably accurate redshift guess $\zinit$
for each cluster, we can estimate the cluster richness and determine the high probability cluster
members.  We then simultaneously
fit our red sequence model to all high probability cluster members to derive an improved redshift
estimate, and iterate this procedure through convergence.  We now describe this full procedure
in detail, including the construction of our initial cluster redshift guess $\zinit$.


\subsection{Redshift Initialization: $\zred$}
\label{sec:zred}

For the full SDSS DR8 survey, we have multiple photometric redshifts based on
large training sets~\citep[e.g.,][]{cdthj07,scmbw12}.  However, these methods
have certain limitations.  First, they require training sets that span a broad
range of magnitudes, which although abundant at $z\lesssim0.5$ for SDSS data,
will be much sparser at higher redshifts for large surveys such as DES.
Second, these methods --- in particular $p(z)$ methods such as that of
\citet{scmbw12} --- are very good at estimating the ensemble of redshifts for a
broad class of galaxies.  However, our needs are much more specific: we wish to
have a good initial single-value estimate of the redshift of the central galaxy
of galaxy clusters to initialize our cluster photometric redshift estimation
procedure.  To that end, we have developed our own photometric redshift
estimator $\zred$ which is specifically designed to work on red sequence
galaxies.

Given a red-sequence galaxy at redshift $z$ with $i$-band magnitude $\imag$, color vector $\bc$, and
photometric error $\covarerr$, the probability distribution of its color is simply
\be
P(\bc) \propto \exp\left( - \frac{1}{2}\chi^2 \right)
\ee
where $\chi^2$ is given by Eqn.~\ref{eqn:chi2}, i.e.,
\begin{equation}
  \chisq = \left ( \bc - \cmodbf \right ) \left (
  \covarint + \covarerr \right )^{-1} \left ( \bc - \cmodbf \right ).
  \label{eqn:chi2zred}
\end{equation}
The corresponding log-likelihood is therefore simply $\ln \lk = -0.5\chisq$. 
In practice, we also
include an additional volume prior that accounts for the fact that there is more
volume at higher redshifts.  Assuming that the luminosity function does not evolve
over the redshift uncertainties,
the probability that a galaxy of a given luminosity is at redshift $z$ is
\be
P_0(z) \propto \frac{dV}{dz} = (1+z)^2D_A^2(z) cH^{-1}(z),
\ee
which leads us to the likelihood
\be
\ln\lk_{\mathrm{red}} = - \frac{\chi^2}{2} + \ln \left| \frac{dV}{dz} \right|.
\ee

The redshift estimator $\zred$ is that which maximizes the above likelihood.
We use the ``red'' subscript to indicate that the redshift estimator assumes a red
sequence galaxy model.  We maximize the likelihood along a redshift
grid with $\delta z = 0.005$, and then use parabolic interpolation to find the correct maximum.
This search is restricted to galaxies with $\imag < m_*(z)+2.5$, since galaxies fainter
than this fall well below the luminosity threshold used to define cluster richness (recall
$m_*(z)$ is defined in Sec.~\ref{sec:nfwandlumfilter}).
The error estimate for $\zred$ is estimated as the standard deviation of the redshift
over its posterior, i.e.
\be
\sigzred^2 = \avg{z^2} - \avg{z}^2
\ee
where 
\be
\avg{z^n} = \frac{ \int dz\ \lk_{\mathrm{red}}(z) z^n }{ \int dz\ \lk_{\mathrm{red}}(z) }.
\ee
We could, of course, store the posterior of the redshift distribution, but we have chosen
not to do so since the only use of $\zred$ in the \redmapper\ algorithm is that of providing
an initial redshift estimates for galaxy clusters.  

\begin{figure*}
  \begin{center}
    \scalebox{1.1}{\plottwo{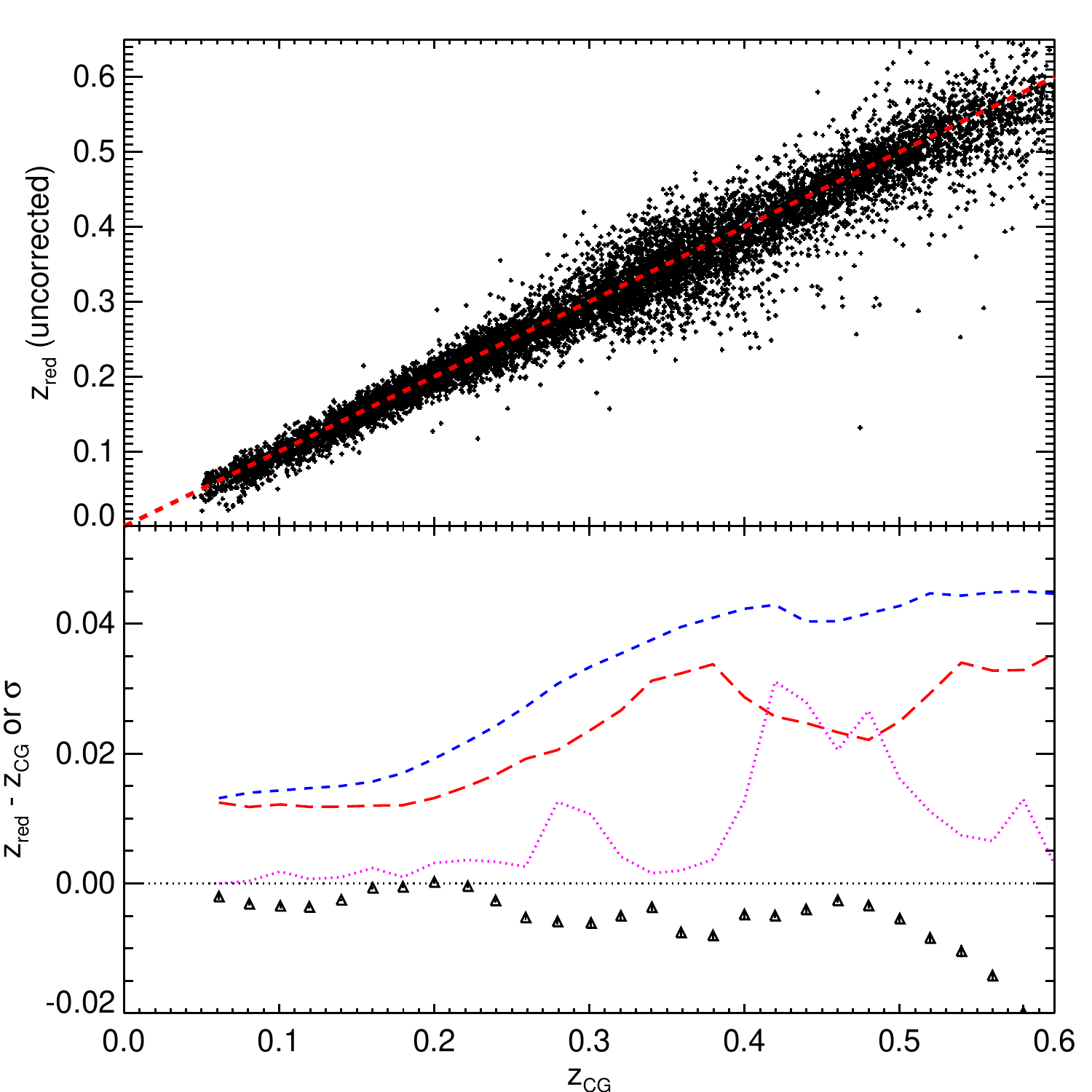}{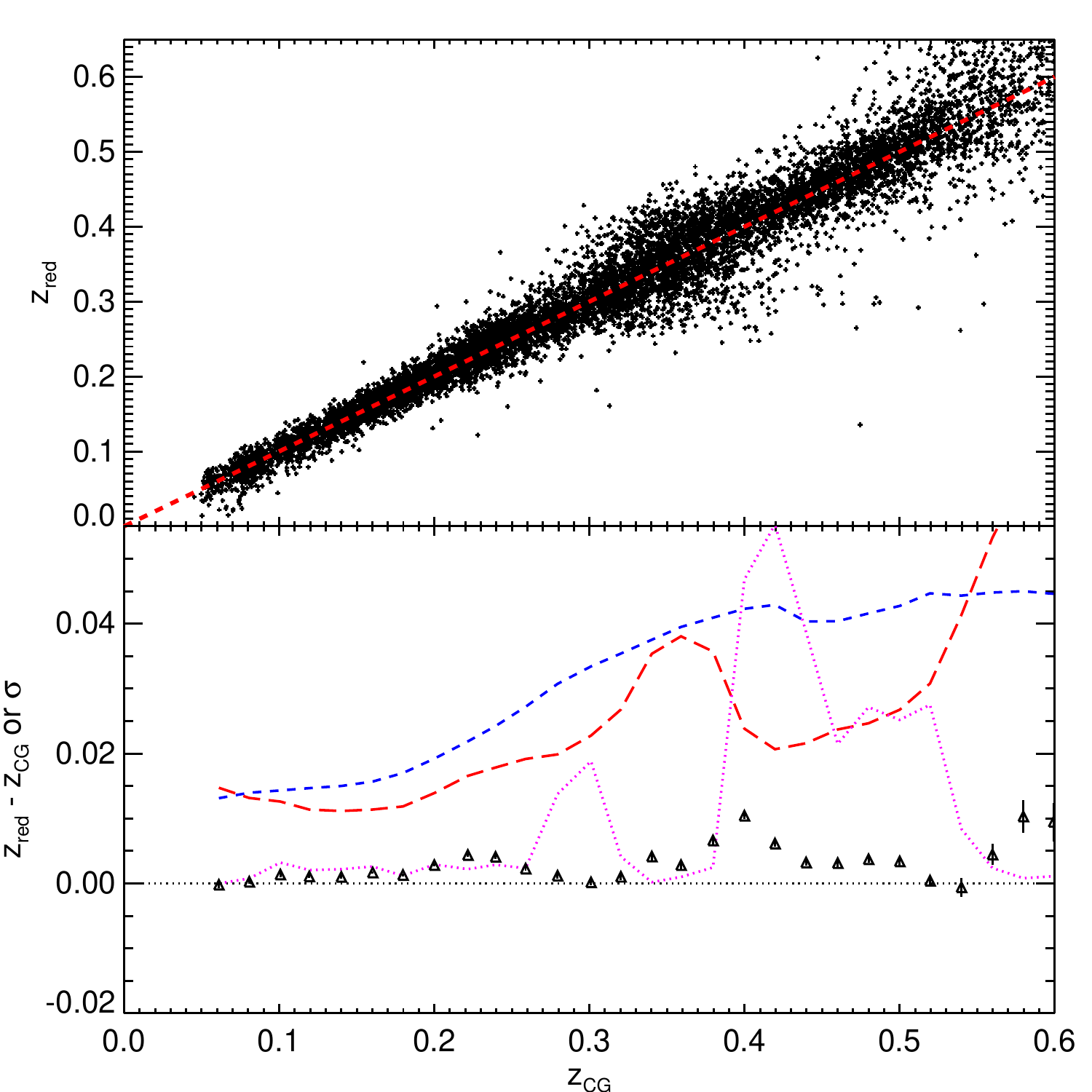}}
    \caption{\emph{Left, top:} Uncorrected photometric redshift
      $\zred$ for cluster member galaxies in DR8 with $\pmem>0.9$, as
      a function of the central galaxy spectroscopic redshift $\zcg$.
      \emph{Left, bottom:} The black triangles show the mean redshift
      offset $\zred - \zcg$ in several redshift bins.  The red
      long-dash line is the rms of these offsets, while the blue
      short-dash line is the average estimated redshift error.  The
      dotted magenta line is the fraction of $4\sigma$ outliers as a
      function of redshift. \emph{Right, top:} Corrected photometric
      redshift $\zred$, using \ref{eq:zcorr}, for cluster member
      galaxies in DR8, as in left panel.  \emph{Right, bottom:} Bias,
      scatter, and outlier fraction, as in left panel, now for the
      corrected redshift.}
    \label{fig:zred}
  \end{center}
\end{figure*}

The top-left panel of Figure~\ref{fig:zred} shows $\zred$ for DR8 cluster
training galaxies with $\pmem \geq 0.9$ versus the spectroscopic
redshift of the corresponding central galaxy $\zcg$.  We see that $\zred$
performs very well, with low bias and scatter, and very few gross outliers.
The ``flare-up'' of the points around $\zcg \sim 0.35$ is due to the $4000\,\mathrm{\AA}$ break
moving from the $g-r$ to the $r-i$ color.

The performance of $\zred$ is better illustrated in the bottom-left panel of the same
figure.  The black triangles show the mean offset $\zred - \zcg$
in redshift bins, the blue dashed line shows the average error in $\zred$
as estimated above, while the red-dashed line shows the observed
rms of the redshift offset in each of the redshift bins.  The magenta dotted
line shows the fraction of $4\sigma$ outliers.  It is clear from the figure
that our errors are somewhat overestimated, and that there is a small
redshift bias in $\zred$.  

We correct for the deficiencies revealed in the left panel of Figure~\ref{fig:zred} by
applying an afterburner.  Specifically, for the above cluster sample we define
the mean redshift offset as a function of redshift,
\begin{equation}
  dz(z) = \avg{(\zred^0 - \zcg )|\zcg }
\end{equation}
where $\zred^0$ is the original, uncorrected redshift estimate defined above.  
That is, $dz(z)$ is the curve traced by the black triangles in the bottom-left panel of Fig.~\ref{fig:zred}.
We define a corrected $\zred$ redshift, as the solution to the equation
\be
\zred = \zred^0 + dz(\zred)
\label{eq:zcorr}
\ee
In practice, the above treatment is slightly simplified, since our correction afterburner allows for the
redshift bias to be a function of magnitude.  For details, we refer the reader to Appendix~\ref{app:zredcorr}.

In the right panel of Figure~\ref{fig:zred} we show the corrected value of $\zred$ as a
function of $\zcg$ after applying our afterburner, again for a sample of galaxies with $\pmem >
0.9$.  The notation is the same as for the left panel.  The biases
are improved at high redshift, although there are still some residual issues at
$z\sim0.4$ where $\zred$ is biased by $\sim0.3\sigma$.  The reason the biases
are not completely removed is due to the asymmetric and non-Gaussian nature of
the scatter at the filter transition.  We also note that the
afterburner removes residual biases observed as a function of $\imag$ (not
shown).  
The overall small bias and scatter in $\zred$ allows us to use this photometric
redshift estimate as a good initial guess with which to initialize our photometric
cluster redshift estimator.



\subsection{Cluster Redshift Estimation: $\zlambda$}
\label{sec:zlambda}

Our approach to computing the cluster photometric redshift $\zlambda$ is 
essentially an iterative extension of $\zred$.    Specifically,
given a central galaxy candidate, we:

\begin{enumerate}
  \item{Start with a cluster redshift $z_{\lambda,i}$, where $i$ indexes the iteration.
  In the first iteration, we set $z_{\lambda,0}=\zred$.}
  \item{Calculate the richness $\lambda$ around the candidate central galaxy setting $z_{\mathrm{cluster}}=z_{\lambda,i}$, and get the associated set of 
  membership probabilities $\pmem$.\label{step:calcrich}}
  \item{Select high membership-probability galaxies to estimate a new redshift $z_{\lambda,i+1}$ by maximizing the likelihood
    function given by Eqn.~\ref{eqn:zlambdalike} below.}
  \item{Repeat from step \ref{step:calcrich} until convergence, such that
    $|z_{\lambda,i+1} - z_{\lambda,i}| < 0.0002$.}
\end{enumerate}

All that remains then is the definition of a suitable likelihood function.
To begin with, let us assume that we have a sample of known cluster member galaxies.
Then, the log-likelihood of the observed colors for these galaxies would be 
\be
\ln \lk = \sum -\frac{\chisq_i}{2} - \frac{\ln|{\bf C}|}{2}.
\ee
In Eqn.~\ref{eqn:chi2zred} we take into account the log of the determinant of the
covariance matrix, $\ln|{\bf C}|$.  We have found that, unlike the case of
$\zred$, including this term improves the performance of $\zlambda$ when the
intrinsic scatter is varying rapidly.  This makes sense, given that when
utilizing multiple galaxies, one can directly probe the scatter in the red
sequence, which is an observable that is inaccessible when estimating
single-galaxy \photozs.

Of course, in practice, we do not have a list of known members, but rather a list of
likely members with membership probabilities.  One might be inclined to adopt a
sharp cut $\pmem \geq p_{\mathrm{min}}$ in order to define a likelihood that
can be used to estimate the cluster redshift.  However, we find that a
sharp cut in $\pmem$ leads to numerical instabilities in the iterative process because
galaxies can scatter in and out of the sample in the course of the iteration.

To overcome this problem, we adopt instead a soft cut, and define a new likelihood
\begin{equation}
   \ln \lk = \sum -\frac{w \left[\chisq + \ln|{\bf C}| \right]}{2},
  \label{eqn:zlambdalike}
\end{equation}
where each galaxy contributes a weight $w$ that smoothly varies from $w=1$ at $\pmem=1$ to $w=0$ at $\pmem=0$.

The assignment of these weights is somewhat ad-hoc.  We assume $w(\pmem)$ follows a Fermi-Dirac
distribution.  The transition from
$w=0$ to $w=1$ occurs at $p_{70}$, which is the probability
threshold that accounts for $70\%$ of the total richness, i.e.,
\be
0.7\lambda = \sum_{\pmem \geq p_{70}} \pmem.
\ee
The advantage
of defining the probability threshold in this way --- as opposed to a redshift independent
threshold $p_{\mathrm{cut}}$ --- is that $p_{70}$ varies
with cluster redshift in such a way that one always uses
the same fraction of cluster galaxies when estimating redshifts. 
Were we to take a constant $\pmem$ cut, the number of galaxies contribution to
$\zlambda$ would decrease with increasing redshifts, since galaxy $\pmem$ 
values decrease as the photometry becomes noisier. 
The width of the distribution is set to $0.04$, which we found is sufficient to regularize
the iterative process.  Thus, our galaxy weights are defined via
\begin{equation}
  w(\pmem) = \frac{1}{\exp \left [ (p_{70}-\pmem)/0.04 \right ] + 1}.
\end{equation}

\begin{figure}
  \begin{center}
    \hspace{-0.15in} \scalebox{1.2}{\plotone{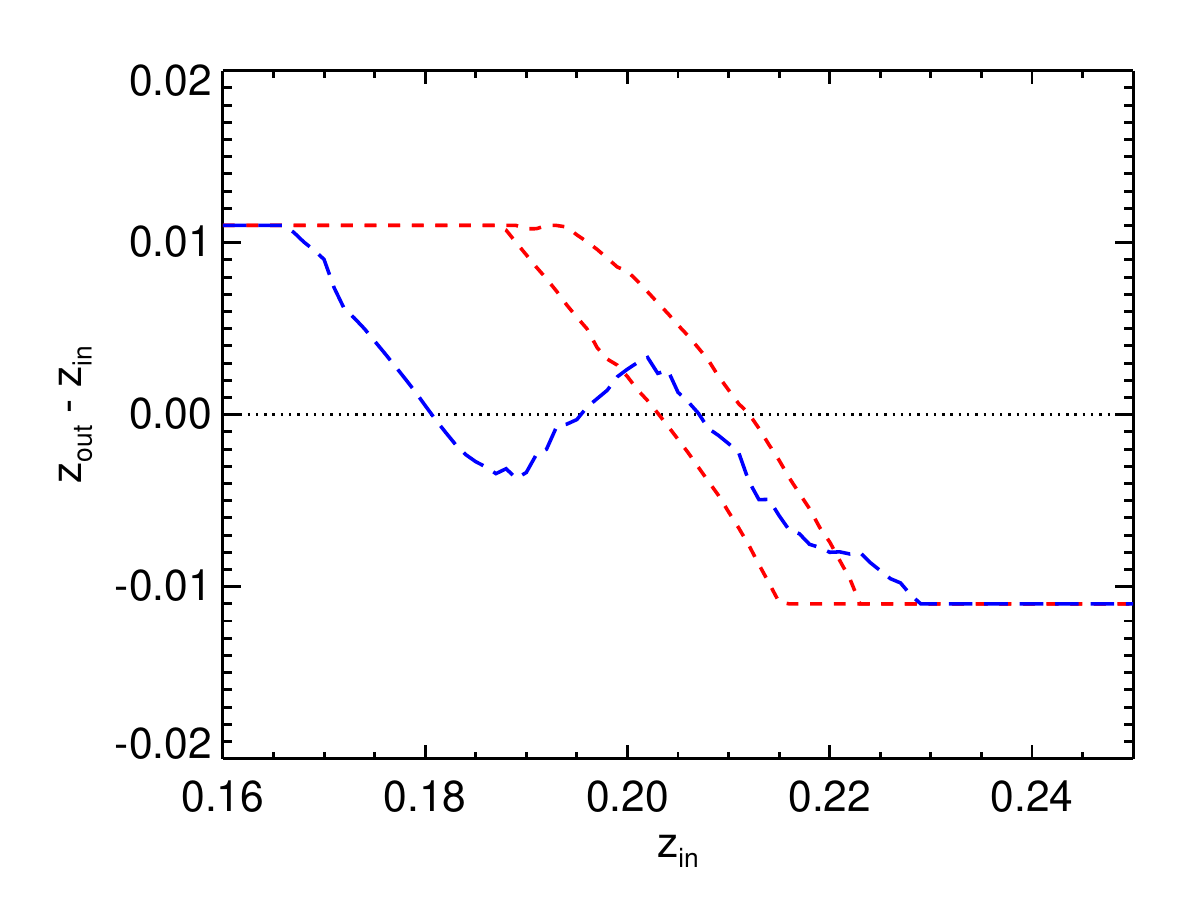}}
    \caption{The redshift difference $\zout-\zin$ for one loop of our iterative photometric
    redshift estimator, as a function of the input redshift $\zin$.
   	Two typical, well behaved clusters are shown with red
      short-dashed lines.  However, $\sim1-2\%$ of clusters have
      convergence curves like the blue long-dashed line.  These appear to be
      projection effects between multiple nearby structures. 
      }
    \label{fig:zlambdaconvergence}
  \end{center}
\end{figure}

In Figure~\ref{fig:zlambdaconvergence} we illustrate how the iterative
process in our redshift estimate works.  Fundamentally, each loop in the
iteration takes a $\zin$ value for the redshift, and produces a
redshift $\zout$, and we wish to find the stable point where
$\zout=\zin$.  In the figure, we show $\zout(\zin)$ for three sample
clusters.  For the two typical clusters denoted with red short-dashed lines,
this function is well behaved, and we quickly achieve convergence.  However,
there are also $\sim1\%-2\%$ of clusters that have convergence curves like the
blue long-dashed line.  These appear to be projection effects between multiple
nearby structures.  As detailed in Section~\ref{sec:cf:percolation},
\redmapper{} often fragments these clusters along the line-of-sight, as it
should.  However, which cluster is ``dominant''
and which is a satellite depends on the initial photometric redshift estimate
($z_{\lambda,0}$).

Given an estimate for $\zlambda$, we can also map out the posterior $P(\ztrue|\zlambda)$.
Defining $\chi^2_{\mathrm{norm}}$ via
\be
\chi^2_{\mathrm{norm}} = \sum w [\chi^2 + \ln|{\bf C}|] - \mathrm{min}\left( \sum
w [\chi^2 + \ln|{\bf C}|] \right),
\ee
we adopt the posterior
\begin{equation}
  P(\ztrue|\zlambda) = \frac{\exp(-\chisq_{\mathrm{norm}}/2) \left| dV/dz
    \right| }{\int d\chisq_{\mathrm{norm}}
    \exp(-\chisq_{\mathrm{norm}}/2) \left| dV/dz \right| }
\end{equation}
where $dV/dz$ is the comoving volume per unit redshift.  The above expression
defines our estimate of the redshift probability distribution of each cluster.
In addition, we fit this distribution with a Gaussian to estimate the redshift
error $\sigma_{\zlambda}$.

Finally, in order to ensure that $\zlambda$ is unbiased, we apply an afterburner
correction, much in the same way as was done for $\zred$, only now we demand
that the redshift be unbiased in the sense that $\avg{\ztrue|\zlambda}=\zlambda$.
We relegate the details to Appendix~\ref{app:zlambdacorr}.

\begin{figure}
  \begin{center}
    \hspace{-0.15in} \scalebox{1.2}{\plotone{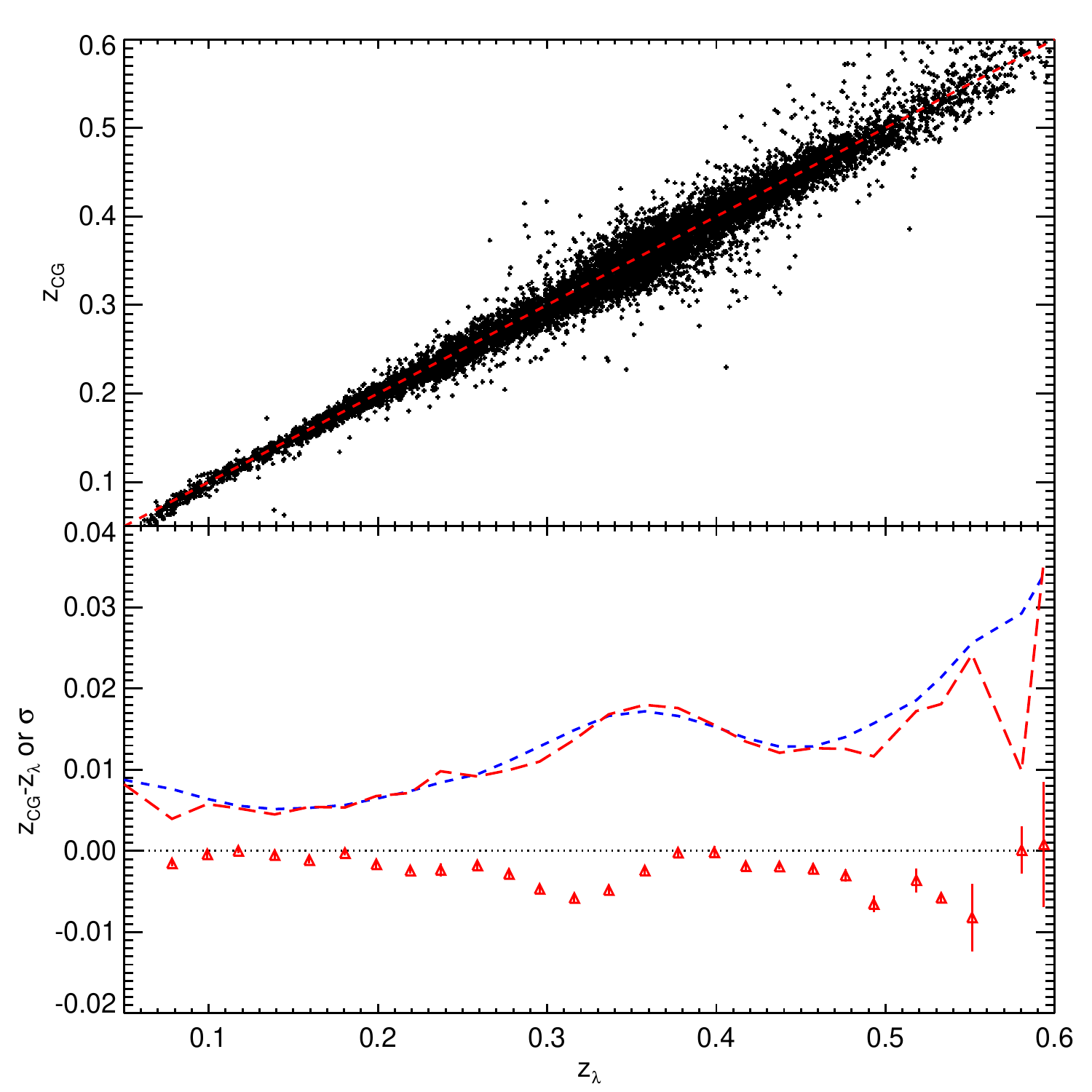}}
    \caption{\emph{Top:} $\zlambda$ vs. spectroscopic redshift of
      the assigned cluster central galaxy (CG) for \redmapper{} clusters in DR8
      with $\lambda/S(z) > 20$.  \emph{Bottom:} Red triangles show
      the mean offset $\zlambda - \zcg$ in various redshift bins.  The blue
      short-dashed line shows the average redshift error on $\zlambda$, while
      the red long-dashed line shows the measured rms of the redshift offset
      distribution.  The vast majority of outliers are due to errors in cluster
      centering, i.e., the offset $\zlambda - \zcg$ is large not because
      $\zlambda$ is incorrect, but rather because the chosen central galaxy is
      not actually a cluster member.  }
    \label{fig:zlambda-dr8}
  \end{center}
\end{figure}

In the top panel in Figure~\ref{fig:zlambda-dr8} we compare our photometric redshift
estimates to the spectroscopic redshift of the central galaxy (where available)
for all clusters in DR8 with $\lambda/S(z)>20$ (i.e., every cluster must have
20 galaxy detections).  The bottom panel shows the
residuals (red triangles), as well as the rms of the distribution (red
long-dashed line) and average estimated error $\sigma_{\zlambda}$ (blue
short-dashed line).  There are small biases that are nevertheless detected with
high confidence.  We do not yet fully understand the origin of these
  biases, but intend to return to this problem in a future paper.  We see too
that there is a feature at $0.35 \lesssim z \lesssim 0.45$, both in the bias and
scatter, reflecting the additional difficulties introduced by the fact that the
$4000\mathrm{\AA}$ break goes from being sampled by $g-r$ to $r-i$.  This is
also the redshift range where we start running into the limit of the DR8
photometry, which further aggravates these failures.  Indeed, these features
are greatly reduced when \redmapper\ is run on deeper 
data~\citep[e.g., SDSS Stripe 82 coadds,][not shown]{assbd11}.

\begin{figure}
  \begin{center}
    \hspace{-0.15in} \scalebox{1.2}{\plotone{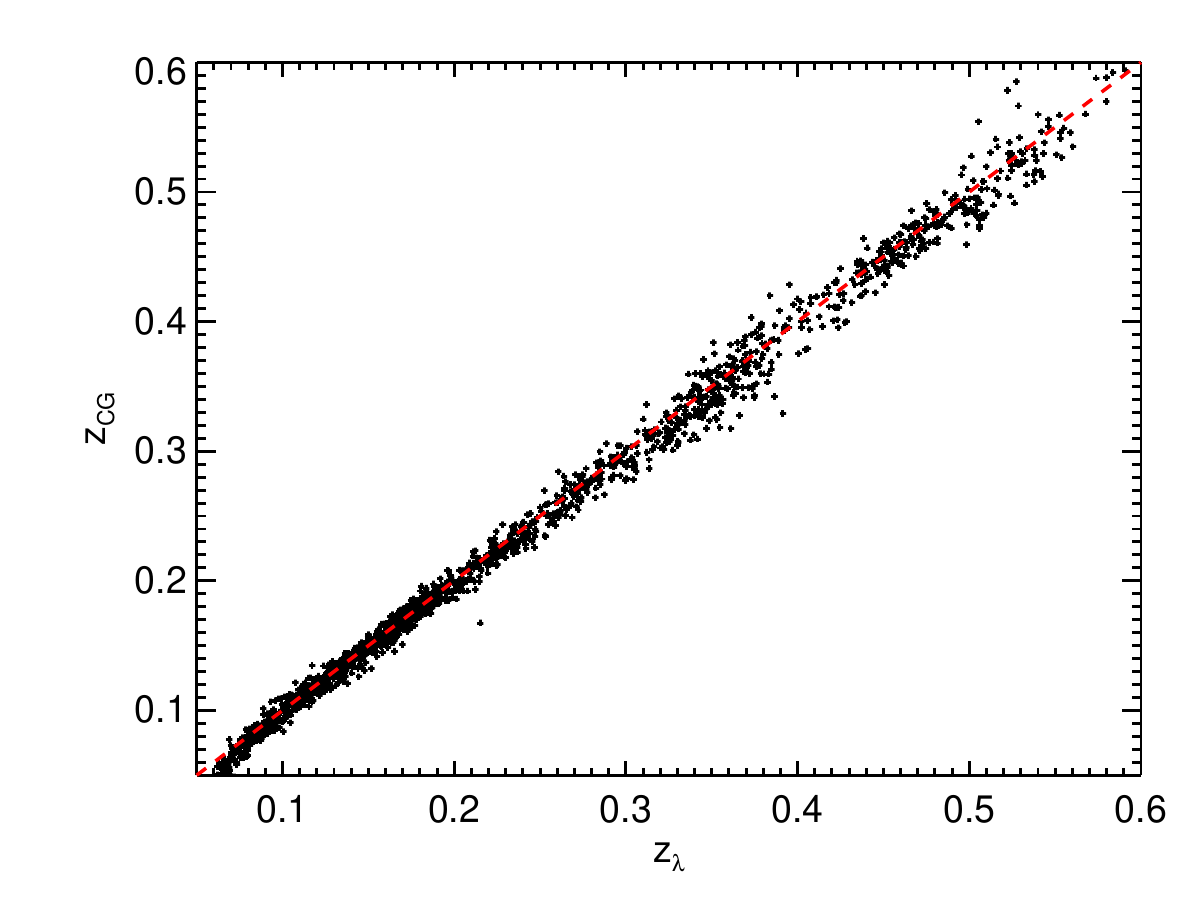}}
    \caption{$\zcg$ vs. $\zlambda$ as in Figure \ref{fig:zlambda-dr8}, but demanding that the cluster
      contain at least two cluster members with $\pmem \geq 0.8$ with
      spectroscopic redshifts within $1000\ \mbox{km/s}$ of the redshift of the
      assigned central galaxy.  This removes clusters centered on
      non-cluster-member galaxies.  Of the $13,178$ \redmapper{} clusters in DR8 with
spectroscopic redshifts, $1,829$ (or $14\%$) meet this criterion.  
      The few remaining outliers ($\lesssim0.2\%$) appear to be
      redshifts failures from multiple systems in projection.}
    \label{fig:zlambda-dr8_clean}
  \end{center}
\end{figure}

One interesting thing to note about the top panel in
Figure~\ref{fig:zlambda-dr8} is that the ``large'' ($\Delta z \sim 0.1$)
redshift offsets in this plot do not reflect errors in the cluster redshift
estimates, but rather cluster miscentering.  That is, when we compare
$\zlambda$ to the redshift of the central galaxy, large offsets are primarily
due to our selection of a central galaxy that is not, in fact, a cluster
member.  To demonstrate this, we have created a ``clean'' sample of clusters
where we demand that there be at least two spectroscopic cluster members with
$\pmem>0.8$ within $1000\,\mathrm{km}\,\mathrm{s}^{-1}$ of the spectroscopic
redshift of central galaxy, thereby ensuring that the central galaxy is in fact
a cluster member.  Of the $13,178$ \redmapper{} clusters in DR8 with
spectroscopic redshifts, $1,829$ (or $14\%$) meet this criterion.  The
corresponding comparison of $\zlambda$ to $\zcg$ in this case
is shown in Figure~\ref{fig:zlambda-dr8_clean}.  We see that this photometric
redshift plot is very clean.  The few
outliers left ($\lesssim0.2\%$) are likely multiple systems in
projection.  In particular, the obvious outlier cluster at $\zlambda \approx
0.22$ correpsonds to the cluster represented by the blue long-dashed line shown
in Figure~\ref{fig:zlambdaconvergence}.

\begin{figure}
  \begin{center}
    \hspace{-0.15in} \scalebox{1.2}{\plotone{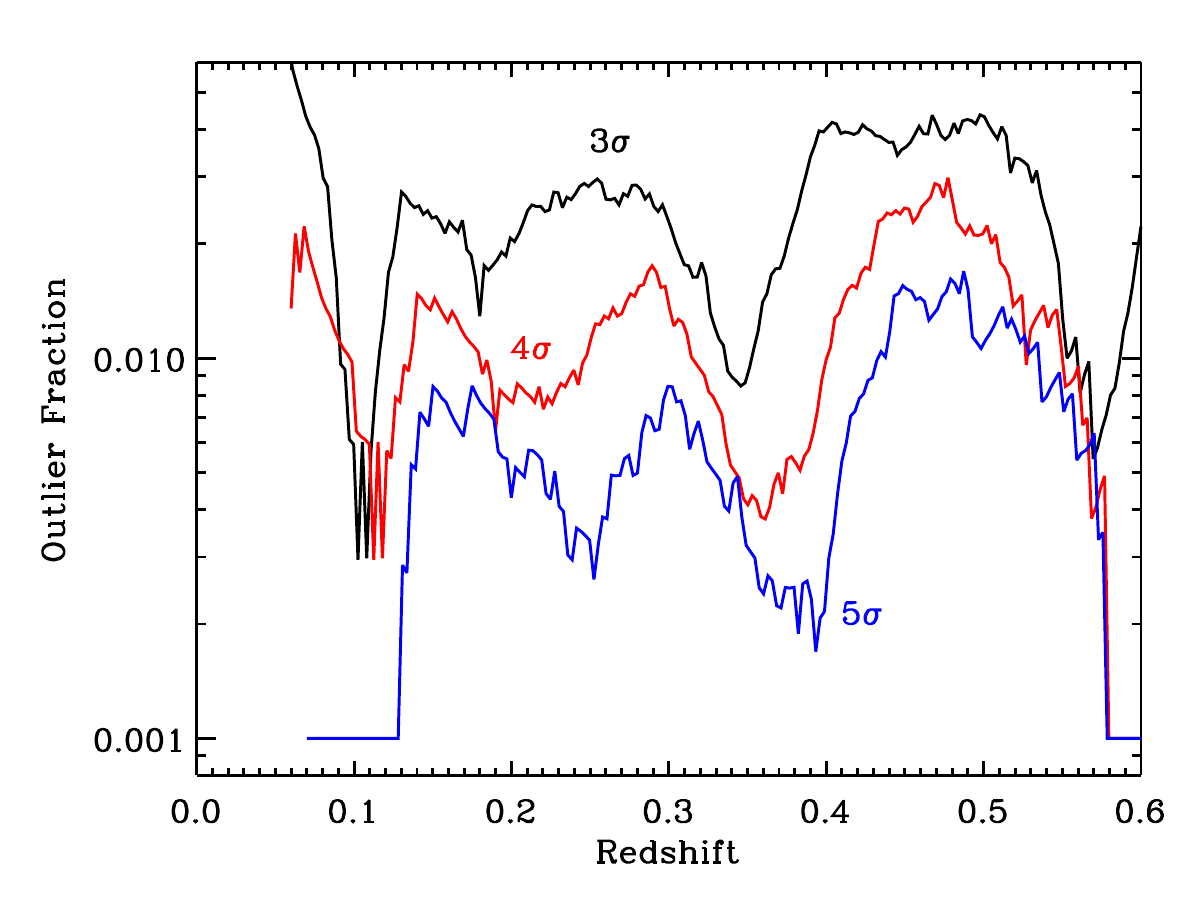}}
    \caption{Fraction of redshift outliers, as a function of photometric cluster redshift.
    A cluster is said to be an $N\sigma$ outlier if $|\zlambda - \zcg| \geq N\sigma_{\zlambda}$.
    We show the fraction of $3\sigma$, $4\sigma$, and $5\sigma$ outliers, as labelled.  These
    are computed using the full \redmapper\ cluster sample, with no additional spectroscopic requirements
    on member galaxies (unlike in Fig.~\ref{fig:zlambda-dr8_clean}).
    }
    \label{fig:zouts}
  \end{center}
\end{figure}

We can get a better sense of the fraction of gross redshift outliers from
Figure \ref{fig:zouts}, where we show the fraction of $3\sigma$, $4\sigma$, and
$5\sigma$ outliers.  A cluster is considered an $N\sigma$ outlier if $|\zlambda
- \zcg| \geq N\sigma_{\zlambda}$.  To estimate the fraction of outliers as a
function of redshift, for each redshift $z$ we collect all clusters with
redshift $\zlambda \in [z-0.025,z+0.025]$, and directly measure the fraction of
$N\sigma$ outliers.  By moving the window $[z-0.025,z+0.025]$ we recover the
outlier fraction as a function of redshift.  We see that $\approx 1\%$ of our
galaxy clusters are $4\sigma$ redshift outliers.  We note that the outlier
fraction is considerably larger than expected if the errors were simply
Gaussian.  We emphasize that this fraction is measured using the full cluster
sample, not the cleaned version used to produce
Figure~\ref{fig:zlambda-dr8_clean}.

\begin{figure}
  \begin{center}
    \scalebox{1.2}{\plotone{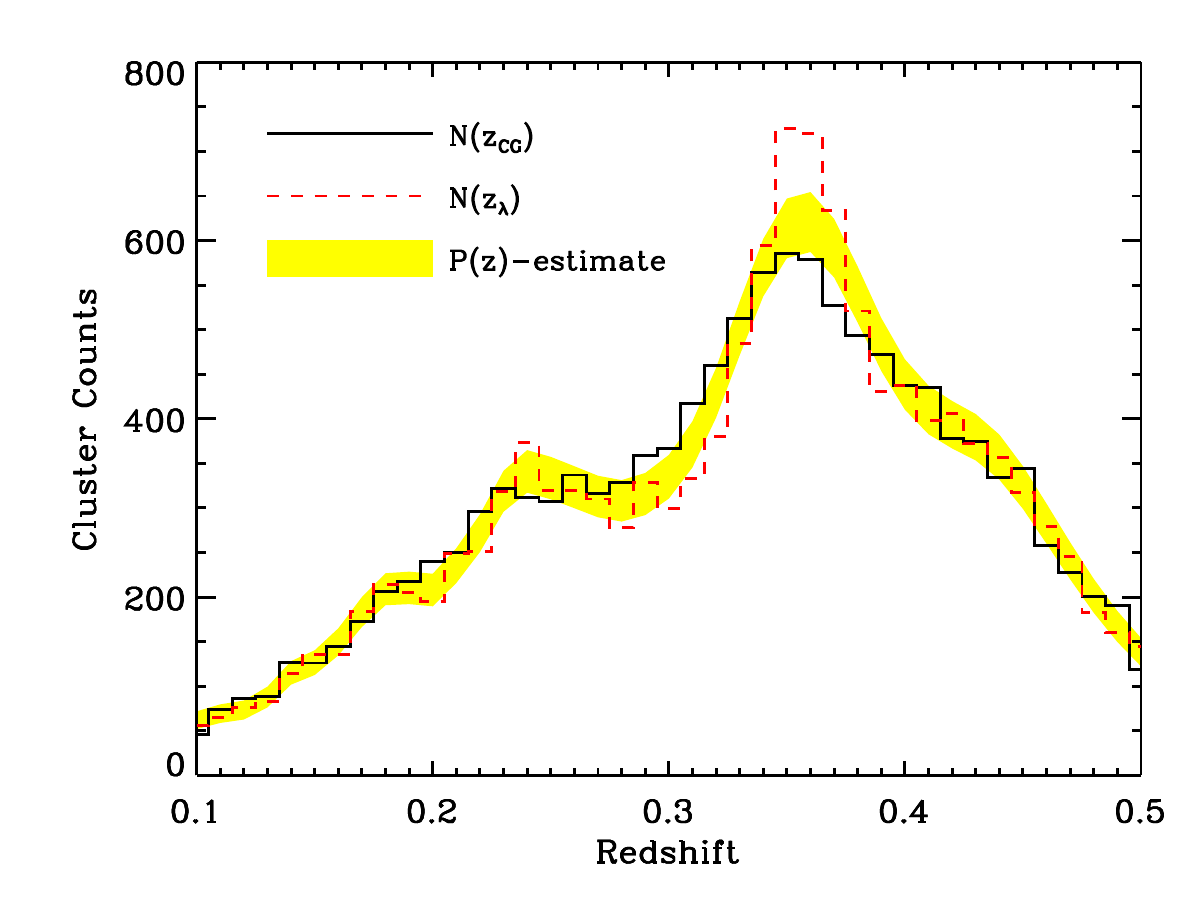}}
    \caption{Comparison of the true and predicted $N(z)$ distribution for
      \redmapper{} DR8 clusters with central galaxies with spectroscopic
      redshifts and $\lambda/S>20$.  We note that this is \emph{not}
      representative of the full cluster population due to uneven spectroscopic
      sampling.  The black solid histogram shows the ``true'' $N(\zcg)$.  The
      red-dashed histogram shows the results of binning the central values of
      $\zlambda$ for the same clusters, leading to obvious biases.  The yellow
      band ($\pm 1\sigma$ errors) shows the results of summing the cluster
      $P(z)$ values, and provides a good fit to the data.}
    \label{fig:zlambdapz}
  \end{center}
\end{figure}

Finally, in Figure~\ref{fig:zlambdapz} we test whether the \redmapper{}
estimates for the cluster redshift probability distributions
$P(\ztrue|\zlambda)$ are accurate.  First, we select all clusters with
spectroscopic central galaxies to create a ``true'' $N(\zcg)$, shown with a black
solid histogram.  We note that this is \emph{not} representative of the full
cluster population due to uneven spectroscopic sampling.  We compare this to
two estimates of $N(z)$ using the same set of clusters.  First, we bin clusters
using the central values of $\zlambda$, shown with the red-dashed histogram.
Second, we integrate $\sum P(\ztrue|\zlambda)$ over the appropriate redshift
bins, shown with a yellow band (including the expected measurement errors and
Poisson sampling, $\pm 1\sigma$).  The red-dashed histrogram is obviously not a
good fit to the spectroscopic redshift distribution.  In particular, there is
an artificial peak near the filter transition at $z=0.35$.  This is properly
smoothed out by our probability distribution estimate (yellow band), which is a
good fit to the spectroscopic data ($\chi^2/dof=45.0/40$).

\section{Cluster Centering}
\label{sec:centering}

The issue of galaxy cluster centering is very important for constraining
  cosmology with photometric surveys.  In particular, miscentered clusters are
  a leading source of systematic error in stacked weak-lensing mass
  estimates~\citep[e.g.,][]{johnstonetal07,mandelbaumetal08,rozoetal10a}, as
  well as mean velocity dispersions~\citep[e.g.,][]{bmkwr07}.  In addition, the
  cluster richness estimates themselves depend on the choice of
  center \citep[][]{lccgd06}.  Thus,
  a well-characterized centering model is essential for precision cosmology.

We assume every galaxy cluster halo has a bright, dominant galaxy residing at its 
center \citep[e.g.,][see also paper II]{mms64,oemler75,schombert86,vbkw07,vonderlindeetal12,menanteauetal13,mahdavietal13,songetal12,stottetal12}.
In our current implementation, we
also assume that the central galaxy is red, which is the case for the vast
majority of massive clusters.  The exceptions are strong cool-core clusters
such as Abell 1835~\citep{allen95}, where there is enough star formation for the broadband color of
the central galaxy to no longer be consistent with that of a red sequence galaxy~\citep[e.g.,][]{mrbsw06}.
Although blue central galaxies are more common (although still rare) at the
group scale~\citep[e.g.,][]{glbft11,mvcsm11,glbfm12,tglbf12}, the \redmapper{} clusters are much more
massive than the scale at which this is an issue.

Miscentering of galaxy clusters wherein the central galaxy is undergoing strong
star formation is a known failure of the \redmapper{} centering algorithm~(see
Section~\ref{sec:bluecent} and Paper II).  Simply removing the requirement that central galaxies be consistent
with the red sequence --- i.e. relying solely on luminosity and proximity ---
can fix some of these clusters, but at the expense of miscentering $\sim 10\%$
of the clusters on foreground galaxies\footnote{We note that foreground
  galaxies are much more likely to be confused as centrals than background
  galaxies because they tend to be brighter in apparent magnitude}.  Likewise,
our tests have shown that both galaxy centroids and luminosity weighted galaxy
centroids result in worse centering properties than the algorithm currently
implemented below~\citep[e.g., see also][]{glbfm12}. Thus, centering on red galaxies is,
as far as we can tell, the ``least bad'' option.  In its current
implementation, the centering success rate is $\approx 85\%$ (see Paper II).
We intend to continue working on improving our centering model for future data
releases, as this is currently the dominant source of systematic failures in
the \redmapper\ cluster catalog.


\subsection{Basic Framework}
\label{sec:centframework}

We introduce a fundamentally new way of thinking about identifying the
central galaxy of a cluster: rather than specifying a unique cluster center,
\redmapper\ estimates the probability that a given galaxy is the central galaxy
of the cluster.  Some clusters have well defined cluster centers, exhibiting a
single galaxy with a centering probability
$\Pcen\approx 1$, whereas others can have two or more reasonable central
candidates, with the most likely center having $\Pcen\approx 50\%$.  We note
that these centering probabilities are the angular-position equivalent
of the standard \photoz{} distributions $P(z)$.  That is, just as a cluster has an
uncertain redshift position characterized by a redshift probability
distribution, so too does the cluster have an uncertain angular position on the
sky, characterized by the probability of any given galaxy of being the correct
cluster center.  The importance of this new way of treating cluster centering
is that it opens up the possibility of a statistical treatment of cluster
centering akin to the statistical treatment of photometric redshifts, allowing
us to improve our estimates of the cluster richness functions and cluster
correlation functions.  A detailed description of this framework will be
presented and tested in a future work.

The key insight that allows us to estimate centering probabilities is that 
there are \emph{three} different types of galaxies in a cluster: a central
galaxy (``CG''), satellite galaxies, and unassociated foreground and background
galaxies.  Let $\bx$ be an observable vector for a galaxy, e.g., color (via
$\zred$), luminosity ($\imag$), and position of each galaxy.
We define $\ucen$, $\usat$, and $\ufg$ as the distribution
of $\bx$ for central, satellite, and background galaxies respectively.
The $\ucen$ and $\usat$ filters are assumed to depend on cluster redshift and richness,
while $\ufg$ depends only on cluster redshift (via $\zred$).
We use the subscript ``fg'' as we expect foreground galaxies will be more likely to be
misidentified as CGs.  Given a galaxy with observable $\bx$, the probability
that it is the central galaxy of a cluster is
\begin{equation}
  \pcen(\bx|\lambda,\zlambda) = \pfree \frac{\ucen}{\ucen + \lsat\usat + \ufg},
  \label{eqn:pcen}
\end{equation}
where $\pfree$ is the probability that a galaxy has not been partially masked
by a higher ranked cluster (as described in Section~\ref{sec:cf:percolation};
typically $\pfree \approx 1$), and $\lsat = \lambda - 1$ is the total number of satellite galaxies.  This
formula can be thought of as the simple definition of probabilities, or it can
be interpreted as a Bayesian classification algorithm.  

Note, however, that the probability $\pcen$ is not the same thing as the probability
$\Pcen$ that the galaxy is the \emph{unique} central galaxy of the cluster.  
By assumption, there can be only one
central galaxy, so if galaxy $i$ is the central galaxy, then every other galaxy
$j\neq i$ must not be a central.  Consequently, the probability that galaxy $i$
be \emph{the} central galaxy of a cluster is
\begin{equation}
  \Pcen \propto \pcen(\bx_i) \prod_{j \neq i} (1 - \pcen(\bx_j)).
  \label{eqn:Pcen}
\end{equation}
The proportionality constant is set by the condition that there is just one
central galaxy in the cluster,
\begin{equation}
  1 = \sum_i \Pcen(\bx_i).
\end{equation}

In addition to the central galaxy probability, we can also calculate the
probability that a cluster is centered on a satellite galaxy, given by
\begin{equation}
  \Psat =
  (1-\Pcen)\frac{\lambda_{\mathrm{sat}}\usat}{\lambda_{\mathrm{sat}}\usat + \ufg}
  \label{eqn:Psat}
\end{equation}
All that remains for us to be able to estimate centering probabilities is the definition of the filters
$\ucen$, $\usat$, and $\ufg$.


\subsection{Centering Filters}

With the basic formalism laid out, we need to specify the observable $\bx$ and
the corresponding filters.  There are three observables that we use to select
the CG: the galaxy $i$-band magnitude $\imag$; the red-sequence photometric
redshift $\zred$ of the galaxy; and a weight $w$ that characterizes the local
cluster galaxy density around the proposed central galaxy.   We also explored
replacing our photometric redshift $\zred$ with $\chi^2$, the ``distance'' in color
space to the red sequence.  However, we have found empirically that $\zred$
works better for estimating central probabilities, in that small amounts of
star formation and/or small color errors due to deblending have a much smaller
impact on $\zred$ than they do on $\chi^2$. 
We consider each of the filters in turn.

\subsubsection{Luminosity Filter: $\phicen$}

The magnitude of the CG is correlated with both richness and redshift, so we
define the CG magnitude filter
\begin{equation}
  \phicen(\imag|\bar{m}_i,\sigma_m) =
  \frac{1}{\sqrt{2\pi}\sigma_m}\exp\left ( -
  \frac{(\imag-\imagbar)^2}{2\sigma_m^2} \right ),
  \label{eqn:phicen}
\end{equation}
where in principle both $\imagbar$ and $\sigma_m$ depend on richness and
redshift.  In practice, we expect $\sigma_m$ to be roughly redshift
independent, whereas $\imagbar$ obviously depends on redshift.  
We assume that $\imagbar$ traces $m_*(z)$, so that the full richness
and redshift dependent parameterization of $\imagbar$ is
\begin{equation}
  \imagbar(\zlambda,\lambda) = m_*(\zlambda) + \Delta_0 + \Delta_1\ln\left [
    \frac{\lambda}{\lambda_p} \right ],
\end{equation}
where $\Delta_0$ and $\Delta_1$ are redshift independent constants, and
$\lambda_p$ is the median richness of the sample.  Our algorithm for fitting for
$\Delta_0$ and $\Delta_1$ is detailed below.

\subsubsection{Photometric Redshift Filter: $\Gcen(\zred)$}

For the photometric redshift filter, we use the red-sequence photometric
redshift $\zred$ for each galaxy in the field.  We model this as a Gaussian
function, with the form:
\begin{equation}
  \Gcen(\zred|\zlambda) = \frac{1}{\sqrt{2\pi}\sigzred}\exp\left ( \frac{-(\zred -
    \zlambda)^2}{2\sigzred^2} \right ).
\end{equation}
As the error in the single galaxy photometric redshift dominates that from
the cluster photometric redshift, we have set the scatter in $\Gcen(\zred)$ to
that of the individual galaxy.  In addition to the photometric redshift filter,
we employ a hard cut such $\chisq(\zred) < 100$ (with 4 degrees of freedom).  Investigations of DR8
spectroscopic galaxies have shown that galaxies with $\chisq>100$ are
\emph{all} catastrophic outliers in $\zred$, which is not surprising
considering the bad fit to the red-sequence template.  By allowing galaxies
with $\chisq<100$, we allow some flexibility for galaxies that have slightly
offset colors to still be considered as central galaxies.  This is especially
an issue for SDSS DR8 for bright, nearby central galaxies that may have color
shifts caused by deblending problems.

\subsubsection{Local Galaxy Density Filter: $\fcen(w)$}

The motivation behind the local galaxy density filter is to define an
observable $w$ that is a pseudo-gravitational potential connecting each galaxy
to every other cluster member.  The weight $w$ assigned to a given central
candidate is
\begin{equation}
  w = \ln \left [ \frac{\sum (\pmem(\bx_i)L_i
      [r_i^2+r_c^2]^{-1/2})}{\Rc(\lambda)^{-1} \sum (\pmem(\bx_i) L_i)} \right ],
\end{equation}
where the sum is over all galaxies within the scale radius $\Rc(\lambda)$
around the candidate central, $r_c = 50\,h^{-1}\mathrm{kpc}$ is a core radius
used to soften the $1/r$ dependence, $L_i$ is each galaxy's $i$-band
luminosity, and $\pmem$ are the usual $\lambda$ membership probabilities.  The
denominator is chosen to make the argument of the natural log dimensionless,
and to remove the obvious dependence of the numerator of $w$ on the total
number of terms in the sum.  Normalized in this fashion, we expect that $w$
does not scale with cluster richness nor redshift.

We assume that for central galaxies, $w$ follows a log-normal distribution
$\fcen(w)$,
\begin{equation}
  \fcen(w) = \frac{1}{\sqrt{2\pi}\sigma_w} \exp \left [ -
    \frac{(\ln(w)-\wcenbar)^2}{2\sigma_w^2} \right ].
  \label{eqn:fcen}
\end{equation}
As noted above, we expect $\wcenbar$ to be richness and redshift independent.
On the other hand, $\sigma_w$ will certainly depend on richness.  The noise in
$w$ should scale with raw galaxy counts $(\lambda/S)^{1/2}$, where $S(z)$
is the redshift-dependent factor that relates the raw-galaxy counts to a richness
estimate when the survey is not sufficiently deep to reach $0.2L_*$ at
the redshift of the cluster (see Eqn.~\ref{eqn:sz}).
For Poisson noise, we set
\begin{equation}
  \sigma_w = \sigwcen \left ( \frac{\lambda}{S\lambda_p} \right
  )^{-1/2}
\end{equation}
where $\sigwcen$ is a constant that we fit for.  As above, the
pivot point $\lambda_p$ should be chosen to match the median richness of the
sample.

With these definitions, the product
\begin{equation}
  \ucen = \phicen(\imag|\zlambda,\lambda) \Gcen(\zred)
  \fcen(w|\zlambda,\lambda)
  \label{eqn:ucen}
\end{equation}
is the filter characterizing the distribution of central galaxies.

\subsubsection{Satellite Filter: $\usat$}

Satellite galaxies on the red sequence can be described by a filter function
analogous to Eqn.~\ref{eqn:ucen}.  Therefore, we have
\begin{equation}
  \usat = \phisat(\imag|\lambda,m_*) \Gsat(\zred)
  \fsat(w|\zlambda,\lambda),
  \label{eqn:usat}
\end{equation}
where $\fsat(w)$ is defined in the same way as Eqn.~\ref{eqn:fcen}, except
with parameters appropriate for the satellite galaxies,
$\wsatbar$ and $\sigwsat$.  The satellite
luminosity function, $\phisat$, is a Schechter function as described in
Eqn.~\ref{eqn:lumfilter}.  The redshift filter $\Gsat(\zred)$ is identical
to $\Gcen(\zred)$.

\subsubsection{Foreground Filter: $\ufg$}

The foreground filter is defined as the expected number of unassociated
galaxies within the cluster radius $R_c(\lambda)$,
\begin{equation}
  \ufg = \barsigz(\imag,\zred) \ffg(w) \frac{\pi R_{c}^2}{d_A^2},
\end{equation}
where $\barsigz$ is the background density per $\mathrm{deg}^2$ per $\imag$ per
$\zred$, calculated in a similar fashion as the red sequence background
described in Section~\ref{sec:background}.  In addition, the
area subtended by the cluster in $\mathrm{Mpc}^{2}$ must be converted to
$\mathrm{deg}^2$ via the angular diameter distance $d_A$, with $d_A$ measured
in $\Mpc/\deg$.  Finally, the $\ffg$
filter describes the local galaxy density filter from Eqn.~\ref{eqn:fcen}, with
parameters appropriate for random points ($\wfgbar$ and $\sigwfg$) as
described below.


\subsection{Implementation}

Implementing this formalism requires that we calculate the parameters that
describe the filters for central galaxies, satellite galaxies, and
foreground/background galaxies.  Of course, calibrating these parameters depends
on having a training sample to start with.  As usual, we approach this problem
in an iterative fashion,
where the centering model is constrained at the same time as the \redmapper{}
red-sequence model.  In the first iteration, we generate a catalog with roughly
correct centering, and use this to provide an initial calibration of the
filters.  In subsequent iterations we make use of the centering filters and use
the output to recalibrate.  This procedure is iterated until convergence.

\subsubsection{First Iteration and Initial Filter Calibration}
\label{sec:centfirstiter}

First, we implement a rough centering algorithm: for every cluster, we simply select
the brightest high probability ($\pmem>0.8$) member galaxy as the central
galaxy of the cluster.  In this fashion, we obtain a full training
catalog with a set of central galaxies that should be roughly correct.

We now use this first iteration of a central galaxy (CG) catalog to determine the filter
parameters that we will use in subsequent iterations.  Note that because
the initial CG catalog contains some miscenterings, when calibrating the CG filters
it is important to account for this contamination. 

If a cluster is improperly centered on a satellite galaxy, it is most often
centered on the brightest satellite. 
Consequently, the luminosity distribution of satellites
which are mistaken as central is \emph{not} simply a Schechter function
($\phisat$).  Rather, the luminosity distribution of satellite galaxies in the
CG catalog is given by $\phi_1(\imag|\lambda,m_*)$, the magnitude distribution
of the brightest satellite in clusters of richness $\lambda$.  The expected
magnitude and $\zred$ distribution of the galaxies in our CG catalog is
\begin{equation}
  \begin{split}
  \rho(\imag,\zred|z) =& \Pcen\phicen(\imag)G(\zred|z) \\&+
  \Psat\phi_1(\imag)G(\zred|z) \\&+ (1-\Pcen-\Psat)\barsigz\frac{\pi R_c^2}{d_A^2},
  \end{split}
  \label{eqn:rhocen}
\end{equation}
where $\Pcen$ is the probability that the galaxy in question is the central
galaxy, as in Eqn.~\ref{eqn:Pcen}, and $\Psat$ is the probability that the
cluster is centered on the brightest satellite galaxy.  The redshift, $z$, is
the spectroscopic redshift of the ``seed'' galaxy used in the training step.

Our primary goal is to constrain the parameters $\Delta_0$, $\Delta_1$, and
$\sigma_m$.  However, we also have the parameters $\Pcen$ and $\Psat$, which are
unknown in the first iteration.  For these parameters, we have found that
setting them in the first iteration at any reasonable initial estimate ($\Pcen
\in [0.7,0.9]$; $\Psat \in [0.05,0.2]$) has no marked effect on the final
calibration of the filter parameters.  Therefore, for simplicity we set $\Pcen
= 0.9$ and $\Psat=0.05$ for each individual cluster in this first iteration.

Before we can continue, we must estimate the parameters for
$\phi_1(\imag|\lambda,m_*)$.  This is modeled as a Gaussian distribution with
central value
\be
\mbarsat = m_*(z) + c_{\phi_1} +
s_{\phi_1}\ln(\lambda/\lambda_p),
\ee
where $m_*(z)$ is obtained from
Eqn.~\ref{eqn:mstar} and $\lambda_p$ is the median richness of the sample.  The
width of the distribution is similarly modeled as 
\be
\sigma_{\mathrm{sat}} = c_{\sigma,\phi_1} + s_{\sigma,\phi_1}\ln(\lambda/\lambda_p).
\ee 
The central value must scale with richness because as we sample more galaxies from the
luminosity function, we are more likely to find a very bright galaxy.  In order
to obtain these parameters, we run a simple Monte Carlo with the luminosity
function parameters from Section~\ref{sec:nfwandlumfilter}.  For $\alpha=-1.0$
with $\lambda_p=30$, we find that $c_{\phi_i} = -0.95$, $s_{\phi_i} = -0.32$,
$c_{\sigma,\phi_1} = 0.40$, and $s_{\sigma,\phi_1} = -0.09$.  We note that
these parameters depend only on $\alpha$ and $\lambda_p$, and thus do not need
to be updated in subsequent iterations.

Finally, in order to constrain the parameters $\Delta_0$, $\Delta_1$, and
$\sigma_m$, we define our likelihood based on Eqn.~\ref{eqn:rhocen}
\begin{equation}
  \ln\lk = \sum_i \ln\rho(\imag,\zred|z).
\end{equation}
By maximizing this likelihood with respect to $\Delta_0$ and $\Delta_1$ 
we can use our training clusters to estimate the $\ucen$ filter parameters.

\subsubsection{Calibrating the $w$ Filters}
\label{sec:centcalw}

We now turn to calibrating the $w$ filters.  We begin by calibrating the
foreground and satellite filters, $\ffg(w)$ and $\fsat(w)$.  For these purposes
we assume that all satellites follow the same spatial profile independent of
brightness.  Both $\ffg(w)$ and $\fsat(w)$ can be calibrated in a Monte Carlo
fashion.  We note that foreground galaxies will uniformly sample the cluster
area $\pi R_c^2(\lambda)$, and thus to evaluate $\ffg(w)$ we draw random points
uniformly within the disc of every training cluster, and compute the
corresponding $\ffg(w)$ parameters, $\wfgbar$ and $\sigwfg$.

The satellite filter $\fsat(w)$ is computed in a similar fashion.  First, for
every training cluster we randomly select a cluster member with a probability
$p$ that is proportional to the membership probability.
For this randomly selected member we compute $w$
at the location of that satellite.  After computing $w$ for all the training
clusters, we compute the corresponding $\fsat(w)$ parameters, $\wsatbar$ and
$\sigwsat$.

We can now turn our attention to the distribution $\fcen(w)$.  Consider now the
distribution $f(w)$ for all central galaxies.  This total distribution is then
\begin{equation}
  f(w) = \Pcen\fcen(w) + \Psat\fsat(w) + (1-\Pcen-\Psat)\ffg(w).
\end{equation}
The only unknowns in this equation are the mean $\wcenbar$ and rms
$\sigwcen$ for the central filter, so we write the likelihood
\begin{equation}
  \ln\lk(\wcenbar,\sigwcen) = \sum_i\ln f(w),
\end{equation}
where the sum is over all training clusters.  We then maximize this
likelihood to find $\wcenbar$ and rms $\sigwcen$.

\begin{figure}
  \begin{center}
    \scalebox{1.0}{\plotone{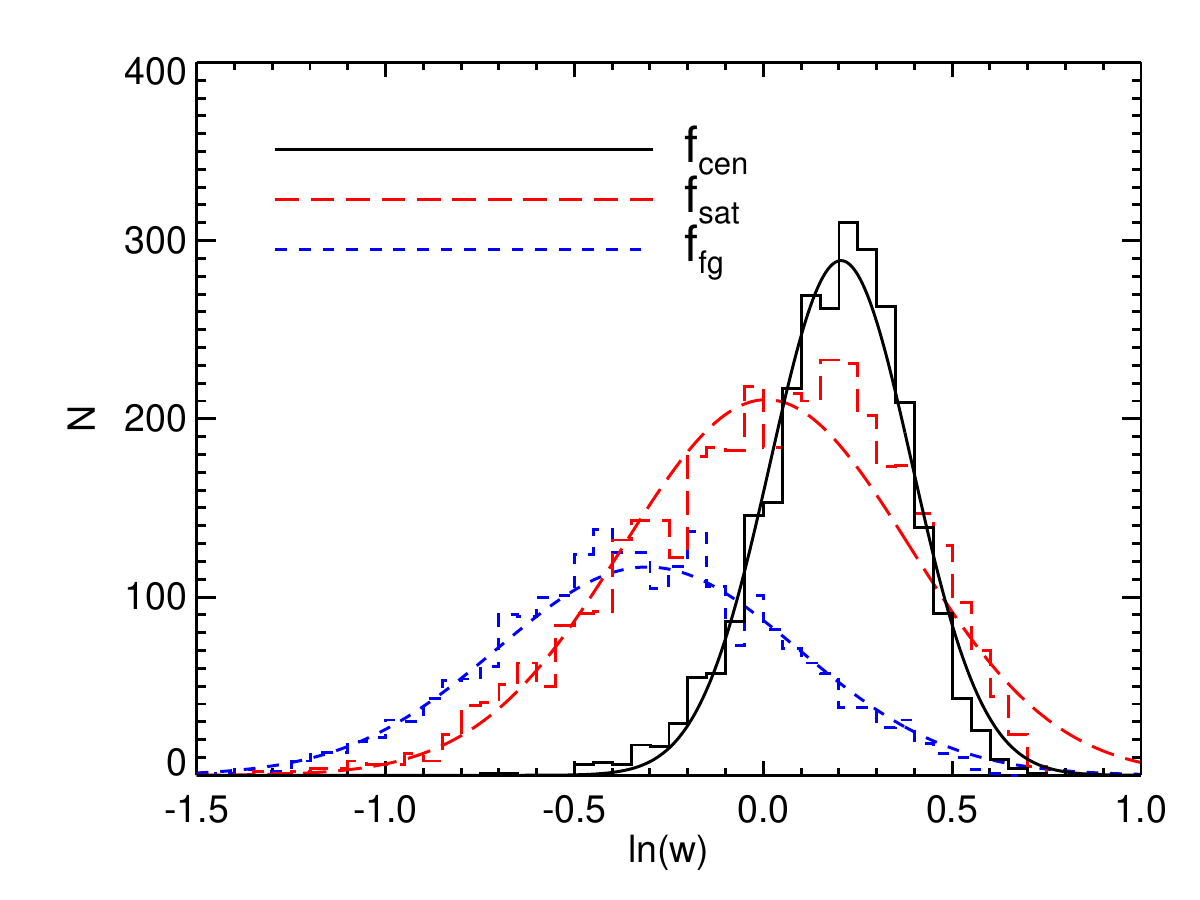}}
    \caption{Histograms of $w$ parameter for central (black solid), satellite
      (red long-dashed), and foreground/random (blue short-dashed) galaxies. It
      is clear that there is some power here to differentiate between central
      galaxies and satellites and foregrounds, but it is not perfect.  However,
      there is some advantage in being able to reject a bright
      galaxy with a low local density as a likely interloper that does not fit the central galaxy
      model.}
    \label{fig:fw}
  \end{center}
\end{figure}

In Figure~\ref{fig:fw} we show the central, satellite, and foreground $f(w)$ filters
for the final training iteration.  It is clear that there is some power here to
differentiate between central galaxies and satellites and foregrounds, but it
is far from perfect.  In particular, satellites are only slightly less well
connected than central galaxies.  

\subsubsection{Subsequent Iterations}

As noted above, in our first iteration our catalog of CG galaxies
is constructed using a simple centering algorithm: i.e., select the
brightest high-probability member as the cluster center.
In subsequent iterations of the cluster finder calibration we use the
probabilistic centering algorithm described in Section~\ref{sec:centframework}.
After application of our centering algorithm we have the important advantage 
that each cluster now comes tagged with $\Pcen$ and $\Psat$.  Therefore, we can now
repeat the calibrations from Sections~\ref{sec:centfirstiter} and
\ref{sec:centcalw} while using the correct $\Pcen$ and $\Psat$ for each individual
galaxy.  In this way we continuously improve our centering model with multiple
iterations of the cluster finding algorithm.

\subsection{A Note on Blue Central Galaxies}
\label{sec:bluecent}

\begin{figure}
  \begin{center}
    \scalebox{1.0}{\plotone{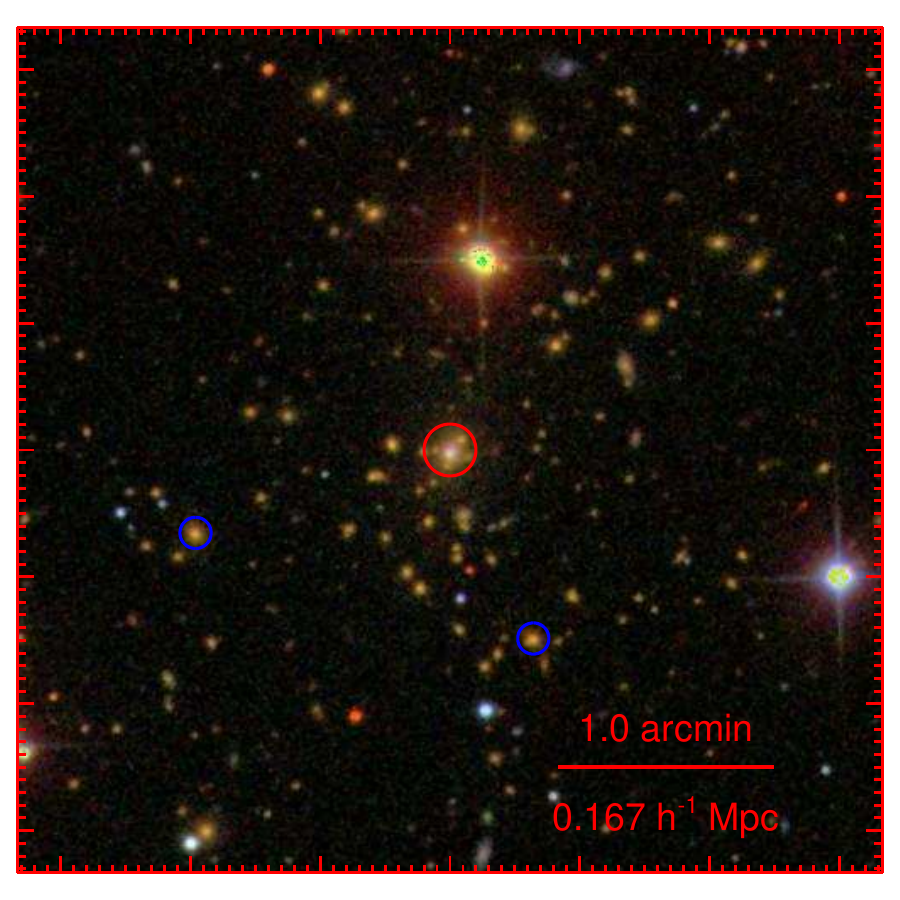}}
    \caption{SDSS image of Abell 1835, a well-known cool-core cluster with
      strong star formation in the central galaxy (marked with red circle).
      The two candidate \redmapper{} centers are denoted with blue circles, and
      the true center is missed due to its strong deviation from the red
      sequence.}
    \label{fig:a1835}
  \end{center}
\end{figure}

At the beginning of this section, we noted that central galaxies undergoing
strong star formation pose a particular problem for the \redmapper{} centering
algorithm.  In Figure~\ref{fig:a1835} we show an SDSS image of Abell~1835, a
well-known strong cool-core cluster with star formation in the central galaxy.
The true center is denoted with a red circle, while the top two candidate \redmapper{}
centers are circled in blue.  As expected, when the color of the central galaxy
is far from the red sequence, our centering algorithm instead chooses one of
the bright red satellites as a possible center.  However, we emphasize that the
cluster will not be missing, it will simply be miscentered.

We have also made use of SDSS spectra to get an initial estimate on the rate of
bad centers due to galaxies that do not agree with our color model (both very
blue and very red).  We have taken every cluster with a spectroscopic central,
and a second brighter spectroscopic galaxy that is within $r<R_\lambda$ and
$|\zcg - z_{\mathrm{spec}}| < 1.5\sigma_v$, where $\sigma_v$ is the expected
velocity dispersion based on the cluster richness (Rozo et al., in prep).  If
this galaxy is not in the top 5 members list for the cluster, it is considered
a possible catastrophic center.  In all, $4.6\%$ of the clusters fit this
category.  Visual inspection of the richest clusters shows that the
majority of these galaxies are in the outskirts of the cluster and are
not centrals, confirming that the \redmapper{} centering algorithm is not
simply taking the brightest member as the center.  Alternatively, we can look
at the subset of these galaxies that are not in the members list and thus were
not considered as candidate centrals at all.  Only $0.7\%$ of clusters contain
such galaxies.  From these estimates, we expect the rate of miscentering due to 
galaxies that are too blue to be selected as central galaxies is $\lesssim 2\%$,


\section{The Cluster Finder}
\label{sec:clusterfinder}

\begin{figure}
\begin{center}
\scalebox{1.1}{\plotone{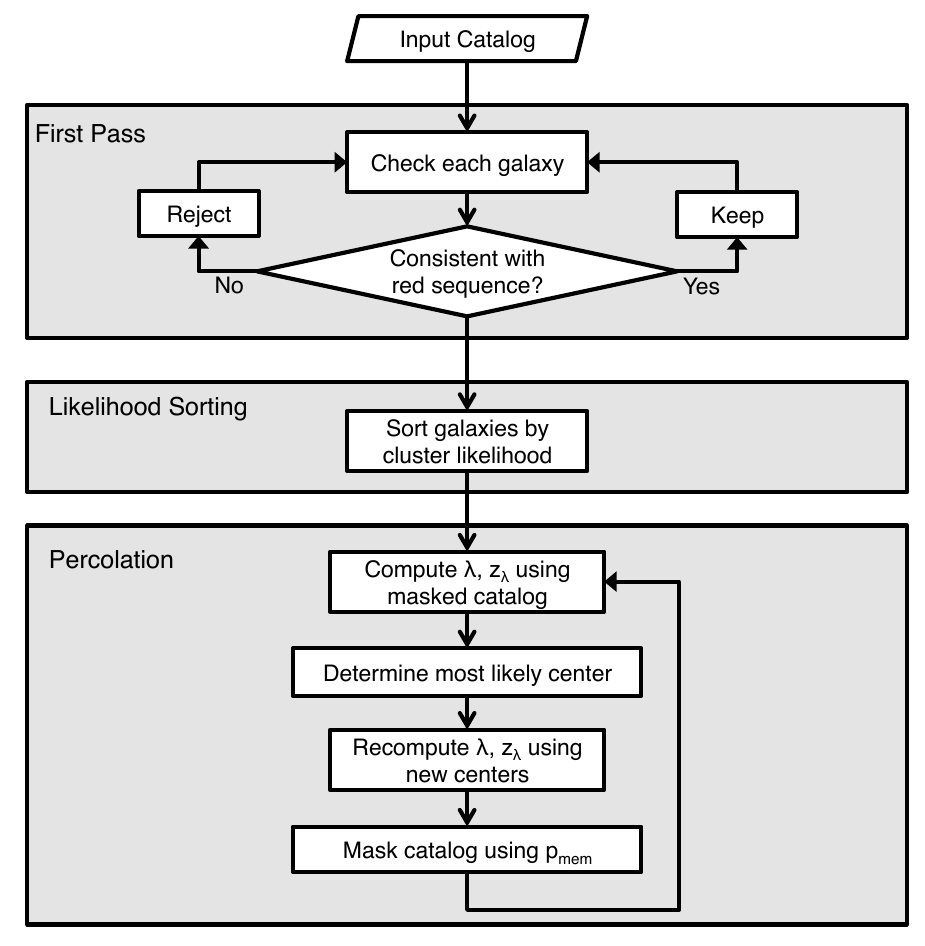}}
\caption{Process flowchart for the \redmapper{} cluster finder, as described in
  this Section}
\label{fig:flowchart}
\end{center}
\end{figure}

We have now described in detail all the ingredients that go into the
\redmapper\ cluster finder.  Here, we focus on how these ingredients
are blended within the context of the cluster finder to produce a catalog.
In particular, we discuss how clusters are ultimately defined and percolated
to ensure that every cluster is found once and only once.
From a practical perspective, the cluster finding is broken into three
stages.    First, we look for overdensities around each
individual galaxy using $\zred$ as an estimate of the cluster redshift.
Second, we calculate the cluster likelihoods
for each of the galaxies that have a sufficient overdensity.  Third, after
sorting by cluster likelihood, we percolate through the full catalog while
probabilistically masking out cluster members.  A process flow-chart for
reference in this section is shown
in Figure~\ref{fig:flowchart}.

\subsection{First Pass}
\label{sec:cf:firstpass}

In the first pass we wish to identify galaxies that are
credible centers of galaxy clusters.
This task involves a lot of data handling, and so we wish to make it as
efficient as possible.

We begin by taking every galaxy in the input catalog with $\chisq(\zred)<20$
and brighter than $0.2L_*$ at the red-sequence photometric redshift
$\zred$.\footnote{In the case of the training runs, we take every ``seed''
  galaxy at the spectroscopic redshift.}  These are very generous
  cuts, yet they reduce the input DR8 catalog
from 56 million galaxies to 23 million possible cluster centers in the redshift
region $0.05<\zred<0.6$.  Next, we take all galaxies
within $0.5\hMpc$ of a candidate center and measure $\lambda$, setting the
cluster redshift to $\zred$.  Candidate centers with
$\lambda/S<3$ are rejected.  The scale value $S=1/(1-C)$
(from Section~\ref{sec:corrterm}) is used to ensure that we have detected at least
three red galaxies above the magnitude limit.  This cut rejects
a further $\sim60\%$ of the catalog of candidate centrals.  
Finally, for all centers that pass these cuts we
calculate $\zlambda$ as described in Section~\ref{sec:zlambda} to better refine
the redshift of the possible cluster.


\subsection{Likelihood Sorting}
\label{sec:cf:sort}

Given our list of possible clusters from the first pass, we now calculate the
cluster likelihood for each of these clusters.  The total likelihood is a
combination of the $\lambda$ likelihood
and the centering likelihood.  To calculate the $\lambda$ likelihood, we first
calculate the richness $\lambda$ using the optimized radial scale parameters with
$R_0 = 1.0\,\hMpc$ and $\beta=0.2$ as described in Section~\ref{sec:lambdachi}.
The $\lambda$ likelihood is then given by
\begin{equation}
  \ln\lk_\lambda = -\frac{\lambda}{S} - \sum \ln(1 - \pmem),
  \label{eqn:lambdalike}
\end{equation}
where $\lambda$ is evaluated at the cluster photometric redshift $\zlambda$.

Next, following Section~\ref{sec:centering} and Eqn.~\ref{eqn:ucen}, the
centering likelihood is given by
\begin{equation}
  \ln\lk_{\mathrm{cen}} = \ln [ \phicen(\imag|\zlambda,\lambda) \Gcen(\zred)
  \fcen(w|\zlambda,\lambda) ],
  \label{eqn:cenlike}
\end{equation}
where we combine the luminosity, $\zred$, and local galaxy density $w$ of each
galaxy. The total likelihood used in the ranking of possible cluster centers is
\begin{equation}
  \ln\lk = \ln\lk_\lambda + \ln\lk_{\mathrm{cen}}.
  \label{eqn:clusterlike}
\end{equation}

Note that the amplitude of the $\lambda$ likelihood function is
typically much larger than that of the centering likelihood.  Thus, to
zeroth order, clusters are first ranked by $\lambda$ likelihood.  Two candidate
centrals with similar $\lambda$ likelihoods are then ranked according to the
central likelihood.  As will be described below, we refine the choice of
cluster center in the percolation step, so the initial centering likelihood is not
especially influential in determining the final cluster center.


\subsection{Percolation}
\label{sec:cf:percolation}

Having rank-ordered the cluster candidates according to likelihood, we now need
to percolate the list to assign galaxies to clusters and ensure that no cluster is
counted multiple times.  The basic outline of the percolation proceeds as
follows.
\begin{enumerate}
  \item{Given cluster number $i$ in the list, recompute  
    $\lambda$ and $\zlambda$ based on the
    percolated galaxy catalog.  At the beginning of the percolation, the percolated
    galaxy catalog is simply the input galaxy catalog.}
    \item{Determine the cluster center and centering probability via the method
      outlined in Section~\ref{sec:centering}.}
    \item{Perform a final calculation of $\lambda$ and $\zlambda$ with respect
      to the new central galaxy.}
    \item{Update the percolated galaxy catalog by masking out galaxies based on their membership probabilities.}
    \item{Remove all lower-ranked possible centers that have a membership
      probability $\pmem>0.5$ of being a member of cluster $i$.
      Note these galaxies are still allowed to provide membership weight
      to lower-ranked clusters as part of the percolated galaxy catalog.}
    \item{Repeat Step 1 for the next cluster galaxy in the ranked list.}
\end{enumerate}

\subsubsection{Masking Galaxies}

Masking galaxies based on their membership probabilities is the ``probabilistic
percolation'' step of the \redmapper{} algorithm.  To perform this step, we
keep track of the ``total probability'' that a galaxy belongs to a cluster,
which we call $\ptaken$.  The probability $\pfree=1-\ptaken$ is the probability
that the galaxy does not belong to any cluster.  Initially, one has $\ptaken=0$
and $\pfree=1$ for all galaxies.  Upon finding a galaxy cluster, the entire
galaxy catalog is percolated by updating the probability $\ptaken$ via
\be
p_{\mathrm{taken},i+1} = p_{\mathrm{taken},i} + p_{\mathrm{free},i}\pmem
\ee
where $\pmem$ is given by Eqn.~\ref{eqn:pmem}.  

Now, when we re-estimate the richness of cluster $i+1$, we must take into account
the fact that some of the galaxies have a non-zero probability of belonging to
a cluster $j < i+1$.  We do so by modifying the richness calculation from
Eqn.~\ref{eqn:lambdadef} via
\begin{equation}
  \lambda = \sum \pfree \pmem(\bx|\lambda).
  \label{eqn:lambdadefpfree}
\end{equation}
The first factor above is simply the probability that a galaxy is ``free'' to
belong to the new cluster, and $\pmem$ is the standard membership probability
from Eqn.~\ref{eqn:pmem}.  For instance, suppose a galaxy has a probability
$\pmem=0.3$ of belonging to the first cluster in the rank-ordered list.  In
this case, the galaxy still has $70\%$ of its probability to give to a cluster
lower in the list.  In practice, when quoting cluster membership probabilities
$\pmem$ we report not the raw $\pmem$ value as given by Eqn.~\ref{eqn:pmem}, but
rather the product $\pfree\pmem$ for that galaxy--cluster pair.  That is, the
reported value is the correct probability that the given galaxy belongs to the
cluster under consideration.  For galaxy clusters that are sparse in the sky (e.g.,
at high richness) these corrections are negligible.


\subsubsection{Extent of Clusters and Percolation Radius}
\label{sec:cf:extent}

As noted above, cluster richness is measured within a radius $R_c(\lambda)$
that optimizes the signal-to-noise ratio of the richness measurements (R12),
but is not in any way chosen to be related to standard definitions of the extent
of a halo, say $R_{200c}$.  For cosmological purposes, it is useful to differentiate
between the radius $R_c(\lambda)$ which defines the richness measurement, and
the \emph{percolation radius} that is used to mask out cluster members and
blend or deblend nearby systems.  In particular, ideally one selects the
percolation radius so that it matches as best as possible the percolation radii
employed in the halo definition used to calibrate the corresponding halo mass
function.

In the appendix of R12 we used maxBCG clusters to obtain an approximate scaling
of mass to $\lambda$ richness\footnote{As shown in Appendix~\ref{app:lamcol},
  $\lcol$ used in R12 is within $\sim10\%$ of the multi-color $\lambda$ used in
  this work.}.  We found that the slope of the mass--$\lambda$
relation is consistent with 1, and that $R_{200c} \approx 1.5R_c(\lambda)$.
Consequently, we have adopted $1.5R_c(\lambda)$ as our default percolation
radius.  That is, galaxies are masked out to this radius.  We note that while
galaxies outside the $R_c(\lambda)$ radius are not used in the summand in
Eqn.~\ref{eqn:lambdadef}, we can still estimate $\pmem$ in exactly the same way
as we do with all other galaxies out to an arbitrary radius, which is how we
implement the large percolation radius above.

In practice, for the $\lambda/S>20$ richness threshold we have employed, changing
the mask radius by $\pm 50\%$ has a very small impact on the resulting cluster
catalog.  Only a small number of clusters ($\sim5\%$) -- primarily satellites
of the richest $\lambda>100$ clusters -- are affected at all by making this change.
We expect to return to the question of what the optimal masking radius is in
future work, particularly within the context of cosmological constraints from
galaxy clusters.

\subsection{A Sample Cluster}
\label{sec:sample}

\begin{figure*}
  \begin{center}
    \scalebox{1.0}{\plottwo{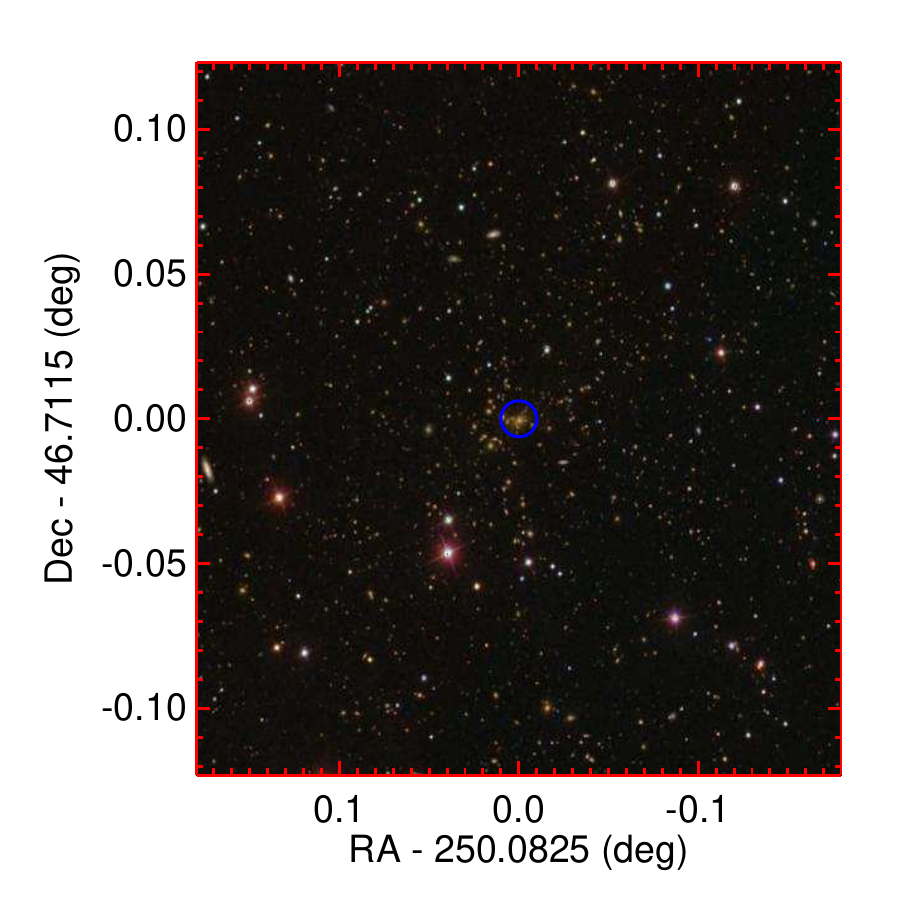}{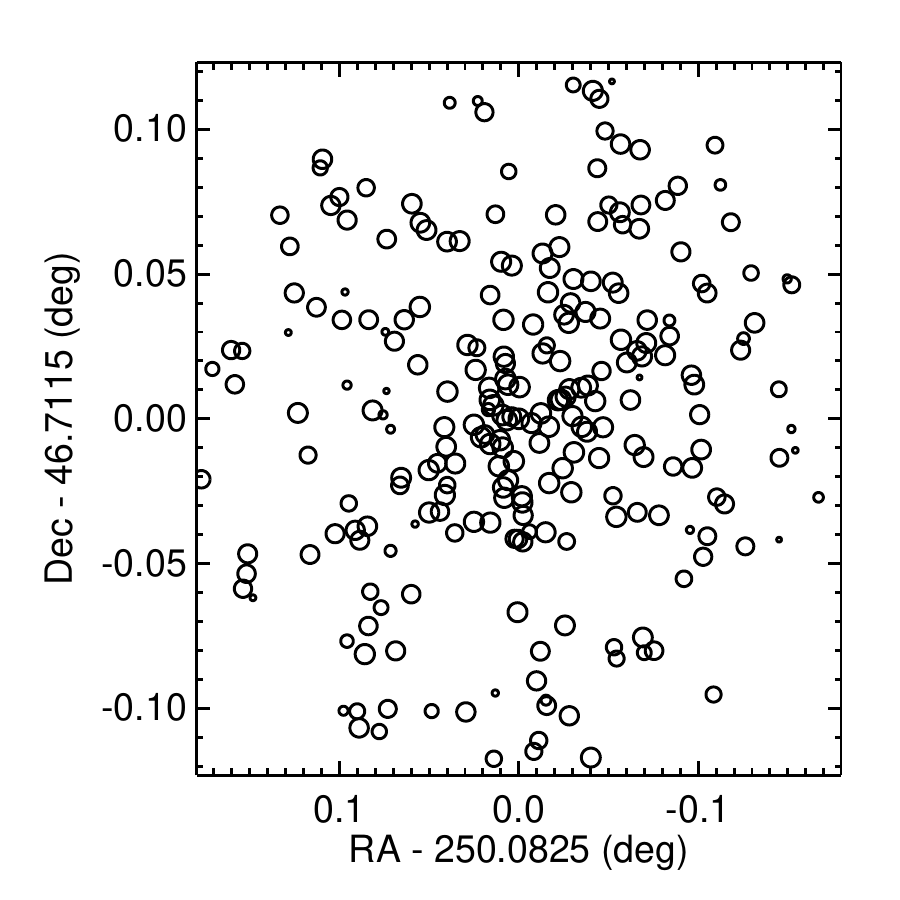}}
    \scalebox{1.0}{\plottwo{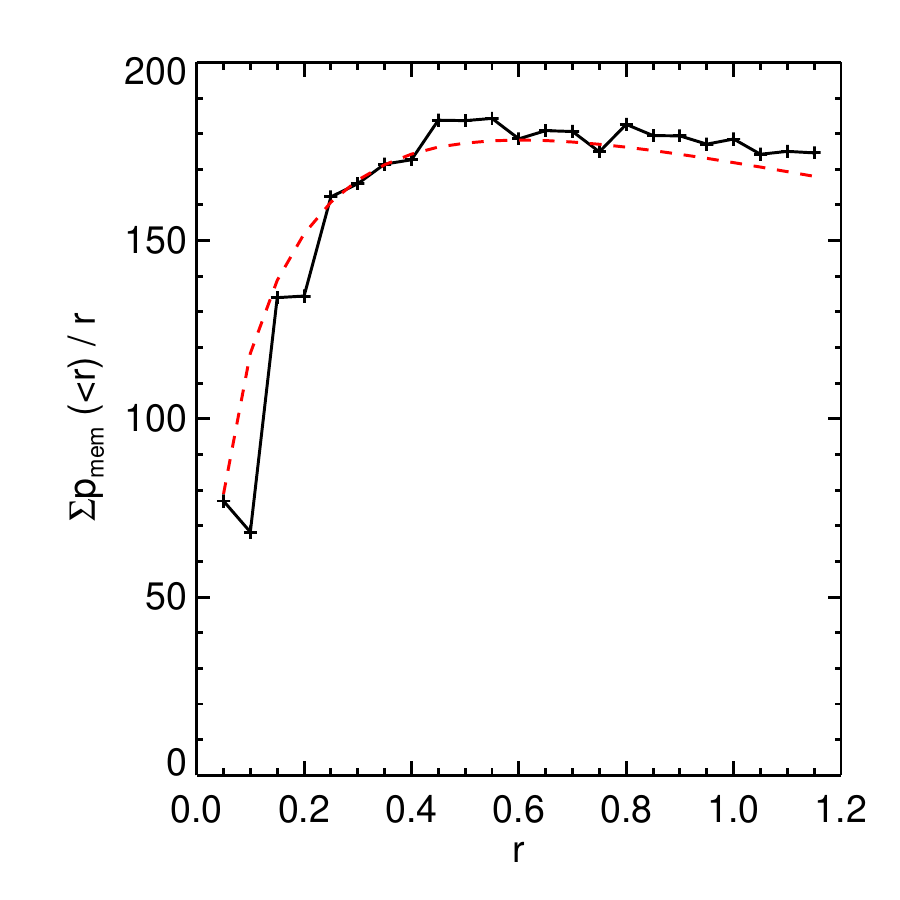}{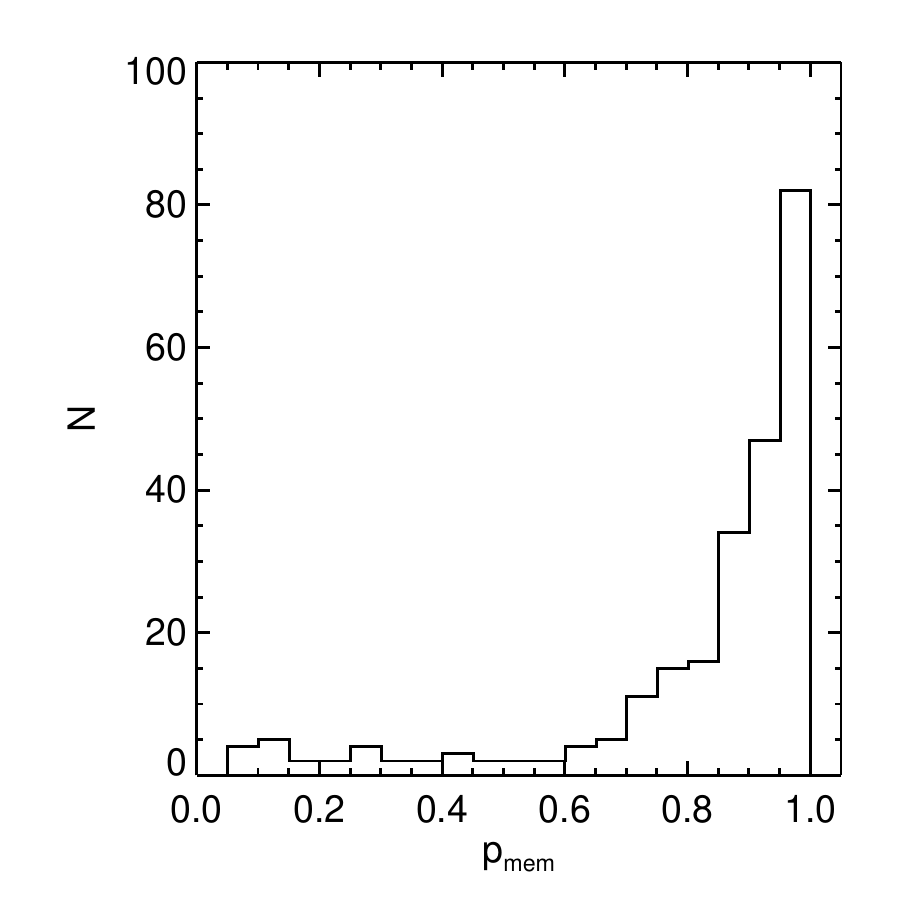}}
    \caption{\emph{Top Left:} SDSS image of RM~J164019.8+464241.5 (Abell 2219).  \emph{Top Right:}
      Locations of cluster members with $\pmem>0.05$, with the symbol size
      proportional to the membership probability. \emph{Bottom Left:}
      Distribution of cluster members as a function of radius.  The dashed red line
      shows the expected distribution of galaxies for an NFW profile with
      $r_s=0.5\,h^{-1}\,\mathrm{Mpc}$.  \emph{Bottom Right:} Distribution of
      $\pmem$ for cluster members.  This is strongly peaked near $\pmem>0.9$.  We note that we do not show another, larger peak at $\pmem=0$,
for the vast majority of galaxies in the field that are not
red-sequence cluster members.}
    \label{fig:illustration}
  \end{center}
\end{figure*}

At this point, it would be useful to investigate a sample rich cluster to
explore the distribution of cluster members.  We have selected
RM~J164019.8+464241.5 (Abell 2219), one of
the richest clusters in SDSS, at a redshift of $z=0.23$.  In
Figure~\ref{fig:illustration} we show four aspects of this cluster.  In the
top-left panel, we show the SDSS image of the cluster.  In the top-right panel,
we show the distribution of cluster members with $\pmem>0.05$, with the symbol
size proportional to the membership probability.  As can be seen in the histogram
of $\pmem$ in the bottom-right panel, this distribution is strongly peaked near
$\pmem>0.9$.  We note that we do not show another, larger peak at $\pmem=0$,
for the vast majority of galaxies in the field that are not
red-sequence cluster members. Finally, in the bottom-left panel we show the
distribution of cluster members as a function of radius.  The dashed red line shows
the expected distribution of galaxies for an NFW profile with
$r_s=0.5\,h^{-1}\,\mathrm{Mpc}$, thus showing that the radial distribution of
cluster galaxies is broadly consistent with the NFW model.

\section{The redMaPPer SDSS DR8 Cluster Catalog}
\label{sec:catalog}

\begin{figure}
  \begin{center}
    \hspace{-0.2in} \scalebox{1.2}{\plotone{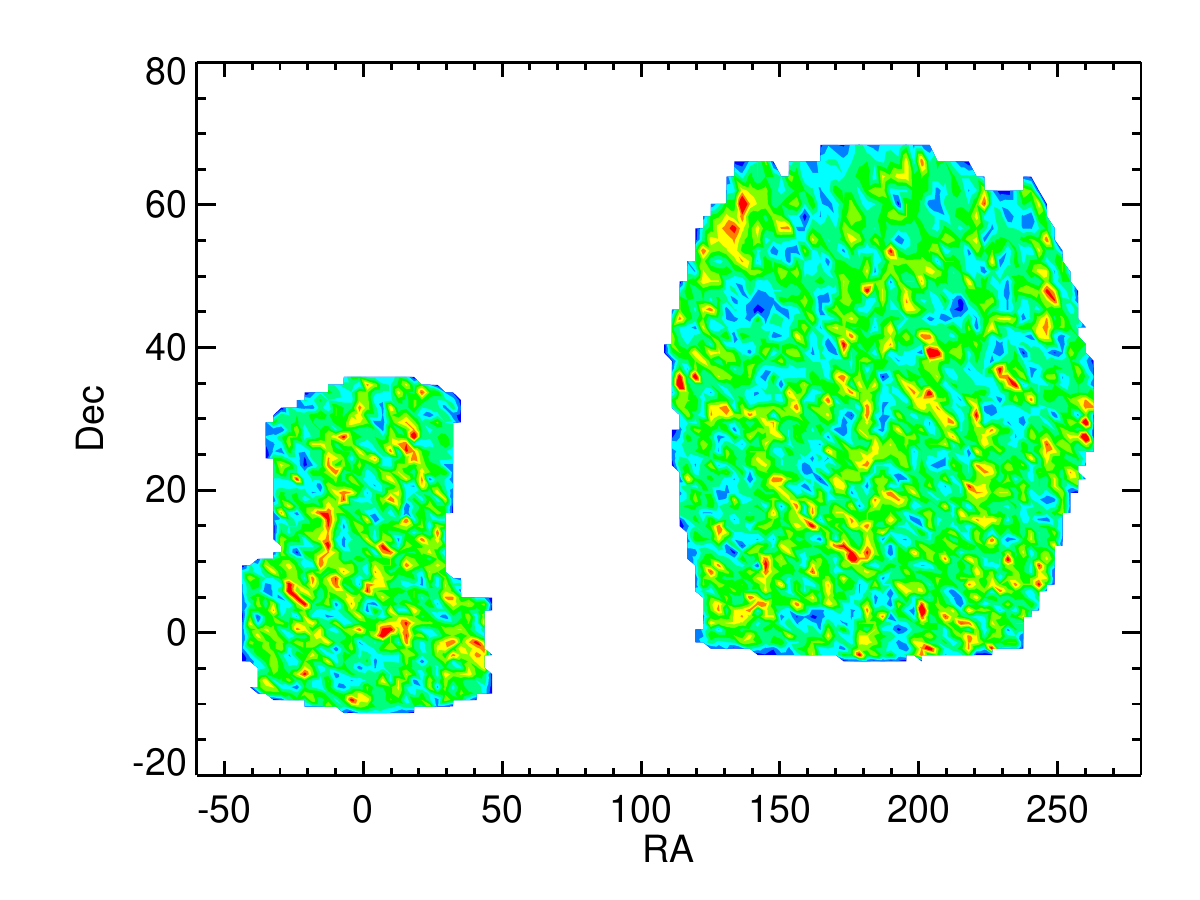}}
    \caption{Footprint of the \redmapper{} DR8 catalog, with clusters binned
      into a Mangle simple pixelization scheme of depth 7.  All clusters with $\lambda>5$ and
      $\zlambda\in[0.1,0.3]$ are shown to better illustrate the large-scale
      structure in the catalog.}
    \label{fig:footprint}
  \end{center}
\end{figure}

We have run the \redmapper\ cluster finding algorithm in the SDSS DR8
photometric catalog described
in Sec.~\ref{sec:data}. The full cluster finder run contains all clusters with
$\lambda \geq 5S(\zlambda)$ and $\zlambda\in[0.05,0.6]$.  However, 
we have chosen to apply very conservative cuts to our catalog .  The
cuts we apply are:
\begin{enumerate}
\item{The richness is cut to $\lambda \geq 20S(\zlambda)$.  Roughly speaking,
	this requires that every cluster have at least 20
    galaxy counts above the flux limit of the survey or $0.2L_*$ at the cluster
    redshift, whichever is higher.  From R12, we estimate that this results in
    an effective mass cut of $M_{200} \gtrsim 10^{14}\,M_\sun$.}
\item{The redshift range is cut to $\zlambda\in[0.08,0.55]$, so as to minimize
    edge effects from the training sample.}
\item{Only clusters with $\fmask < 0.2$ are included (see
    Eqn~\ref{eqn:maskfrac}), ensuring clusters are not overly compromised by
    bad fields, bright stars, and survey edges.}  
\end{enumerate}
The resulting cluster catalog contains $25,236$ systems.
In Figure~\ref{fig:footprint} we show the full footprint of the catalog.  The color
scale shows the density contrast relative to the mean cluster density, where
red regions are denser than average and blue regions are less dense, as
estimated using all $\lambda \geq 5$, $\zlambda \in [0.1,0.3]$ clusters so as
to give a better sense of the large scale structure in the survey.  The image was produced by
binning the catalog into a Mangle simple pixelization scheme of depth 7~\citep{sthh08}.

\begin{figure}
  \begin{center}
    \scalebox{1.2}{\plotone{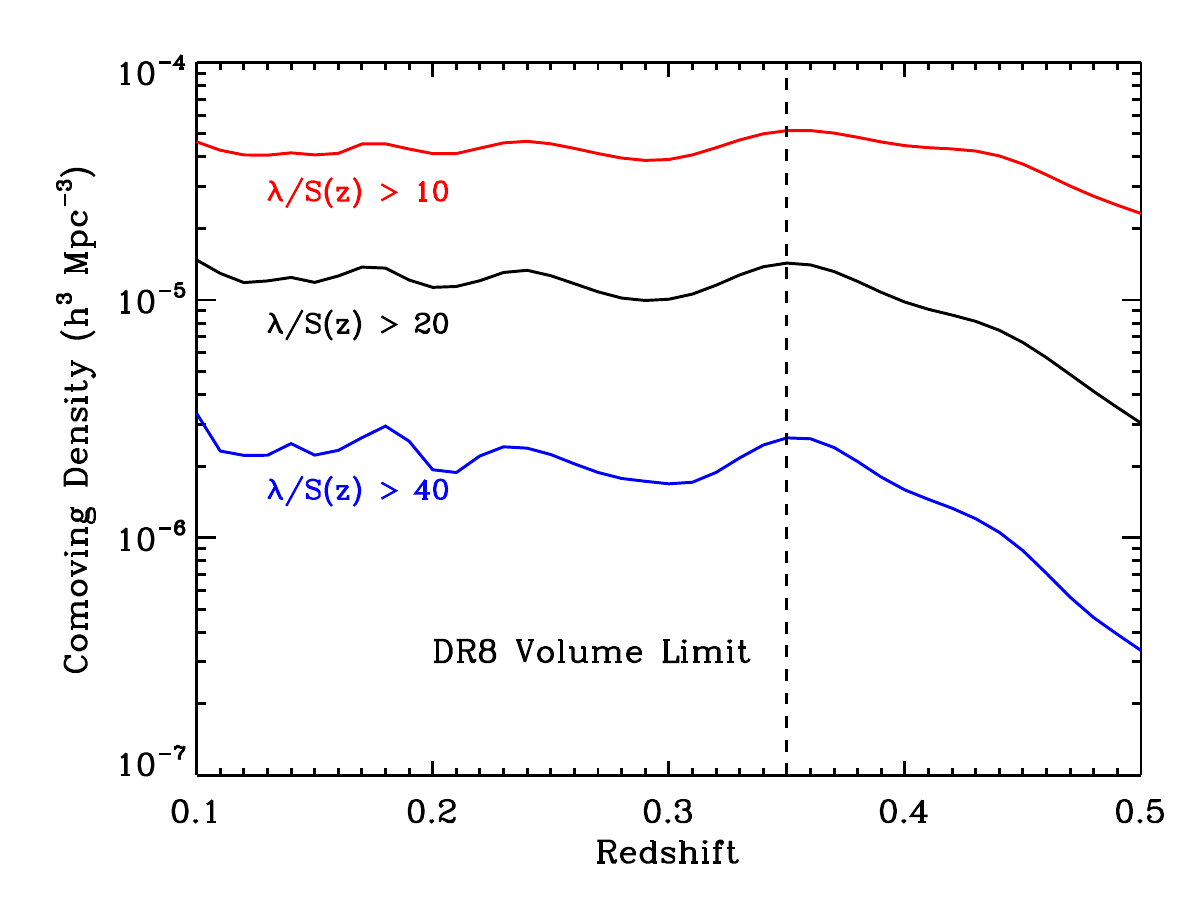}}
    \caption{Comoving density of \redmapper{} DR8 clusters as a function of
      photometric redshift ($\zlambda$) for clusters with $\lambda/S(z) >
      10,20,40$.  All densities have been computed by taking the sum of cluster
      $p(z)$.  The comoving density is roughly constant at $\zlambda<0.35$, where the
      catalog is volume limited (denoted by the vertical dashed line).  
      Above this redshift the comoving density falls off rapidly as the detection
      threshold rapidly increases.}
    \label{fig:codens}
  \end{center}
\end{figure}

\begin{figure}
  \begin{center}
    \scalebox{1.2}{\plotone{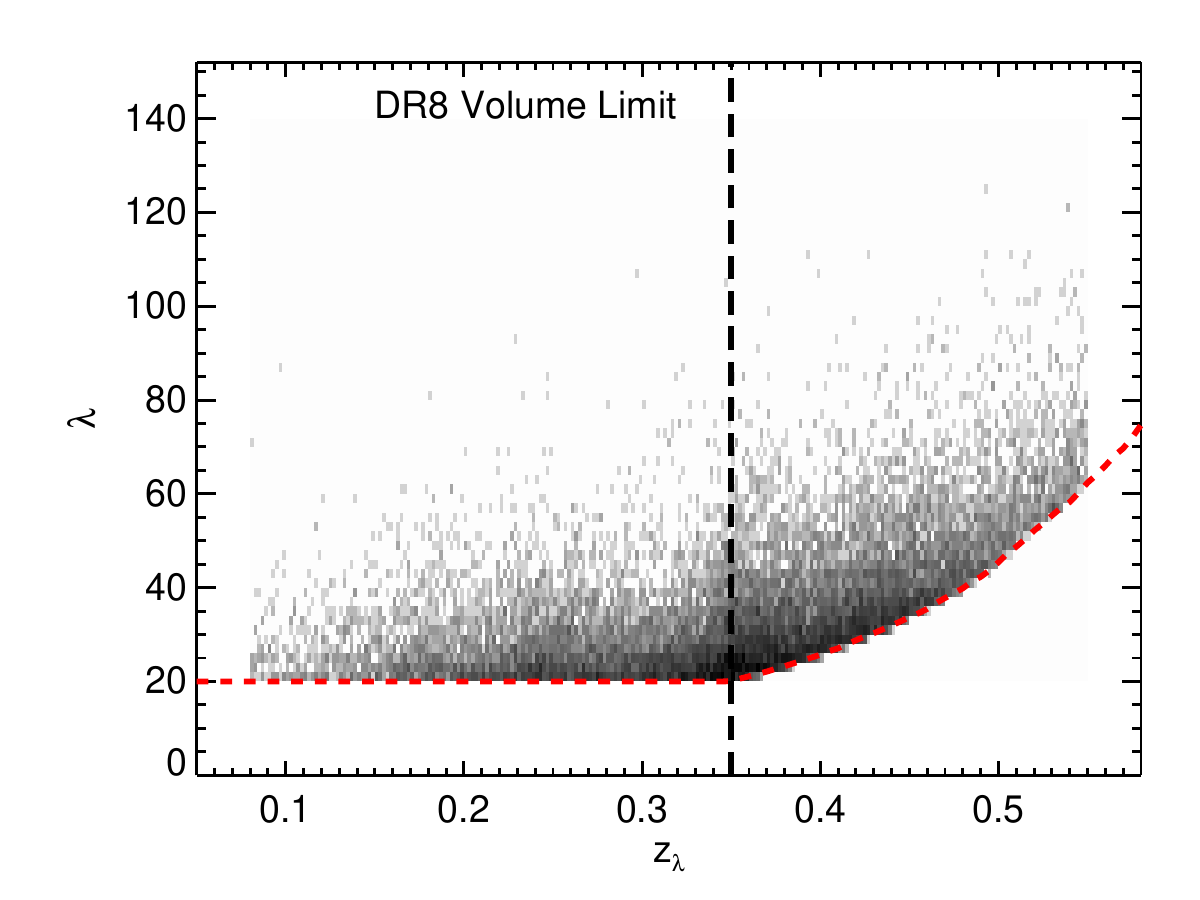}}
    \caption{Two-dimensional histogram of $\lambda$ vs. $\zlambda$ for
      \redmapper{} clusters.  The red dashed line shows the
      redshift-dependent richness cut of $\lambda > 20 S(z)$.
      Although this figure generally shows a smooth distribution, the
      boost in low richness clusters at the transition redshift of
      $z=0.35$ apparent; this redshift is denoted by the vertical
      dashed line.}
\label{fig:lamvzlam}
\end{center}
\end{figure}

In Figure~\ref{fig:codens} we show the comoving density of \redmapper{} clusters
with $\lambda/S(z)>20$ over the full redshift range of interest.  The comoving
density is roughly constant at $\zlambda\lesssim0.35$ where the catalog is
volume limited.  At $\zlambda\sim0.35$ the richness and redshift scatter are
significantly boosted by both the $4000\,\mathrm{\AA}$ break filter transition
and the magnitude limit of the survey reaching $0.2L_*$.  Therefore, the
comoving density is boosted by low richness clusters scattering up into our
sample.  A full accounting for this scatter must be made in order to precisely
calculate the \redmapper{} abundance function, which we leave to future work.
Above this redshift the magnitude limit starts to kick in (via the scale factor
$S(z)$), and we only observe the most massive clusters.  As an illustration of
this effect, in Figure~\ref{fig:lamvzlam} we plot the richness $\lambda$ vs the
photometric redshift $\zlambda$ for the final \redmapper{} catalog.  The red
dashed line shows the redshift-dependent richness cut $\lambda > 20 S(z)$.  

\begin{figure}
  \begin{center}
    \hspace{-0.1in} \scalebox{1.1}{\plotone{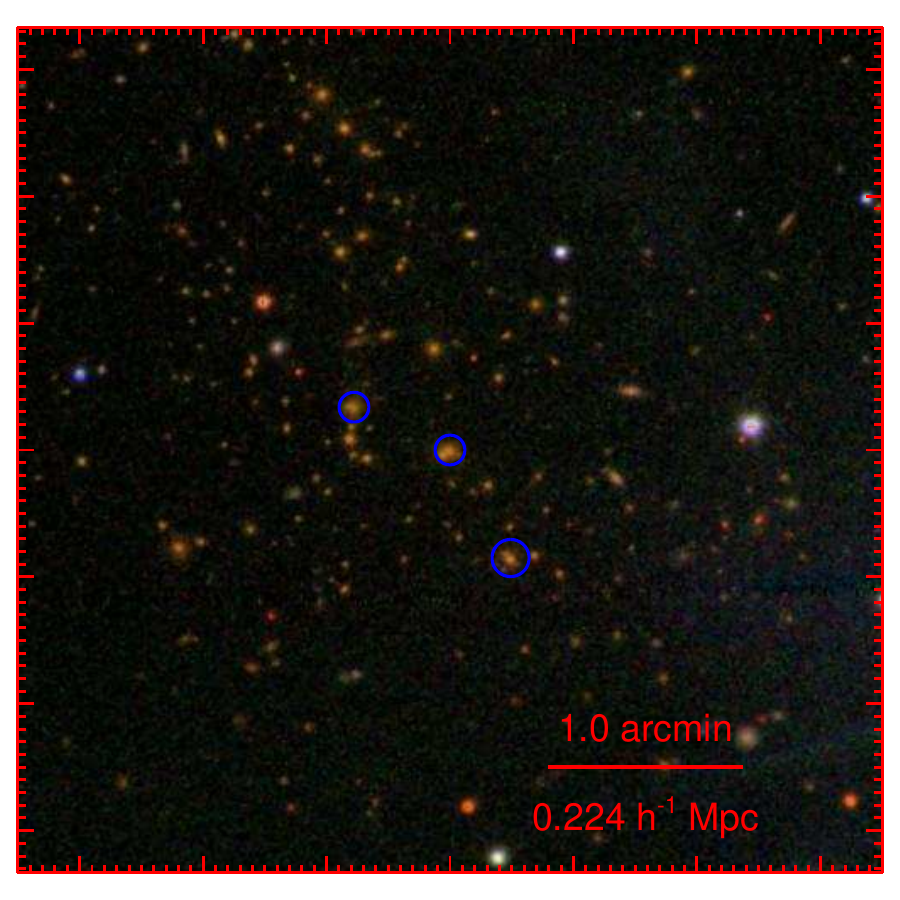}}
    \caption{SDSS composite image of the cluster RM J003208.2+180625.3, the
      richest \redmapper{} cluster not found within the MCXC cluster catalog.
      This system has $\lambda=236\pm12$ at a redshift of
      $\zlambda=0.396\pm0.013$, and is associated with a source in the ROSAT
      Bright Source Catalog.  This particular cluster has three candidate
      centers, denoted with blue circles, with $\Pcen=\{0.5,0.25,0.25\}$.}
\label{fig:sample_cluster}
  \end{center}
\end{figure}

 Finally, we show a sample
\redmapper\ cluster. In Figure~\ref{fig:sample_cluster}
we show REDM~J003208.2+180625.3, the richest \redmapper\ cluster not found within the MCXC cluster catalog \citep{piffarettietal11},
a system with $\lambda = 236\pm12$ at redshift
$\zlambda=0.396\pm0.013$. We note that this cluster is associated with a source
in the ROSAT Bright Source Catalog~\citep{vogesetal99}.  The specific data
available for each of the clusters and members are described in
Appendix~\ref{app:catalog}, and a detailed comparison of the \redmapper{}
clusters to X-ray and SZ catalogs is presented in Paper II.

\section{Purity and Completeness}
\label{sec:purcomp}

Purity and completeness can mean many different things depending on the
context.  There is a tendency to think of purity as the probability that a
cluster in the catalog is a real cluster, and to think of completeness as the
probability that a real cluster is in the catalog.  However, it is often
incorrect to think of these quantities as calibrating failure rates of the
algorithm.  Here, we adopt specific definitions of purity and completeness and
discuss their implications.  For an alternative definition of purity and
completeness by comparing \redmapper{}-detected clusters to X-ray catalogs, we
refer the reader to Paper II.

For cluster cosmology, the relevant quantity is the probability of detecting a
halo of mass $M$ with richness $\lobs$, which can be decomposed into a
convolution of two components,
\begin{equation}
  P(\lobs|M) = \int d\ltrue P(\lobs|\ltrue) P(\ltrue|M).
  \label{eqn:plobsm}
\end{equation}
The probability $P(\ltrue|M)$ is a feature of the Universe, and must be
properly marginalized over in any cosmological study that relies on the cluster
number function.  Constraining this probability distribution can also be
supplemented by utilizing realistic mock catalogs~\citep[e.g.,][Wechsler et
  al., in preparation]{smbwd12} which we return to in future work.  On the
other hand, $P(\lobs|\ltrue)$ is a feature of the cluster finding algorithm
itself.  This probability fully contains all of the information associated with
measurement error in our catalog.  In the present work, we define completeness
and purity as specific integrals over this distribution.

Purity and completeness, used as a simple parametrization of $P(\lobs|\ltrue)$,
can be estimated in several ways.  Perhaps the simplest
consists of removing galaxy clusters, randomizing galaxy positions, and then
re-inserting galaxy clusters.  The cluster finding algorithm can then be rerun,
and one can determine which clusters are detected, and how many ``false''
clusters are detected~\citep[e.g.,][]{gotoetal02,kmawe07a,hmkrr10}.  However, as
shown in the literature \citep{lopesetal04,rrknw11}, such an algorithm is fundamentally flawed because
background galaxies are not uniformly distributed.  Consequently, we take a
somewhat different approach, as described below.

\begin{enumerate}
  \item{\emph{Generate random points:} We generate a list of random points
    uniformly sampling the input survey mask.  These points are to be the centers
    of mock clusters that will be inserted into the data set. 
    This procedure ensures we sample all the
    systematics in the survey, as well as the effect of masked regions.  We
    note that these locations are not sampled from the DR8 galaxy positions.}
  \item{\emph{Sample ``true'' cluster richness and redshift:} Using the full
    cluster catalog ($\lambda>5$), we randomly sample galaxy clusters
    to generate pairs of parameters $(\ltrue,\ztrue)$.
    This ensures our model cluster distribution has the same
    richness and redshift distribution as the final catalog, including covariances.}
  \item{\emph{For each pair of sampled values $(\ltrue,\ztrue)$, assign them a spatial
  location using a random point from Step 1, and
  sample galaxies using the cluster model:} Using the
  same method as in Section~\ref{sec:correval}, we use Monte Carlo sampling to
  generate 5000 galaxies with the model radial and luminosity profiles.  
  From this sample of 5000 cluster galaxies, we randomly sample
  $\ltrue$ galaxies from within $R_c(\ltrue)$, as well as $k\ltrue$
  galaxies from $R_c(\ltrue) < r < R_c(2\ltrue)$, where $k$ scales with
  $\ltrue$ as appropriate for the radial profile.  This ensures that our fake
  clusters do not have artificial hard edges.}  
  \item{\emph{Measure $\lobs$ for the generated fake cluster at the random location, 
    and repeat 100 times.}  When measuring $\lobs$, we mask out galaxies
  according to the bright stars and edges in the survey mask, as well as
  applying any necessary magnitude limits.  We do not, however, make
  corrections for higher order effects such as blending of galaxies.}
\end{enumerate}

In this way, we generate a map over the full sky of the detectability of
clusters as a function of redshift and richness, while taking into account the
large-scale structure that is already imprinted on the galaxy catalog.  This is
a more stringent test than a random-background test, but still does not capture
the additional effect that correlated large scale structure can have on the
galaxy clusters.  However, we expect that these additional effects are
subdominant: while the correlation length of galaxy clusters can be as large as
$\sim 20\ \Mpc$, the typical length-scale over which projections are effective
is $\sim 100\ \Mpc$ or more, so for most of this volume $1+\xi \approx 1$.
Roughly speaking, we would expect no more than $20\%$ corrections to our
estimated impurity from these effects, so for example, if 5\% of our clusters
suffer from projection effects in this analysis, it likely that this fraction
is underestimated by by $\sim 0.2\times 0.05=1\%$.  A more detailed treatment
of projection effects will be presented in a future work 
\citep[see also the discussion on the impact of the background on $\lambda$ in][]{rrknw11}.

One additional concern with regards to the above test is that the clusters that we 
input into our data set are produced with the same filters that we used to find galaxy
clusters.  As discussed in \citet{rkrae12}, the choice of luminosity and radial filters
have negligible impact on the richness measurements.  \citet{rrknw11} also demonstrated
that cluster ellipticity is largely irrelevant.  Thus, we expect our results to be robust to
changes to the cluster model used to generate the inserted clusters.  We have verified
this by inserting real \redmapper{} clusters as above, where the cluster galaxies
are sampled from the members list according to their membership probabilities.  
Our results are nearly identical to those derived from the model clusters above.

\begin{figure}
  \begin{center}
    \scalebox{1.2}{\plotone{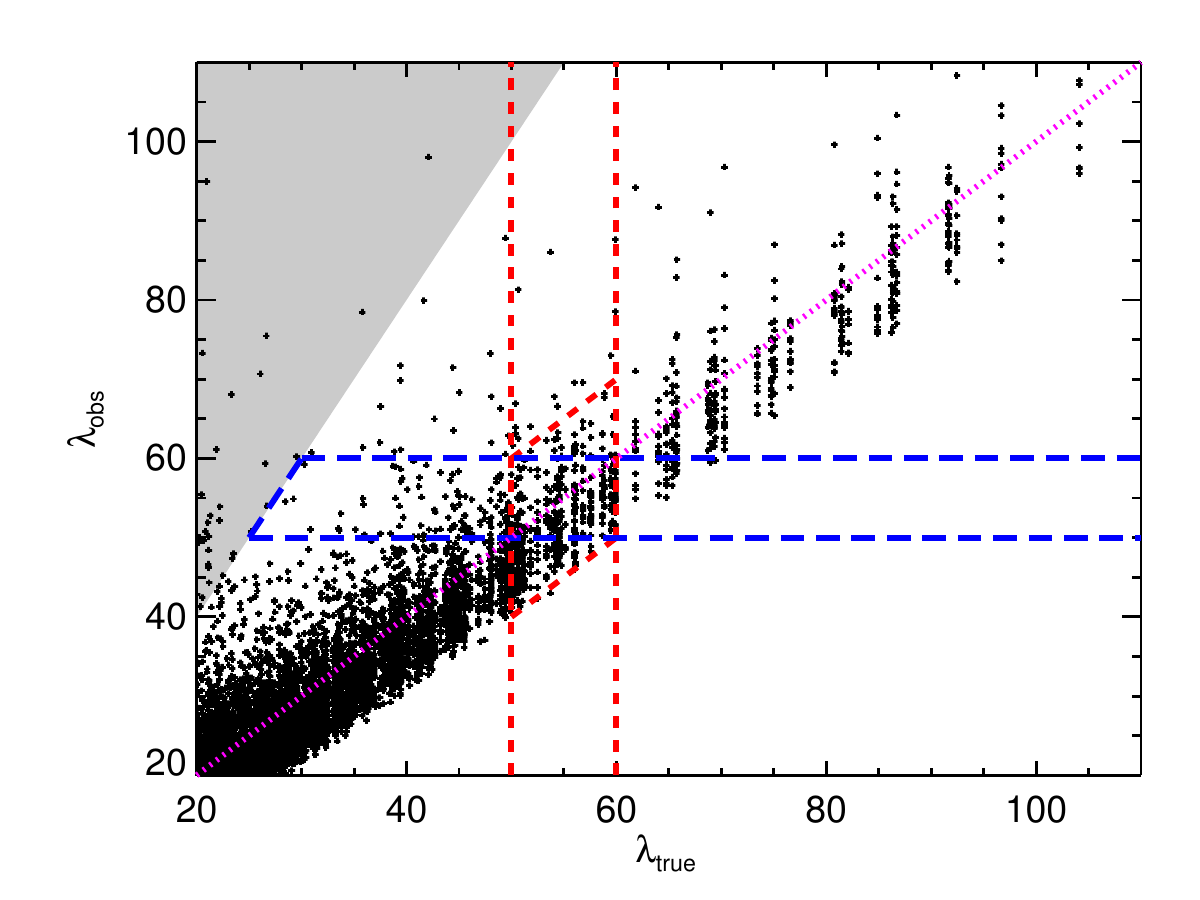}}
    \caption{Expectation value of the measured richness ($\left < \lobs \right
      >$) vs. input richness ($\ltrue$) for simulated clusters in the narrow
      redshift range $0.2<z<0.22$.  Note that although $\ltrue$ is a fixed
      value, each cluster has a distribution of $\lobs$ and we have plotted the
      mean value.  To measure completeness, we consider the sub-sample of
      clusters in a richness bin in $\ltrue$, defined by the vertical red
      short-dashed lines.  While most of the cluster sample falls within a
      tight locus around $\lobs\approx\ltrue$, there are some cluster that fall
      above the $4\sigma$ contours defined by the diagonal red short-dashed
      lines.  These outliers are projection effects, where we placed a fake
      cluster atop an existing richer structure, and are counted toward
      incompleteness.  For reference, the one-to-one line is shown with the
      magenta dotted line.  To measure purity, we consider the sub-sample of
      clusters in a richness bin in $\lobs$, defined by the blue long-dashed
      lines.  The clusters that are significant outliers with low $\ltrue$ are
      impurities where the measured cluster is the result of a projection
      effect of multiple systems.  The grey region denotes ``unphysical''
      projections where $\lobs \geq 2\ltrue$, and as such the fake cluster with
      richness $\ltrue$ is the secondary rather than the primary halo.}
    \label{fig:purcompill}
  \end{center}
\end{figure}

In Figure~\ref{fig:purcompill} we illustrate how we use the above outputs to define
purity and completeness.  The figure shows the expectation value $\left< \lobs
\right >$ for the observed richness of a galaxy cluster vs. $\ltrue$ for a
narrow redshift slice ($0.2<z<0.22$).  Note that although $\ltrue$ is a fixed
value, each cluster has a distribution of $\lobs$, and we have plotted the mean
value.

To define completeness, consider the sub-sample of galaxy clusters in some
richness bin in $\ltrue$, e.g., that defined by the vertical red short-dashed lines
in Figure~\ref{fig:purcompill}.  The bulk of this cluster sample falls within a
tight locus around the $\lobs\approx\ltrue$ line, with some noise associated with
background fluctuations.  The mean relation can be measured, including its
scatter, using fitting methods robust to outliers (we rely on median
statistics).  The diagonal red short-dashed lines show the $\pm 4\sigma$ scatter, and
points outside this region are gross outliers.  We see that all such outliers
fall {\it above} the main cloud of points: these are projection effects, where
we placed a fake galaxy cluster atop an existing richer structure.  It is
  precisely to characterize these outliers that we do not limit our model
  clusters to avoid the existing structure in the catalog. The
completeness $c(\ltrue)$ is defined as the fraction of the non-outlier points
to the total number of clusters in the richness bin,
i.e., it is the ratio of the number of clusters within the red
dashed parallelogram ($\lobs$ is consistent with $\ltrue$ within the scatter)
to the number of clusters in the $\ltrue$ bin. 

Note that with this definition $c(\ltrue) \leq 1$ does not imply that we are
missing clusters.  Instead, it is simply estimating the fraction of clusters at
a given $\ltrue$ that suffer from severe projection effects.  Similarly, for
clusters with $\ltrue$ near the detection threshold, a fraction of these
clusters with have $\lobs$ less than the detection threshold.  Thus, these
clusters are only ``missing'' due to well understood observational scatter.

Similarly, we can estimate purity by considering clusters in a bin in $\lobs$,
e.g., that defined by the blue long-dashed lines in Figure
~\ref{fig:purcompill}.  Several of the clusters in this bin are clear outliers
compared to their corresponding $\ltrue$.  The fraction of such outliers in the
$\lobs$ bin is the impurity.  However, we note one additional restriction; that
is, we discard all outliers with $\lobs \geq \ltrue/2$, denoted by the grey
region in the figure.  We note that in any projection effect, $\lobs =
\lambda_{\mathrm{main}} + \lambda_{\mathrm{proj}}$, the richness of the main
and projected halo respectively.  By definition, the main halo has a richness
$\lambda_{\mathrm{main}} \geq \lobs/2$, and we are concerned with calculating
the purity of main halos only.  Thus, any fake cluster with $\ltrue \leq
\lobs/2$ is necessarily a projection on a real, significantly richer cluster in
the catalog and should be discarded in this analysis.  As with the completeness
calculation, we emphasize that the resulting purity is the fraction of galaxy
clusters as a function of the observed richness that suffer from projection
effects, and does not represent an absence of galaxies.

To formalize all of the above discussion, we define the completeness in a bin
of richness $\ltrue$ as 
\begin{equation}
  \mathrm{completeness} = \frac{\sum\limits_{\mathrm{in\,\ltrue\,bin}}\int_{\lambda_0}^{\lambda_1} d\lobs\ P(\lobs|\ltrue) }{N(\ltrue)},
  \label{eqn:completeness}
\end{equation}
where the sum is over all clusters in a given bin of $\ltrue$.  We define 
$\lambda_0 = \ltrue - 4\sigma$, with the restriction that $\lambda_0 >
20$; $\lambda_1 = \ltrue + 4\sigma$; $\sigma^2 = \sigint^2 + \sigma_\lambda^2$;
and $N(\ltrue)$ is the total number of clusters in the bin.  We
estimate $\sigint$ directly from the output as the intrinsic scatter in the
$\lobs$--$\ltrue$ relation, and we have chosen to define an outlier
(incomplete) cluster as any cluster that has a measured richness $\lobs$ that
is more than $4\sigma$ discrepant from its true richness $\ltrue$.

By the same token, the purity is defined as
\begin{equation}
  \mathrm{purity} =
  \frac{\sum\limits_{\mathrm{in\,\lobs\,bin}}\int_{\lambda_0}^{\lambda_1}d\lobs\ P(\lobs|\ltrue)}
       {\sum\limits_{\mathrm{in\,\lobs\,bin}}\int_{\lambda_{\mathrm{obs},1}}^{\lambda_{\mathrm{obs},2}}d\lobs\ P(\lobs|\ltrue)},
  \label{eqn:purity}
\end{equation}
where now the sums are over all clusters in a given bin of $\lobs$.  Note
that we have the additional restruction that the sum is restricted to systems
with $\ltrue \geq \avg{\lobs}/2$.  We define $\lambda_0 = \ltrue - 4\sigint$
and $\lambda_1 = \ltrue + 4\sigint$ as before, and $\lambda_{\mathrm{obs},1}$
and $\lambda_{\mathrm{obs},2}$ are the extent of the richness bin in question.

\begin{figure}
  \begin{center}
    \scalebox{1.2}{\plotone{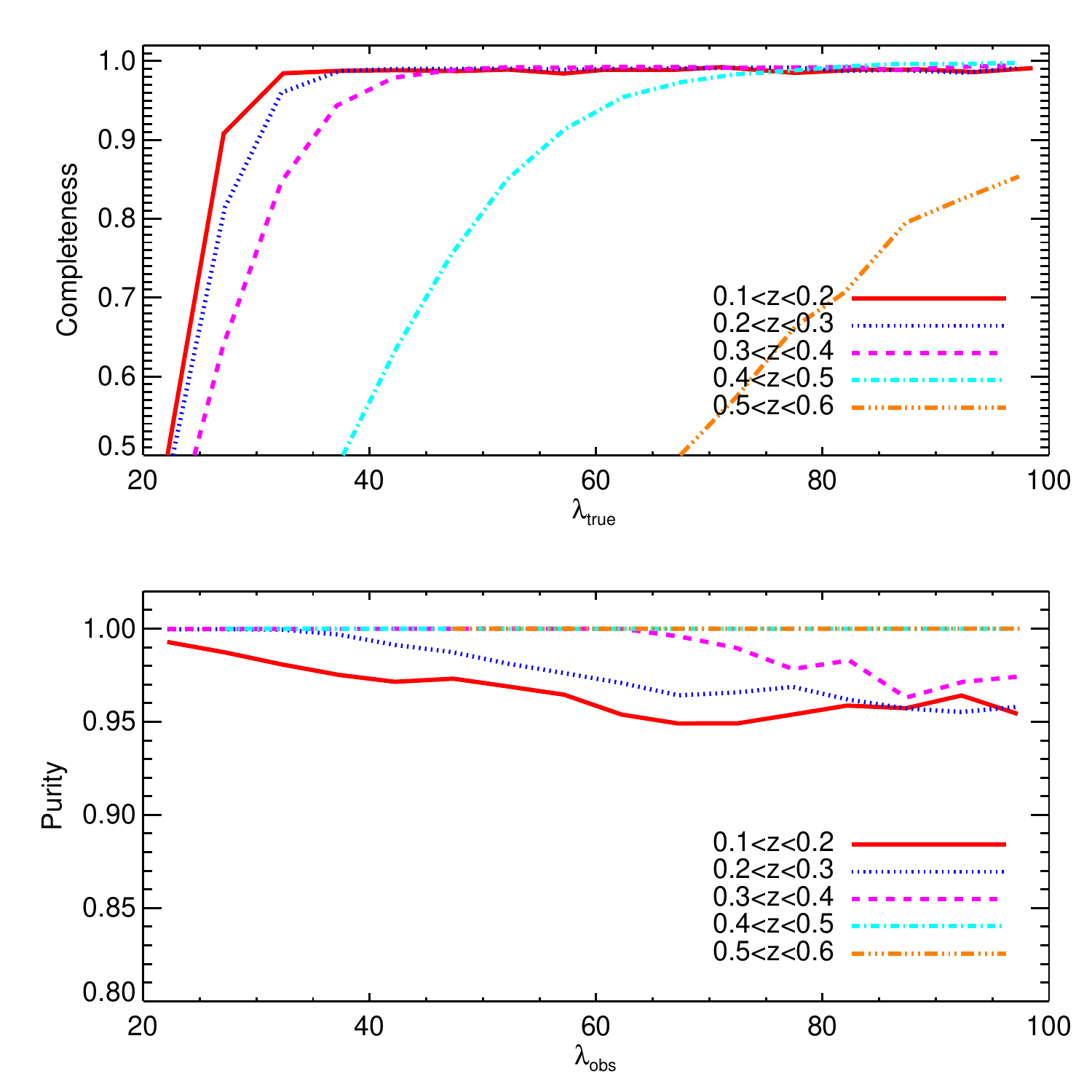}}
    \caption{\emph{Top panel:} Completeness as a function of input richness,
      $\ltrue$, in five redshift bins for the DR8 catalog.  At low redshift, the completeness at
      $\lambda<30$ falls off as measurement errors scatter clusters in and out of our $\lambda \geq 20$
      selection threshold.  At higher redshifts, the selection threshold increases, as does the measurement
      error, leading to a broader decrease extending to higher richness values.  
      \emph{Bottom panel:} Purity --- i.e., fraction of galaxy clusters not affected by projection
      effects  --- as a function of measured
      richness, $\lobs$, in five redshift bins.}
    \label{fig:purcomp}
  \end{center}
\end{figure}

In Figure~\ref{fig:purcomp} we show the completeness and purity as a function
of richness for five redshift bins for the DR8 galaxy and cluster catalog.  At
low redshifts ($z<0.3$) the completeness is essentially $\gtrsim 99\%$ at
$\lambda > 30$, but falls below this threshold due to clusters randomly scattering
in and out of the selection threshold.  At higher redshift, as we encounter the
magnitude limit of the DR8 catalog our richness threshold increases and thus
the richness at which these threshold effects come into play also increases.

Our purity is $>95\%$ for all richness and redshift bins, with the richest systems
being less pure.  This can be understood very simply: consider a chance
superposition of two clusters of richness $\lambda$ leading to a single
detection of richness $2\lambda$.  This factor of two shift has a much more
dramatic impact on the overall abundance function at the rich end, simply
because the richness function is steeper there,  i.e., a constant projection
fraction in $\ltrue$ translates into a projection fraction that decreases with $\lobs$.
Again, all these ``impurities'' actually correspond to real clusters; it's just that
the observed richness has been systematically overestimated.

One curious feature of our purity is that it seems to increase with decreasing richness, and with
increasing redshift.  This is a consequence of our definition: at lower richness and higher redshifts,
the measurements errors in the richness are larger, so a cluster that is a $4\sigma$ outlier needs
to be more and more of an extreme projection, which makes such $4\sigma$ outliers rarer.  That is,
the purity increases not because there are fewer projections, but rather because the projections that
do occur become increasingly less important relative to the observational errors in the richness
estimates.
  
In this context, it is important to emphasize that the purity that we have
defined here is fundamentally different from the purity that is usually defined
for X-ray or SZ cluster finding.  Let us take SZ as a specific example: a false
detection in SZ occurs when random Gaussian noise in the map produces a
fluctuation that can be mistaken for an SZ cluster.  In our galaxy
catalog, galaxies are detected at $>10\sigma$, so a ``false detection'' of a
cluster near our threshold ($\lambda=20$) would require 20 simultaneous
$10\sigma$ noise fluctuations colocated in the sky, and strongly correlated
across all 5-bands to mimic red-sequence colors.  If regions with bad
photometry have been properly culled out (by application of the BOSS mask),
this simply does not happen.  Every galaxy overdensity identified by
\redmapper{} is a true galaxy overdensity: there are no true false detections
driven by observational noise. The key, however, is that the galaxy overdensity
is a cylindrical galaxy overdensity which may contain more than one massive
halo, and is precisely this rate that we have tried to characterize with our
definition of purity.


\section{Cluster Masks}
\label{sec:clustermasks}

One of the great advantages of using model clusters placed randomly on the real
sky is that we can use the same output to map the detectability of \redmapper{}
clusters across the entire survey.  In this way, we can directly construct a
set of random points directly applicable to the \emph{cluster} mask, which is
not the same as the \emph{galaxy} mask that defines the survey.
An appropriate set of random points is essential for cross-correlation studies
for cluster cosmology~\citep[e.g.,][]{landyszalay93}.

As an illustration of the difference between the galaxy mask that defines the
survey and the cluster mask that defines the \redmapper{} catalog, we have run
a dense sample of random points in the vicinity of Arcturus using the methods
described in Section~\ref{sec:purcomp}.  This very bright star contaminates the
SDSS photometry over a large area, and thus effectively masks out a region of
the sky that is $0.8^\circ$ in radius.  To isolate the effect of the survey
mask, all the random points shown are associated with model clusters of the
same true richness $\ltrue=40$, with a redshift distribution appropriate for
\redmapper.

\begin{figure}
  \begin{center}
    \scalebox{1.2}{\plotone{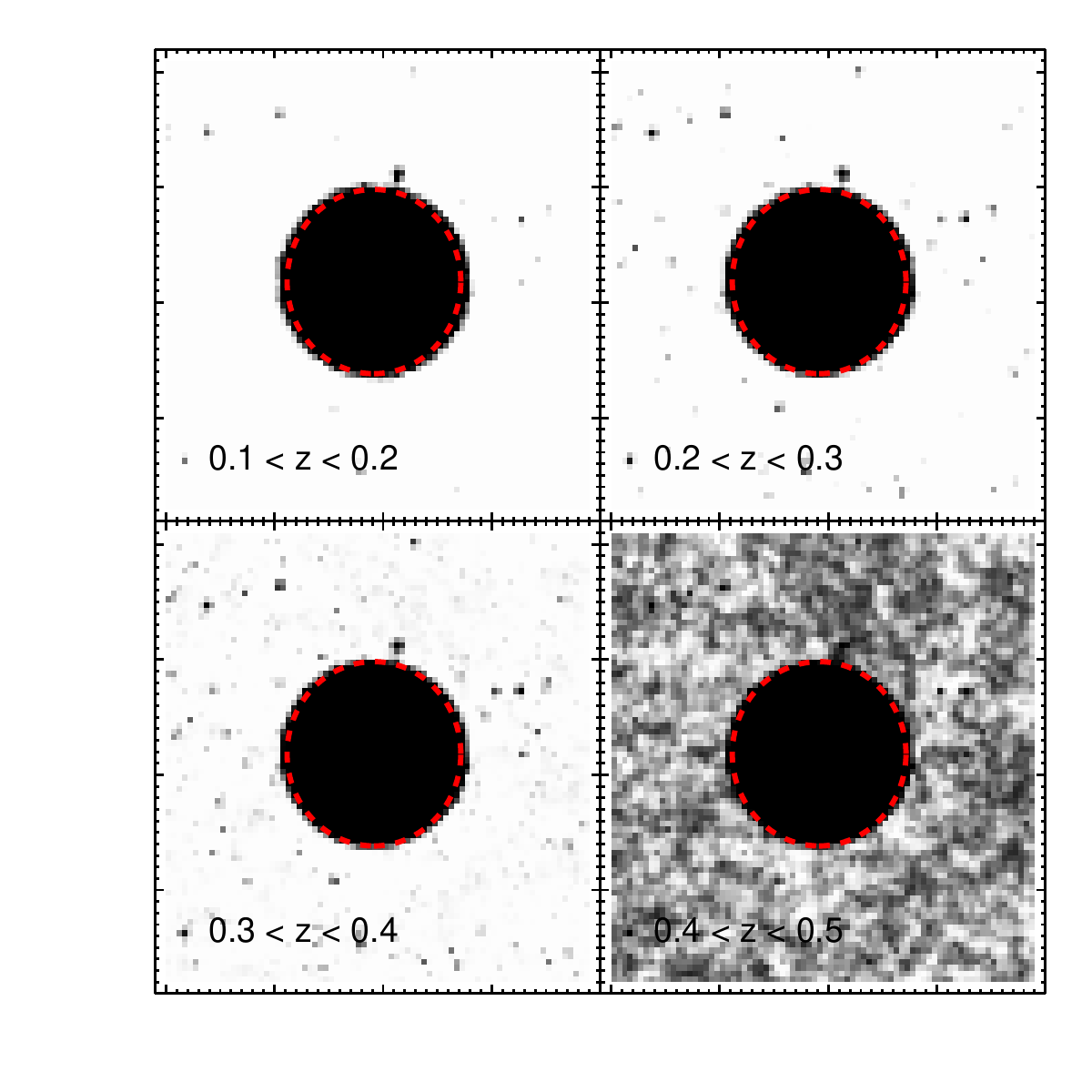}}
    \caption{Map of detectability of \redmapper{} clusters of $\ltrue=40$
      in the region of the DR8 galaxy mask in the vicinity of Arcturus (red
      dashed circle).  Each panel is
      $4^\circ$ on a side.  At low redshift
      ($0.1<z<0.2$, upper left panel) a cluster will be detected essentially
      $100\%$ of the time, except when it falls directly on top of a star
      (including typical Tycho stars, which show up as small black regions in
      the plot).
      Note that due to our requirement than no more than $20\%$ of the area of
      the cluster is masked out that the effective mask from Arcturus is
      slightly broader than that of the survey mask.  At higher redshift this
      effect is smaller because the clusters subtend a smaller physical region
      on the sky.  In the $0.4<z<0.5$ bin (lower right panel) the cluster is
      only detected $\sim60\%\pm30\%$ of the time (see Figure~\ref{fig:purcomp}).}
    \label{fig:arcturus}
  \end{center}
\end{figure}

In Figure~\ref{fig:arcturus} we show the map of the detectability of $\ltrue=40$
clusters in a $4^\circ\times4^\circ$ region around Arcturus.  Each pixel shows
the fraction of time a sample cluster is detected, using
Eqn.~\ref{eqn:completeness}, where black is $0\%$ and white is $100\%$.  In the
low redshift bin ($0.1<z<0.2$, upper left panel) the detectability of
$\ltrue=40$ clusters is essentially 100\% outside the Arcturus mask, 
except for a few pixels around  bright Tycho stars.  Note, however, 
that due to our requirement that
the area of a cluster must not be significantly masked out ($\fmask<0.2$), the
edge for the detectability of a \emph{cluster} at these redshifts is slightly
farther from the center of the Arcturus than the edge of the
\emph{galaxy} mask (denoted by the red dashed line).  At higher redshift these edges
drift closer to each other as the angular extent of the clusters decreases.
However, in the highest redshift bin ($0.4<z<0.5$, lower right panel) the
cluster is only detected $\sim60\%\pm30\%$ of the time (see Figure~\ref{fig:purcomp})
due to the survey depth.  The detectability varies significantly 
when clusters approach the threshold and we see a strong dependence on the local depth 
and structure.

\section{Summary}
\label{sec:summary}

In this paper, we have introduced \redmapper, a red-sequence cluster finder
that is designed to make optimal use of large photometric surveys.  As a case
study in the implementation of the algorithm, we have run on the SDSS DR8
photometric catalog.  We have shown that \redmapper{} improves significantly on
previous cluster finders (see also Paper II), with many features that will be
required to take advantage of upcoming surveys such as DES and LSST.  In
particular:
\begin{enumerate}
\item{\redmapper{} is based on a multi-color extension of the optimized richness
    estimator $\lambda$, which has been shown to be a good mass proxy (R12 and
    Paper II).}
\item{\redmapper{} is self-training, with a modest requirement in the number of
  training spectra, which can themselves be limited to the brightest cluster
  galaxies.  This makes it particularly well-suited to high-redshift surveys.
  Furthermore, the multi-color red-sequence model makes optimal use of all
  color data at all redshifts, with no sharp features as the
  $4000\,\mathrm{\AA}$ break transitions between filters.}
\item{\redmapper{} can handle complex survey masks.  Both mask-corrected
  richness values can be computed, as well as \emph{cluster-appropriate} random
  point catalogs for large-scale structure studies.}
\item{All clusters are assigned a redshift probability distribution $P(z)$,
  which enables a more accurate reconstruction of the redshift distribution
  of the cluster population relative to simple point-redshift estimates.}
\item{The centering of clusters is fully probabilistic.  In this way, the
    uncertainty in the position of the cluster can be handled in an analagous
    way to the redshift uncertainty provided by $P(z)$.}
\item{The algorithm is numerically efficient, and can be run on large surveys
    with modest computing power.}
\end{enumerate}
%


Using the red-sequence model derived in the \redmapper{} calibration phase, we
have derived two red-sequence based photometric redshifts.  The first, $\zred$,
is a red-sequence template-based \photoz, that has been designed to generate a
good ``first-guess'' estimation of the redshift in each cluster.  We have also
shown, in Appendix~\ref{sec:photozcompare}, that $\zred$ compares very well to
existing DR8 photometric redshifts for this specific class of galaxies.  However,
$\zred$ has the advantage that it requires many fewer spectroscopic training
galaxies.  Moreover, these galaxies can be the brightest galaxies in the
clusters, with no penalty to the performance of $\zred$ at the faint end of the
galaxy sample.  The second, $\zlambda$, is a very precise \photoz{} derived from
fitting all cluster members simultaneously to the red sequence model.  In
addition, we derive a $P(z)$ estimator for $\zlambda$, which we show is
superior to point-based photometric redshifts for the purposes of estimating
the redshift distribution of the galaxy clusters.  In DR8, this is
especially true in the region of the filter transition at $z\sim0.35$.

As a case study in the implementation of the algorithm, we have run
\redmapper{} on the $10,400\,\mathrm{deg}^2$ BOSS region from the SDSS DR8
photometric catalog.  Using red galaxy spectroscopic redshifts from 1/5 of the
total area from $z\in[0.05,0.6]$, we are able to constrain a robust
red-sequence model that defines both the richness and photometric redshift
estimators.  The photometric redshifts, $\zlambda$, have small bias and low
scatter, ranging from $\sigma_z=0.006$ at $z\sim0.1$ to $\sigma_z=0.020$ at
$z\sim0.5$, due to increased photometric noise near the survey limit.  The rate
of catastrophic outliers is low, with only $\sim 1\%$ of galaxy clusters
appearing as $4\sigma$ outliers.  Note that because of our high \photoz{}
precision, a cluster at $z=0.1$ with a redshift offset as small as $\Delta
z=0.025$ is considered a catastrophic redshift failure.  Furthermore, we show
that the majority of these outliers are bad \emph{centers} rather than bad
redshifts; when the catalog is cleaned by demanding that central and satellite galaxies with spectroscopy must all be within 
$1,000\ \mbox{km/s}$, the failure rate decreases to $\lesssim0.2\%$.

After running \redmapper{} on the full DR8 photometric catalog, we apply a
conservative selection cut of $\lambda/S(z) > 20$, for a total of $25,236$ clusters in
the redshift range of $z\in[0.08,0.55]$.  As shown in paper II, the comoving
density of \redmapper\ clusters satisfying this cut it {\it lower} than that 
of all other SDSS photometric cluster catalogs.
The scale factor, $S(z)$, given by
Eqn.~\ref{eqn:sz}, defines the correction factor on the richness caused by the
survey depth.  The catalog is volume-limited at $z<0.35$, where $S=1$ and the
survey depth is brighter than the fiducial luminosity cut of $0.2L_*$ used by
the $\lambda$ richness.  Because our selection threshold corresponds to a total
of 20 galaxy detections, as we lose galaxies at high redshift due to the
magnitude limit of the survey, these 20 galaxies must all be due to bright
members.  Therefore, the corresponding richness threshold of $20/S(z)$ is much
higher.  This increased detection threshold results in fewer galaxy clusters at
high redshifts.   Our adopted richness threshold of 20 detected red sequence
galaxies is chosen to provide the most robust cluster catalog possible, with a
mass threshold of $M\gtrsim10^{14}\,M_\sun$ where our catalog is volume limited
at $z\lesssim0.35$~(R12, Paper II).  Although the full
\redmapper{} catalog extends to lower richnesses, we expect expect
performance will worsen as one moves towards lower and lower richness
thresholds.

Finally, we investigate the purity and completeness of our cluster finding
algorithm, focusing on the observationally relevant probability distribution
$P(\lobs|\ltrue)$.  We have defined impurity and incompleteness as the fraction
of clusters for which the observed richness $\lobs$ is significantly different
from the true richness $\ltrue$.  These outliers are caused by projection
effects: when two halos are merged together, this manifests itself as
incompleteness --- a cluster with richness $\ltrue$ is up-scattered significantly,
so it is ``missing'' from where it should have been ---
or impurity --- the richness $\lobs$ of such a cluster is significantly overestimated.
We note that while the completeness of \redmapper{} is near $100\%$, the purity
is $\sim95\%$ at the rich end, increasing at lower richness.  This decrease
simply reflects larger observational error (in a proportional sense) for lower richness
clusters: i.e. ``outliers'' become more rare not because projection effects are less rare,
but because projection effects become sub-dominant to observational uncertainties.
Our estimate of the incidence of projection effects is thus $\sim 5\%$, similar to what was
estimated in \citet{rrknw11}.  A more detailed analysis of projection effects for 
\redmapper\ clusters will be presented in a future work.

In Paper II, we present a detailed comparison of the \redmapper{} cluster
catalog to various X-ray and SZ catalogs with high quality mass proxies.  In
all cases, we show that the \redmapper{} richness $\lambda$ is a low scatter
mass proxy with high completeness and low impurity compared to these ``truth''
tables.  We also compare the performance of \redmapper{} to other photometric
cluster finders that have been run on SDSS data, and show that \redmapper{}
outperforms these other algorithms in all metrics (e.g., \photoz{}
performance; mass scatter; and purity and completeness), though some do perform
equally well in subsets of these categories in specific redshift ranges.

While this present work has focused on the application of \redmapper{} to the
SDSS DR8 catalog, we emphasize that this algorithm was developed specifically
for upcoming large photometric surveys such as DES and LSST.  In particular,
its ability to simultaneously utilizes all available photometric data, its
smooth handling of the filter transition of the $4000\,\mathrm{\AA}$ break across filter
passes, and its ability to self-calibrate using only minimal spectroscopic
training samples of bright cluster galaxies are all specifically designed
to enable cluster finding in these new photometric data sets.
This will be especially advantageous at $z\gtrsim0.7$ in the
Southern Hemisphere, where we do not have the advantages of more than a decade
of survey data from the SDSS spectrograph.  Thus, in short order we expect
\redmapper{} will be capable of producing large, high quality catalogs
of $\sim 80,000$ clusters at $z<1$ with DES, opening a new era of high redshift
cluster cosmology.

\acknowledgements

We thank the anonymous referee for an exceedingly careful read of our
manuscript, and their numerous comments which helped improve the readability
and content of this manuscript.

This work was supported in part by the U.S. Department of Energy contract to
SLAC no. DE-AC02-76SF00515.  AEE acknowledges support from NSF AST-0708150 and
NASA NNX07AN58G.  This work was also supported by World Premier International
Research Center Initiative (WPI Initiative), MEXT, Japan.

Funding for SDSS-III has been provided by the Alfred P. Sloan Foundation, the Participating Institutions, the National Science Foundation, and the U.S. Department of Energy Office of Science. The SDSS-III web site is http://www.sdss3.org/.

SDSS-III is managed by the Astrophysical Research Consortium for the Participating Institutions of the SDSS-III Collaboration including the University of Arizona, the Brazilian Participation Group, Brookhaven National Laboratory, University of Cambridge, Carnegie Mellon University, University of Florida, the French Participation Group, the German Participation Group, Harvard University, the Instituto de Astrofisica de Canarias, the Michigan State/Notre Dame/JINA Participation Group, Johns Hopkins University, Lawrence Berkeley National Laboratory, Max Planck Institute for Astrophysics, Max Planck Institute for Extraterrestrial Physics, New Mexico State University, New York University, Ohio State University, Pennsylvania State University, University of Portsmouth, Princeton University, the Spanish Participation Group, University of Tokyo, University of Utah, Vanderbilt University, University of Virginia, University of Washington, and Yale University.

\appendix

\section{Photometric Redshift Correction Parameters}

\subsection{Constraining $\zred$ Correction Parameters}
\label{app:zredcorr}

Our approach to constraining the $\zred$ correction parameters $\cbarz(z)$ and
$\sbarz(z)$ is similar to that employed for the red sequence calibration (note the $z$
subscript).  As
before, we have chosen to constrain these parameters at a given node spacing,
using cubic spline interpolation between the nodes.  The node spacing we have
chosen for DR8 is $0.05$ for $\cbarz(z)$ and $0.10$ for $\sbarz(z)$, suited to the
characteristic variation scales.

One significant complication that we have to deal with is that we have
membership probabilities for all the galaxies.  In order to properly make use
of the probabilities, as in Eqn.~\ref{eqn:rspdf} we need to know the background
PDF.  Unfortunately, there is no first-principle way of calculating the
$\zred$ background as a function of $\ztrue$.  Therefore, we have chosen to
assume the background is a Gaussian function with zero mean and finite width,
and to marginalize over this background as a set of nuisance terms.  As above,
we assume the background width, $\sigma_b(z)$ is a smoothly interpolated
function with a node spacing of $0.10$.  To ensure that we are calculating the
correction factors appropriate for red galaxies, and not blue cluster members and
interlopers, we limit ourselves to galaxies with $\pmem>0.7$.

Given a model correction, 
\begin{equation}
  \corrmod = \cbarz(\ztrue) + \sbarz(\ztrue)[\imag - \refmag(\ztrue)],
\end{equation}
then we have a Gaussian PDF for the true galaxies,
\begin{equation}
  G_1 = \frac{1}{\sqrt{2\pi}\sigzred}\exp\left ( \frac{-[(\zred-\ztrue] -
    \corrmod]^2}{2\sigzred^2} \right ),
\end{equation}
and for the background,
\begin{equation}
  G_2 = \frac{1}{\sqrt{2\pi}\sigma_b}\exp\left (\frac{-[\zred-\ztrue]^2}{2\sigma_b^2}
  \right ).
\end{equation}

The total likelihood is then
\begin{equation}
  \lk = w[\pmem G_1 + (1-\pmem)G_2],
\end{equation}
where we have made the addition of a weight function, $w$, which is a smooth
function of $\chisq$ that de-weights galaxies with large $\chisq$ and are
possible outliers.  The weight $w$ is
\begin{equation}
  w = \frac{1}{\exp[(\chisq - \chisq_{95})/0.2] + 1},
\end{equation}
where $\chisq_{95}$ is the $95^{\mathrm{th}}$ percentile of all galaxies with
$\pmem > 0.7$.

As before, we find the $\cbarz(z)$, $\sbarz(z)$, and $\sigma_b(z)$
parameters by maximizing $\sum \lk$ using the downhill-simplex method.


\subsection{Constraining $\zlambda$ Correction Parameters}
\label{app:zlambdacorr}

Our approach to constraining the $\zlambda$ correction parameters is analogous
to that used for the $\zred$ parameters in Appendix~\ref{app:zredcorr}.
However, our job is a little easier because we are applying corrections such that $\left
<\ztrue|\zlambda \right >$ is unbiased rather than the converse.  Therefore,
the correction term can be a function of $\zlambda$.  For DR8, we use a cubic spline
interpolation with node spacing of $0.04$ for $\corrmodzl$.  In addition, we
allow an additional variance term as we find that our raw $\zlambda$ errors are
underestimated.  For $\sigma_{\zlambda,\mathrm{int}}$ we use a smooth
function with a node spacing of $0.10$.  To ensure that we are using
well-measured clusters, we limit ourselves to calibration clusters that have
$\lambda/S(z) > 10$, where $S$ is the scale factor defined in Eqn.~\ref{eqn:sz}.
Essentially, this limits us to clusters with at least 10 red galaxies above the
luminosity threshold or magnitude limit.

Given a model correction $\corrmodzl$ and intrinsic scatter correction
$\sigma_{\zlambda,\mathrm{int}}$, we have a Gaussian PDF for the clusters,
\begin{equation}
  G = \frac{1}{\sqrt{2\pi}\sigma_{\mathrm{tot}}} \exp \left ( \frac{- [ (
      \zlambda - \zcg ) - \corrmodzl ]^2}{2\sigma_{\mathrm{tot}}^2} \right ),
\end{equation}
where $\sigma_{\mathrm{tot}} = \sqrt{\sigma_{\zlambda}^2 +
  \sigma_{\zlambda,\mathrm{int}}^2}$.  The total likelihood is then given by
$\ln\lk = \sum \ln G$.  As before, we find $\corrmodzl$ and
$\sigma_{\zlambda,\mathrm{int}}^2$ by maximizing this likelihood using the
downhill-simplex method.

With this parametrization in hand, we can calculate the corrected $\zlambda$
and error as $\zlambda =
\zlambdaraw + \corrmodzl$, and $\sigma_{\zlambda}^2 =
\sigma_{\zlambda,\mathrm{raw}}^2 + \sigma_{\zlambda,\mathrm{int}}^2$.  However,
we find that after applying these corrections there may still be small
residuals in the training sample.  Therefore, we iterate on this solution two
further times to obtain a final corrected redshift $\zlambda$.

After the calibration is complete, we must also apply these corrections to the
$P(\ztrue|\zlambda)$ estimation for each cluster.  To replicate the $\zlambda$
offset represented by $\corrmodzl$, we first offset the central value of the
$P(z)$ distribution.  Next, to replicate the increased scatter we
``expand the space'' between the $P(z)$ bins, so that a Gaussian fit to $P(z)$
will measure the same width as the corrected $\sigma_{\zlambda}$ value.  We
find that this does an adequate job of maintaining asymmetries in the
$P(z)$ distribution which show up near the filter transitions.


\section{How Many Training Clusters?}
\label{app:ntrain}

When calibrating the red sequence in Section~\ref{sec:calibration} on DR8 data,
we make use of all the spectroscopy available in our $2000\,\mathrm{deg}^{2}$
training region.  However, much of this is superfluous.  First, most of the
spectroscopic galaxies --- even the LRG samples --- are not in massive
clusters.  Second, our strategy of leveraging central galaxy spectroscopy to
all the galaxies in a cluster means that we do not require thousands of
clusters to perform the calibration.  In this section, we investigate how many
training clusters --- each represented by a single spectroscopic redshift for
the central galaxy --- are required to create an accurate and unbiased richness
and redshift estimate.

To test the number of required training clusters we follow the method of
Section~\ref{sec:iterating} to measure the bias in the recovered richness and
photometric redshift values on a predetermined set of test clusters.  For our
test suite, we select the $\{5,10,20,40,80\}$ richest clusters per redshift
bin of $\pm0.025$ in the training region.  The redshift binning is used to
ensure we have a relatively uniform coverage over the redshift range of interest.
In practice, of course, the training clusters need not be so uniformly sampled.
For each of these spectroscopic training sets we recalibrate the red sequence
and measure the richness $\lambda$ and redshift $\zlambda$ for each of the test
clusters from Section~\ref{sec:iterating}.

\begin{figure*}
  \begin{center}
    \plottwo{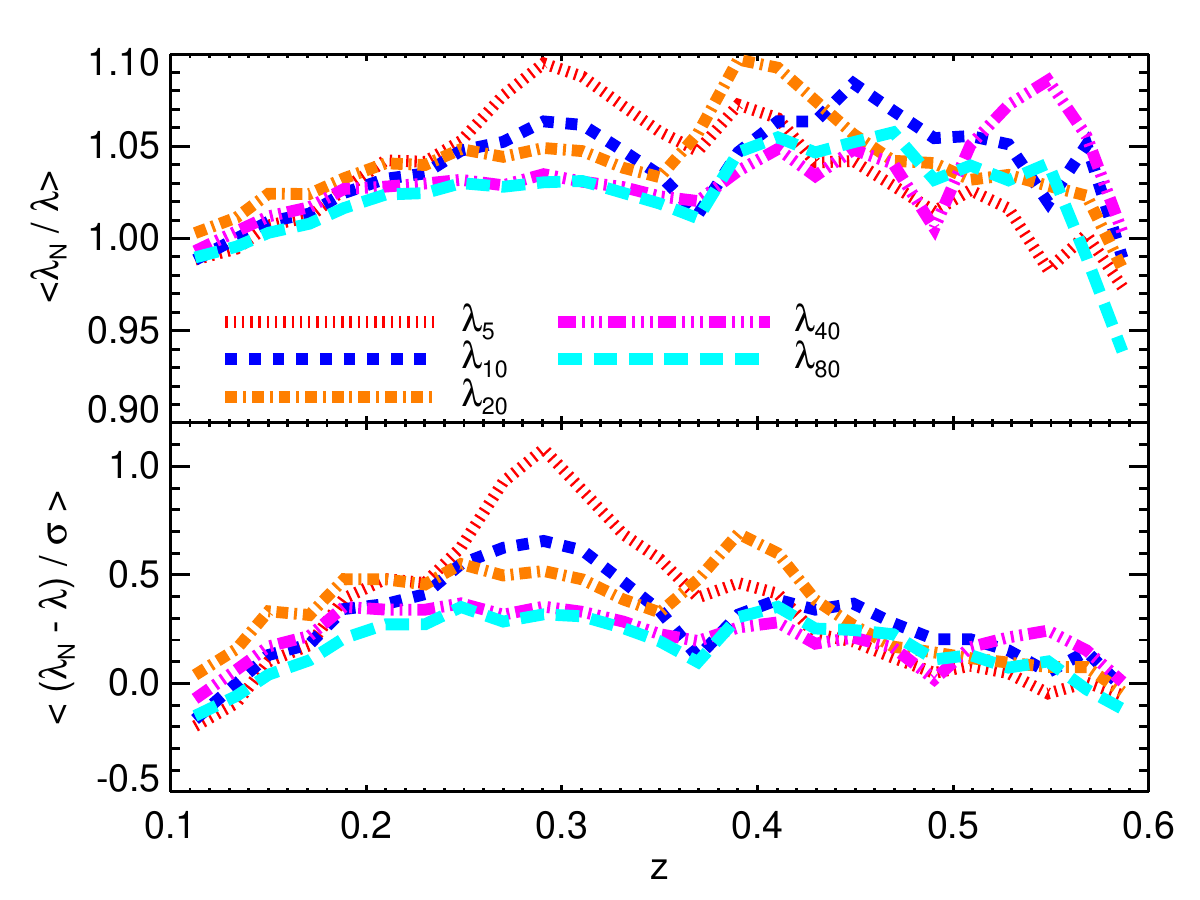}{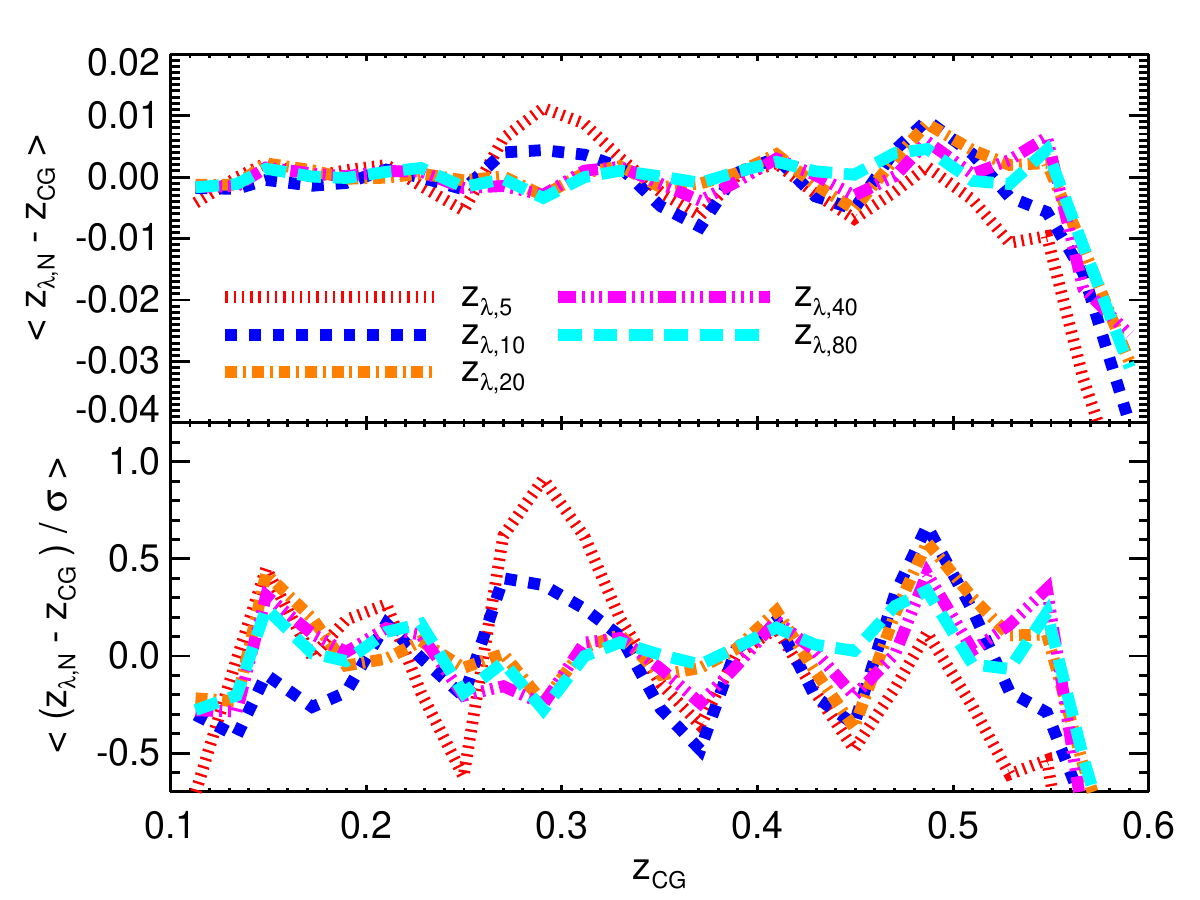}
    \caption{\emph{Left panel, top:} Average richness bias as a function of
      redshift for $\{5,10,20,40,80\}$ training clusters per redshift bin of
      width $\pm0.025$, compared to the richness using the full DR8 training
      sample.  All curves use the same set of 4300 test clusters.      
      \emph{Left panel, bottom:} Error normalized average deviation relative to
      the baseline.  With at least 40 (10) clusters per redshift bin of
      $\pm0.025$, biases are always $<0.3\sigma$ ($<0.5\sigma$).  Thus, with
      only 400 well-chosen spectra of the brightest galaxies, we can achieve
      nearly the same precision as is possible with all the SDSS spectra.     
      \emph{Right panel, top:} Average uncorrected photometric redshift
      ($\zlambda$) bias as a function of redshift for $\{5,10,20,40,80\}$
      training clusters per redshift bin of width $\pm0.025$, compared to the
      spectroscopic redshift of the central galaxy, $\zcg$.
      \emph{Right panel, bottom:} Error normalized average deviation relative
      to the baseline.  With at least 20 clusters per redshift bin of
      $\pm0.025$ we achieve the same redshift performance as is possible with
      all SDSS spectra.}
    \label{fig:ntrain}
  \end{center}
\end{figure*}

In Figure~\ref{fig:ntrain} we show the results of these test runs.  The left panel
shows the richness bias and significance of the bias as a function of redshift
for the various training samples.  Although we can get a reasonable calibration
of the red sequence with as few as five spectra per $\pm0.025$ redshift bin,
the resulting richnesses are significantly biased ($\sim1\sigma$) at the
transition redshift $z\sim0.35$.  In order to achieved unbiased richness
estimates ($<0.3\sigma$) then we require $\sim40$ clusters per redshift bin.
We assume any residual biases are due to the noise in estimating the
off-diagonal elements of the covariance matrix.  This results in a total of
$\sim400$ spectra to achieve essentially the same fidelity of calibration as we
can achieve with millions of SDSS spectra.  The right panel shows the
photometric redshift
bias and significance, similar to the left panel.  For accurate \photoz{} estimation, we require even fewer training spectra: $\sim20$ per
redshift bin, or a total of 200.  

For upcoming photometric surveys such as DES, we can obtain these spectra by
first running a crude run with an approximate red sequence model.  After
selecting bright central galaxies, these can easily be followed up
spectroscopically, as they are the most luminous galaxies at any redshift.  For
example, over $85\%$ of the training spectra required for DR8 training are
brighter than $\imag < 18.5$.  Thus, our method allows for an incredibly
efficient use of limited spectroscopic resources to enable science in
large photometric surveys.


\section{Comparison of $\zred$ to SDSS DR8 Photo-$z$s}
\label{sec:photozcompare}

We consider two sets of photometric redshift estimates available for all of DR8.
The first, ``$\zphoto$'', uses an updated method of
\citet{cdthj07}\footnote{See http://www.sdss3.org/dr8/algorithms/photo-z.php},
and the second, ``$p(z)$'', uses the method of \citet{scmbw12}.  In this
section we make use of high probability cluster member galaxies to compare
these photometric redshifts to $\zred$ at both bright magnitudes (where
training galaxies are plentiful) and at fainter magnitudes.

For our ``pseudo-spectroscopic'' test sample, we start with all clusters with
$\lambda>5$ and a central galaxy with spectroscopic redshift $\zcg$.  We then
select all members with $\pmem>0.9$.  We thus expect a contamination rate of up
to 10\%, although the real rate should be smaller than this.  By assigning each
high probability member to the spectroscopic redshift of the central galaxy 
we can leverage the red sequence to obtain
spectroscopic quality redshifts to much fainter magnitudes than available
in the SDSS main or LRG spectroscopic samples.

\begin{figure}
  \begin{center}
    \scalebox{1.2}{\plotone{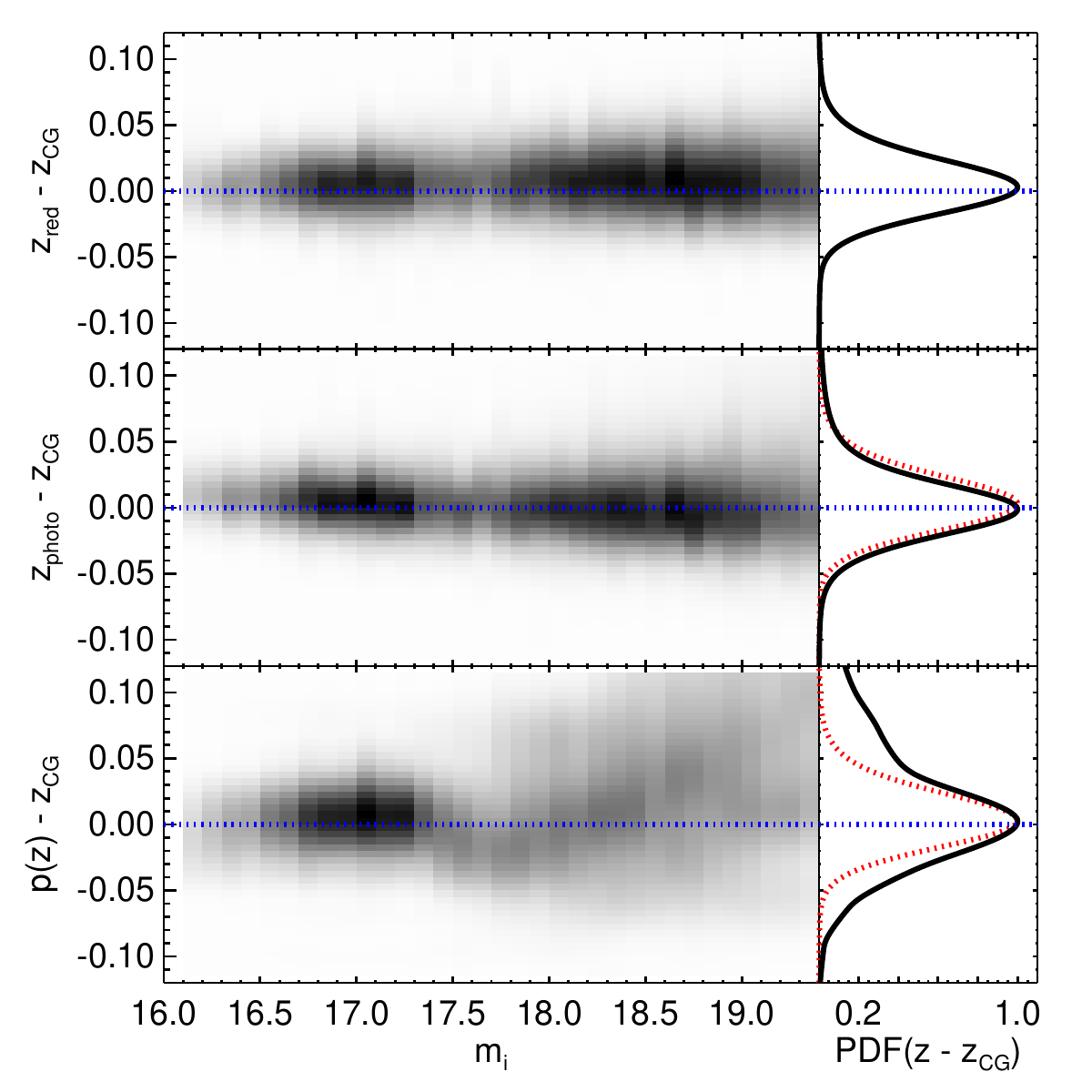}}
    \caption{\emph{Top:} Density of total $p(\zred - \zcg)$ as a
      function of $i$-band magnitude $\imag$ for all cluster members with
      $\pmem>0.9$ and $0.195<\zcg<0.205$ in clusters with $\lambda >
      5$.  The right panel shows the total PDF at all magnitudes. There is a
      small bias in $\zred$, though it is constant with magnitude.
      \emph{Middle:} Same as top panel, with $\zphoto$ calculated
      with the algorithm of \citet{cdthj07}.  The performance is good down to
      $0.2L_*$.  The right panel compares the distribution for
      $\zphoto$ (black line) to $\zred$ (red dotted line).
      \emph{Bottom:} Same as top panel, with $p(z)$ values from
      \citet{scmbw12}.  While the bright galaxy performance is good, there are
      biases at the faint end and the distribution is significantly wider.}
    \label{fig:photoz2}
  \end{center}
\end{figure}

Figure~\ref{fig:photoz2} shows the density map of the photometric redshift
biases as a function of magnitude for $\zred$, $\zphoto$, and $p(z)$ for a
narrow redshift slice of $0.195<\zcg<0.205$.  For $\zred$ and $\zphoto$ we have
assumed a probability distribution function (PDF) that is Gaussian with mean
$\zred$ ($\zphoto$) and width $\sigzred$ ($\sigma_{\zphoto}$).  For the $p(z)$
values we use a spline interpolation to smooth the PDF and normalize the area
to unity.  On the right-hand side are projected histograms from the density
field.  The dotted red lines show the $\zred$ distribution for comparison.  
Note that the density plot clearly shows the separation in magnitude
between central and satellite galaxies. 
Both $\zred$ and $\zphoto$ perform well down to the
$0.2L_*$ limit of the \redmapper{} richness estimation, while the $p(z)$ values
have a broader distribution at the faint end.  These is also obvious structure
in the \photoz{} bias as a function of magnitude.

\begin{figure}
  \begin{center}
    \scalebox{1.2}{\plotone{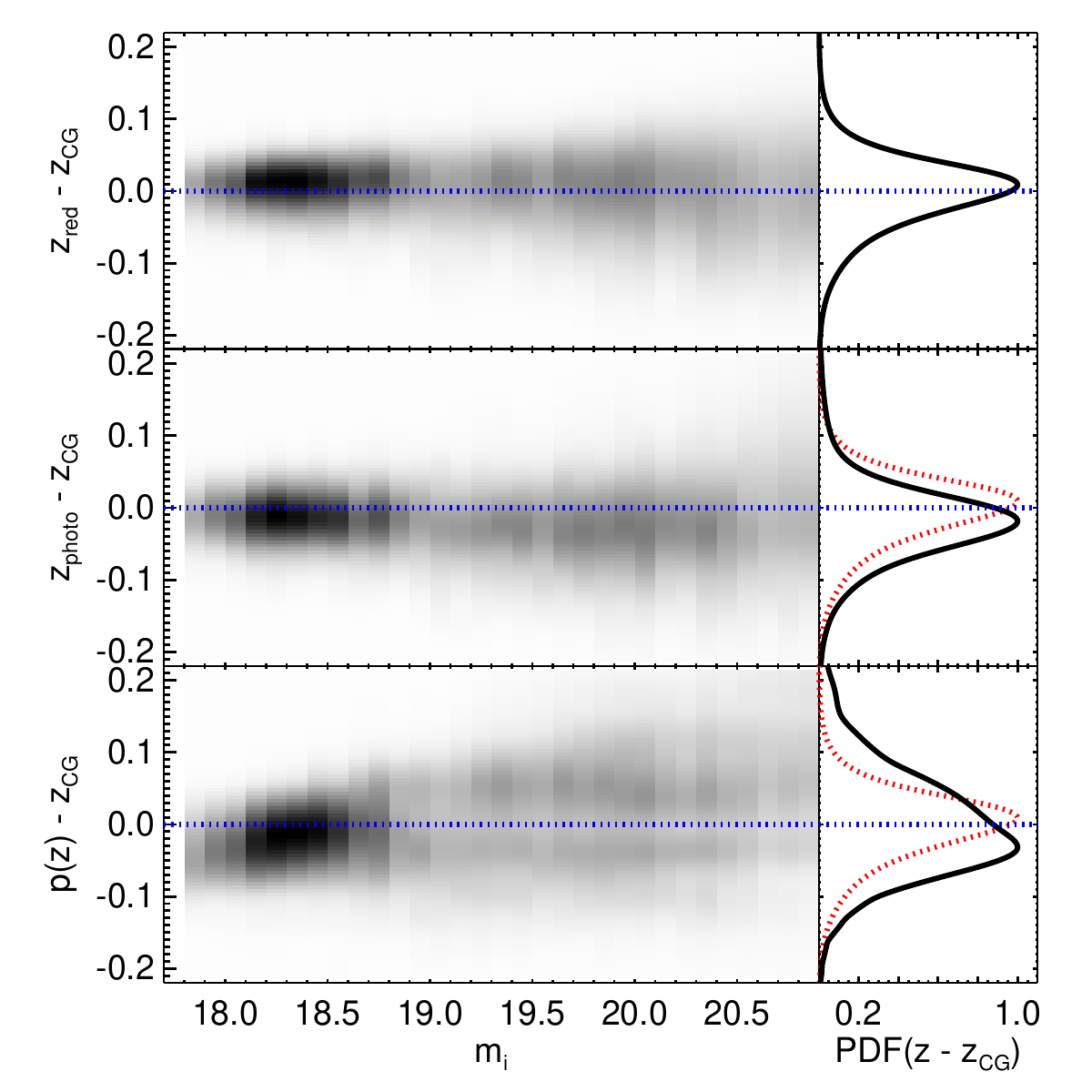}}
    \caption{Same as Figure~\ref{fig:photoz2}, with members selected from
      clusters with $0.395<z_{\mathrm{CG}}<0.405$.}
    \label{fig:photoz4}
  \end{center}
\end{figure}

Figure~\ref{fig:photoz4} shows the same map for a narrow redshift slice of
$0.395<\zcg<0.405$.  While all the photometric redshifts handle the luminous
galaxies very well, the appear to be slight biases at the faint end in the case
of the DR8 $\zphoto$, and a bifurcation of the distribution for the $p(z)$
redshifts.  The evolution of the bias in the $p(z)$ estimates is due to a
combination of effects.  First, the $r$-band magnitude was used as an input to
the \photoz{} estimator.  For a field galaxy, a fainter magnitude correlates
with a higher redshift.  For cluster galaxies, however, galaxies of a wide
range of luminosity occupy the same cluster.  As a result, when using
magnitude-based \photoz{} estimators on galaxies in clusters, one should expect
an increasing bias with magnitude, which is simply a manifestation of the
intra-cluster luminosity function.  The large width of the error distributions
relative to the other estimators are due primarily to the lack of training set
galaxies in that range. As
discussed in \citet{scmbw12}, the main focus was on recovering the
full $r<21.8$ galaxy sample. To avoid biases induced by training set selection,
the authors did not include the most recent BOSS LRG samples in that work and
deferred LRG-optimized $p(z)$ estimates to a future paper. It is also worth
pointing out that, despite the extra width of the error distributions obtained
when using $p(z)$, the recovered redshift distributions obtained by summing the
$p(z)$ of \citet{scmbw12} are still superior to the distributions estimated using
the DR8 $\zphoto$ or single-point $\zred$ estimates.

There are two important take-home messages from this comparison.  First, the
performance of state-of-the-art \photoz{} estimators appears to be sufficiently
accurate for bright galaxies that we would likely be able to use these in the
initialization phase of \redmapper{} without any loss.  Second, $\zred$ appears to
be at least as good --- if not better --- than what is currently achieved, with
{\it much} smaller spectroscopic training samples.  As shown in
Appendix~\ref{app:ntrain}, we can achieve this redshift performance with only
$\approx 400$ of the brightest CG spectra.  With the technique of assigning the
spectroscopic redshift of the central galaxies to the members, we effectively
increase the faint end of our training sample.  This is very useful for future
surveys because of the high cost of obtaining spectroscopic redshifts of faint
galaxies.

\section{Computing Performance Benchmarks}

The \redmapper{} algorithm has been designed to be fast, efficient, flexible,
and trivially parallelized.  As there are two parts to running \redmapper, the
calibration and cluster-finding stages, we split the performance benchmarks
into two parts.

For the calibration phase, the runtime depends on the number of training
spectra and clusters.  For the DR8 training sample on $2000\,\mathrm{deg}^2$,
the full calibration takes $\sim30\,\mathrm{CPU}\,\mathrm{hr}$ on a 3-year old
2.8~GHz AMD Opteron 8389.  Current Intel processors can run the calibration
roughly twice as fast.  For the minimal training sample of 40 clusters per
redshift bin (see Appendix~\ref{app:ntrain}) calibration takes
$\sim13\,\mathrm{CPU}\,\mathrm{hr}$.

The cluster-finding stage is designed to be split into chunks of arbitrary size
on the sky.  For these purposes we use the Mangle simple pixelization
scheme~\citep{sthh08}, although any pixelization scheme will work.  As long as
the overlap region between pixels is wider than twice the largest size of any
cluster in the catalog, then the percolation of clusters within each cell is
guaranteed to be unique.  For the DR8 catalog, this corresponds to a border
region of $1.5^\circ$, corresponding to twice the size of a cluster of richness
$\sim300$ at $z=0.05$, given the mask radius parameters.  In total, running the
cluster finder on the full DR8 catalog requires
$\sim500\,\mathrm{CPU}\,\mathrm{hr}$ including all galaxy mask corrections.  On
a modestly sized compute cluster this can be run in much less than one day.


\section{Validating the Correction $C$}
\label{sec:corrtest}

In Section~\ref{sec:holes} we laid out our methodology for correcting the
richness for survey holes and a magnitude limit that is brighter than $0.2L_*$.
In order to validate the calculation of the correction term $C$ described in
that section, we have taken a subsample of clusters with $0.15<z<0.3$ and
simulated a more restrictive magnitude limit.  We have chosen a magnitude limit
of $\imag<19.6$, which is $0.2L_*$ at $z=0.2$, so that all clusters at higher
redshift will have their richness corrected according to our formalism.  The
average correction for the $z=0.3$ clusters is similar to that for the highest
redshift clusters in our catalog, so this test will sample the full range of
corrections employed.

\begin{figure}
  \begin{center}
    \scalebox{1.0}{\plotone{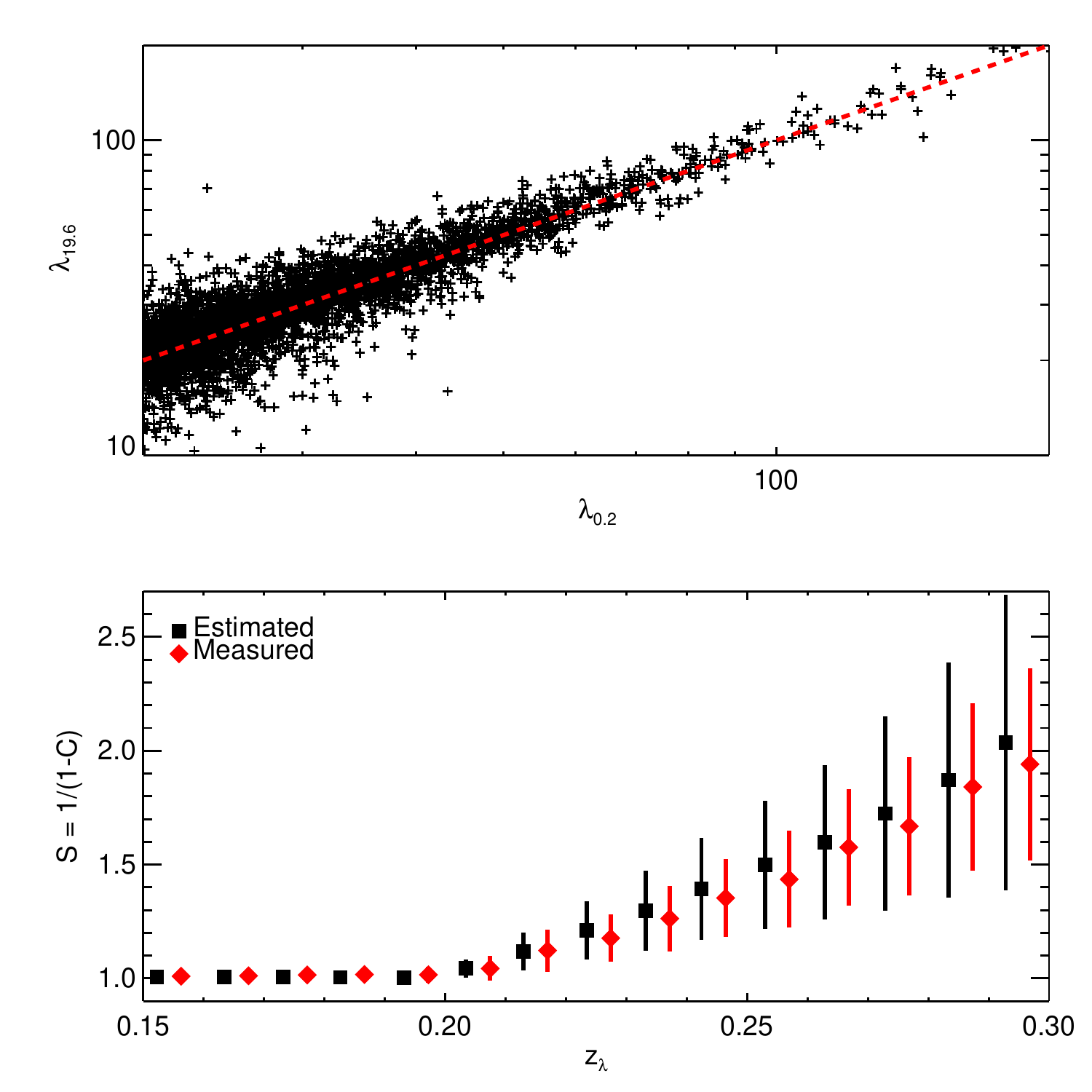}}
    \caption{\emph{Top:} Richness calculated with a $\imag<19.6$ cut vs. full
      $\lambda$ richness, for clusters with $0.15<z<0.3$.  The magnitude cut of
      $\imag<19.6$ is equivalent to $0.2L_*$ at $z=0.2$, so all cluster at
      $z>0.2$ in this test have $S(z) > 1$.  The corrected richness is
      consistent with the full richness.  \emph{Bottom:} Scale factor
      $\frac{1}{1-C}$ vs. photometric redshift.  Black squares show the scale
      factor and uncertainty in the scale factor estimated in the $\imag<19.6$
      run (shifted slightly for clarity).  Red diamonds show the measured shift
      and width.  Our measured values agree with our model.  However at the
      largest corrections we are slightly overestimating the correction term as
      well as the uncertainty in the correction term.}
    \label{fig:corrfactor}
  \end{center}
\end{figure}

In Figure~\ref{fig:corrfactor} we show the results of our test.  In the top panel we
show $\lambda_{19.6}$ vs. $\lambda_{0.2}$ for all clusters with $0.2<z<0.3$,
where $\lambda_{19.6}$ is the richness calculated with a magnitude limit of
$\imag<19.6$ and $\lambda_{0.2}$ is the standard $\lambda$ with a $0.2L_*$ cut.
When calculating $\lambda_{19.6}$ we have re-fit the photometric redshift
$\zlambda$ to ensure that our comparison is as fair as possible.  It is clear
in the top panel that the correction richness scales with uncorrected richness,
with some scatter as expected.  In the bottom panel we show the richness scale
value ($S=\frac{1}{1-C}=\lambda_{\mathrm{scaled}}/\lambda_{\mathrm{raw}}$) as a function of redshift.  The black squares show the
median estimated value of $S$ derived from Eqn.~\ref{eqn:meanc}, while the black error
bars represent the median error in $S$ as derived from
Eqn.~\ref{eqn:siglambda}.  The red diamonds show the median measured value of
$S$, and the red error bars represent the observed width in the distribution of
$S$.  Our predicted correction factor does scale with redshift as expected.
However, our errors are slightly overestimated for the largest corrections.

\section{Comparing $\lambda$ to $\lcol$}
\label{app:lamcol}

We now explore how the richness estimate used in this work, $\lambda$, compares
to the single-color richness $\lcol$ used in R12.  As detailed in
Section~\ref{sec:lambdachi}, the primary difference in richness estimators is
the replacement of the Gaussian color filter with a multi-color $\chisq$
filter.  However, we emphasize that there is also a subtle difference in the
background model, as described in Section~\ref{sec:chi2filter}.  That is, the
$\chisq$ filter does not distinguish between galaxies that are too red or too
blue relative to the model, and while the red sequence model is symmetric, the
background model is not.  

To make our comparisons, we have started with all \redmapper{} clusters with
$\lambda>20$.  We then calculate $\lambda_{g-r}$ and $\lambda_{r-i}$ using
the appropriate color model from the red sequence parametrization.  Our
expectation is that $\lgmr$ should trace $\lambda$ at low redshift where the
dominant signal is from the $g-r$ color, and $\lrmi$ should trace $\lambda$ at
high redshift.

\begin{figure}
  \begin{center}
    \scalebox{1.2}{\plotone{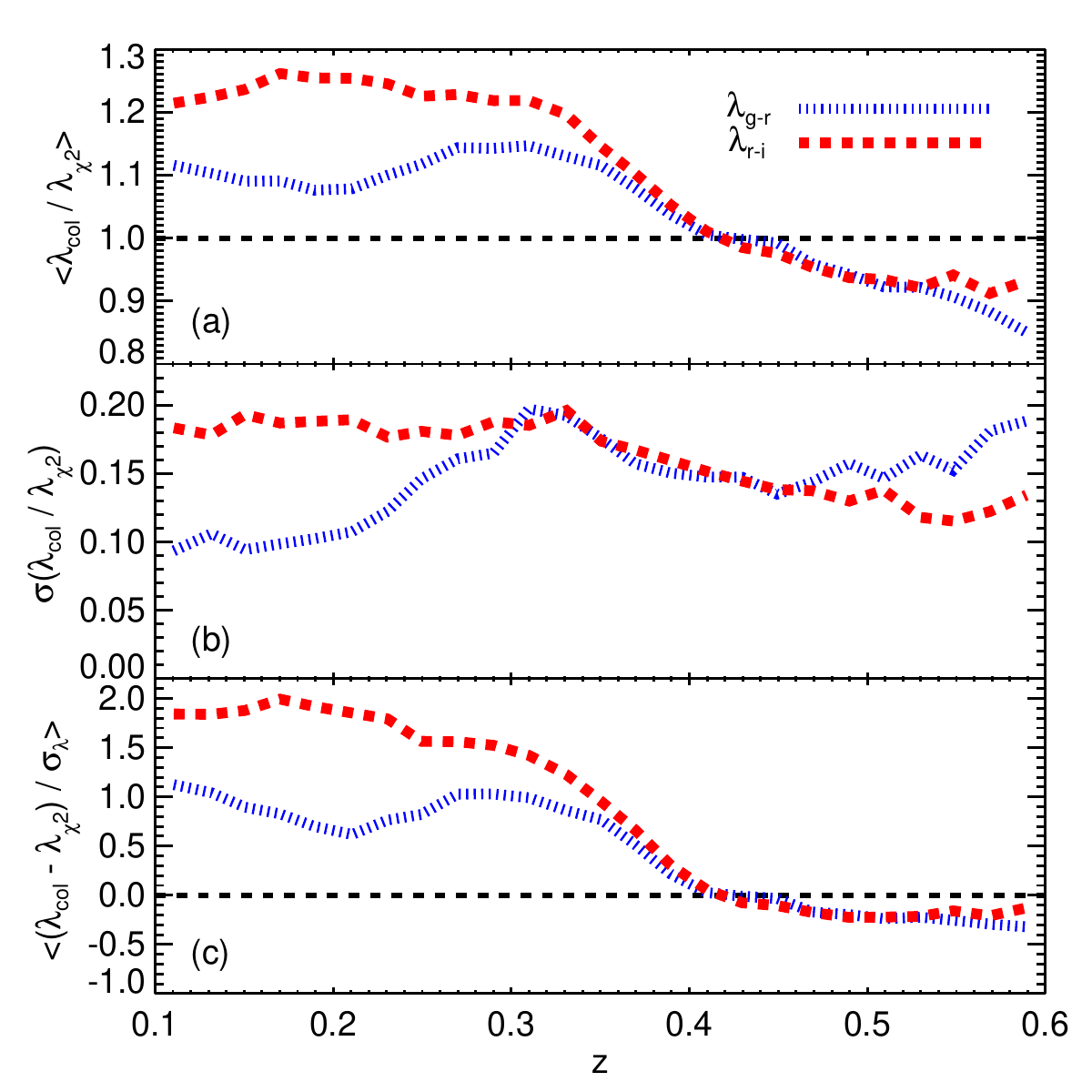}}
    \caption{\emph{(a)} Average ratio between the single-color $\lcol$
      vs. redshift for both $\lambda_{g-r}$ (blue dotted line) and
      $\lambda_{r-i}$ (red dashed line).  As discussed in the text, this offset
      is likely due to different background models.  In all cases the
      difference between the full multicolor $\lambda$ and the color appropriate
      for the redshift range ($g-r$ for $z<0.35$ and $r-i$ for $z>0.35$) are
      less than $10\%$.  \emph{(b)} Width of the $\lcol/\lambda$ distribution as
      a function of redshift.  The scatter is $\lesssim15\%$ for the
      appropriate color except for the transition redshift of
      $z\sim0.35$. \emph{(c)} Average offset normalized by the richness error.
      Thus, using the single color $\lambda_{g-r}$ is systematically biased
      high by $\sim1\sigma$ at low redshift, and $\lambda_{r-i}$ is
      systematically biased low by $0.2\sigma$ at high redshift.}
    \label{fig:colratio}
  \end{center}
\end{figure}

In Figure~\ref{fig:colratio} we show the statistics from comparing $\lgmr$ and
$\lrmi$ to $\lambda$.  In the top panel we show the median ratio as a function of
redshift.  At all redshifts the bias between the appropriate $\lcol$ and
$\lambda$ is $\lesssim10\%$.  In the bottom panel we show the median normalized
deviation, which is $\sim1\sigma$ at low redshift and less so at high redshift
where the richness errors are much larger due to the magnitude limit.  We
attribute this bias at low redshift to the different background model employed,
as galaxies that are redder than the red sequence are down-weighted in the
$\lambda$ model compared to the $\lcol$ model.  These biases are not large, but
they are significant and thus show the importance of using the same color model
and consistent survey data to achieve the best richness estimation.

The middle panel of Figure~\ref{fig:colratio} shows the width of the
$\lcol/\lambda$ distribution as a function of redshift.  The scatter is
$\lesssim15\%$ for the appropriate color except at $z\sim0.35$, where the
$4000\mathrm{\AA}$ break is transitioning from $g$ to $r$.  It is in this
transition region that a single color richness estimator does especially
poorly and we have the biggest advantage of using a multi-color estimator.

\section{Description of Columns in the DR8 Cluster Catalog}
\label{app:catalog}

The full \redmapper{} DR8 cluster and member catalogs are available at {\tt
  http://risa.stanford.edu/redmapper/} in FITS format, and from the online
journal in machine-readable formats.  A summary of the cluster catalog
information is given in Table~\ref{tab:catkey}.  A summary of the member
information is given in Table~\ref{tab:memkey}.

\begin{deluxetable*}{llcl}
\tablewidth{0pt}
\tablecaption{\redmapper{} DR8 Cluster Catalog Format}
\tablehead{
  \colhead{Column} &
  \colhead{Name} &
  \colhead{Format} &
  \colhead{Description}
}
\startdata
1 & ID & I7 & \redmapper{} Cluster Identification Number\\
2 & NAME & A20 & \redmapper{} Cluster Name\\
3 & RA & F12.7 & Right ascension in decimal degrees (J2000)\\
4 & DEC & F12.7 & Declination in decimal degrees (J2000)\\
5 & Z\_LAMBDA & F6.4 & Cluster \photoz $\zlambda$\\
6 & Z\_LAMBDA\_ERR & F6.4 & Gaussian error estimate for $\zlambda$\\
7 & LAMBDA & F6.2 & Richness estimate $\lambda$\\
8 & LAMBDA\_ERR & F6.2 & Gaussian error estimate for $\lambda$\\
9 & S & F6.3 & Richness scale factor (see Eqn.~\ref{eqn:sz})\\
10 & Z\_SPEC & F8.5 & SDSS spectroscopic redshift for most likely center (-1.0 if not available)\\
11 & OBJID & I20 & SDSS DR8 CAS object identifier\\
12 & IMAG & F6.3 & $i$-band cmodel magnitude for most likely central galaxy (dereddened)\\
13 & IMAG\_ERR & F6.3 & error on $i$-band cmodel magnitude\\
14 & MODEL\_MAG\_U & F6.3 & $u$ model magnitude for most likely central galaxy
(dereddened)\\
15 & MODEL\_MAGERR\_U & F6.3 & error on $u$ model magnitude\\
16 & MODEL\_MAG\_G & F6.3 & $g$ model magnitude for most likely central galaxy
(dereddened)\\
17 & MODEL\_MAGERR\_G & F6.3 & error on $g$ model magnitude\\
18 & MODEL\_MAG\_R & F6.3 & $r$ model magnitude for most likely central galaxy
(dereddened)\\
19 & MODEL\_MAGERR\_R & F6.3 & error on $r$ model magnitude\\
20 & MODEL\_MAG\_I & F6.3 & $i$ model magnitude for most likely central galaxy
(dereddened)\\
21 & MODEL\_MAGERR\_I & F6.3 & error on $i$ model magnitude\\
22 & MODEL\_MAG\_Z & F6.3 & $z$ model magnitude for most likely central galaxy
(dereddened)\\
23 & MODEL\_MAGERR\_Z & F6.3 & error on $z$ model magnitude\\
24 & ILUM & F7.3 & Total membership-weighted $i$-band luminosity (units of
$L_*$)\\
25 & P\_CEN[0] & E9.3 & Centering probability $\Pcen$ for most likely central\\
26 & RA\_CEN[0] & F12.7 & R.A. for most likely central\\
27 & DEC\_CEN[0] & F12.7 & Decl. for most likely central\\
28 & ID\_CEN[0] & I20 & DR8 CAS object identifier for most likely central\\
29-32 & \_CEN[1] & & $\Pcen$, R.A., Decl., and ID for second most likely central\\
33-36 & \_CEN[2] & & $\Pcen$, R.A., Decl., and ID for third most likely central\\
37-40 & \_CEN[3] & & $\Pcen$, R.A., Decl., and ID for fourth most likely central\\
41-44 & \_CEN[4] & & $\Pcen$, R.A., Decl., and ID for fifth most likely
central\\
45-65 & PZBINS & F7.4 & Redshift points at which $P(z)$ is evaluated\\
66-86 & PZ & E10.3 & $P(z)$ evaluated at redshift points given by PZBINS\\
\enddata
\label{tab:catkey}
\tablecomments{The full catalog contains 86 columns of information on 25325
  clusters.  This table is presented in its entirety in the online edition of
  the journal, and at {\tt
    http://risa.stanford.edu/redmapper}.}
\end{deluxetable*}

\begin{deluxetable*}{llcl}
\tablewidth{0pt}
\tablecaption{\redmapper{} DR8 Member Catalog Format}
\tablehead{
  \colhead{Column} &
  \colhead{Name} &
  \colhead{Format} &
  \colhead{Description}
}
\startdata
1 & ID & I7 & \redmapper{} Cluster Identification Number\\
2 & RA & F12.7 & Right ascension in decimal degrees (J2000)\\
3 & DEC & F12.7 & Declination in decimal degrees (J2000)\\
4 & R & F5.3 & Distance from cluster center ($h^{-1}\,\mathrm{Mpc}$)\\
5 & P\_MEM & F5.3 & Membership probability $\pmem$\\
6 & IMAG & F6.3 & $i$-band cmodel magnitude (dereddened)\\
7 & IMAG\_ERR & F6.3 & error on $i$-band cmodel magnitude\\
8 & MODEL\_MAG\_U & F6.3 & $u$ model magnitude 
(dereddened)\\
9 & MODEL\_MAGERR\_U & F6.3 & error on $u$ model magnitude\\
10 & MODEL\_MAG\_G & F6.3 & $g$ model magnitude
(dereddened)\\
11 & MODEL\_MAGERR\_G & F6.3 & error on $g$ model magnitude\\
12 & MODEL\_MAG\_R & F6.3 & $r$ model magnitude
(dereddened)\\
13 & MODEL\_MAGERR\_R & F6.3 & error on $r$ model magnitude\\
14 & MODEL\_MAG\_I & F6.3 & $i$ model magnitude
(dereddened)\\
15 & MODEL\_MAGERR\_I & F6.3 & error on $i$ model magnitude\\
16 & MODEL\_MAG\_Z & F6.3 & $z$ model magnitude
(dereddened)\\
17 & MODEL\_MAGERR\_Z & F6.3 & error on $z$ model magnitude\\
18 & Z\_SPEC & F8.5 & SDSS spectroscopic redshift (-1.0 if not available)\\
19 & OBJID & I20 &  SDSS DR8 CAS object identifier\\
\enddata
\label{tab:memkey}
\end{deluxetable*}


\begin{thebibliography}{85}
\expandafter\ifx\csname natexlab\endcsname\relax\def\natexlab#1{#1}\fi

\bibitem[{{Abell}(1958)}]{abell58}
{Abell}, G.~O. 1958, \apjs, 3, 211

\bibitem[{{Abell} {et~al.}(1989){Abell}, {Corwin}, \& {Olowin}}]{aco89}
{Abell}, G.~O., {Corwin}, Jr., H.~G., \& {Olowin}, R.~P. 1989, \apjs, 70, 1

\bibitem[{{Ahn} {et~al.}(2012){Ahn}, {Alexandroff}, {Allende Prieto},
  {Anderson}, {Anderton}, {Andrews}, {Aubourg}, {Bailey}, {Balbinot}, {Barnes},
  \& et~al.}]{dr9}
{Ahn}, C.~P. {et~al.} 2012, \apjs, 203, 21

\bibitem[{{Aihara} {et~al.}(2011){Aihara}, {Allende Prieto}, {An}, {Anderson},
  {Aubourg}, {Balbinot}, {Beers}, {Berlind}, {Bickerton}, {Bizyaev}, {Blanton},
  {Bochanski}, {Bolton}, {Bovy}, {Brandt}, {Brinkmann}, {Brown}, {Brownstein},
  {Busca}, {Campbell}, {Carr}, {Chen}, {Chiappini}, {Comparat}, {Connolly},
  {Cortes}, {Croft}, {Cuesta}, {da Costa}, {Davenport}, {Dawson}, {Dhital},
  {Ealet}, {Ebelke}, {Edmondson}, {Eisenstein}, {Escoffier}, {Esposito},
  {Evans}, {Fan}, {Femen{\'{\i}}a Castell{\'a}}, {Font-Ribera}, {Frinchaboy},
  {Ge}, {Gillespie}, {Gilmore}, {Gonz{\'a}lez Hern{\'a}ndez}, {Gott}, {Gould},
  {Grebel}, {Gunn}, {Hamilton}, {Harding}, {Harris}, {Hawley}, {Hearty}, {Ho},
  {Hogg}, {Holtzman}, {Honscheid}, {Inada}, {Ivans}, {Jiang}, {Johnson},
  {Jordan}, {Jordan}, {Kazin}, {Kirkby}, {Klaene}, {Knapp}, {Kneib},
  {Kochanek}, {Koesterke}, {Kollmeier}, {Kron}, {Lampeitl}, {Lang}, {Le Goff},
  {Lee}, {Lin}, {Long}, {Loomis}, {Lucatello}, {Lundgren}, {Lupton}, {Ma},
  {MacDonald}, {Mahadevan}, {Maia}, {Makler}, {Malanushenko}, {Malanushenko},
  {Mandelbaum}, {Maraston}, {Margala}, {Masters}, {McBride}, {McGehee},
  {McGreer}, {M{\'e}nard}, {Miralda-Escud{\'e}}, {Morrison}, {Mullally},
  {Muna}, {Munn}, {Murayama}, {Myers}, {Naugle}, {Neto}, {Nguyen}, {Nichol},
  {O'Connell}, {Ogando}, {Olmstead}, {Oravetz}, {Padmanabhan},
  {Palanque-Delabrouille}, {Pan}, {Pandey}, {P{\^a}ris}, {Percival},
  {Petitjean}, {Pfaffenberger}, {Pforr}, {Phleps}, {Pichon}, {Pieri}, {Prada},
  {Price-Whelan}, {Raddick}, {Ramos}, {Reyl{\'e}}, {Rich}, {Richards}, {Rix},
  {Robin}, {Rocha-Pinto}, {Rockosi}, {Roe}, {Rollinde}, {Ross}, {Ross},
  {Rossetto}, {S{\'a}nchez}, {Sayres}, {Schlegel}, {Schlesinger}, {Schmidt},
  {Schneider}, {Sheldon}, {Shu}, {Simmerer}, {Simmons}, {Sivarani}, {Snedden},
  {Sobeck}, {Steinmetz}, {Strauss}, {Szalay}, {Tanaka}, {Thakar}, {Thomas},
  {Tinker}, {Tofflemire}, {Tojeiro}, {Tremonti}, {Vandenberg}, {Vargas
  Maga{\~n}a}, {Verde}, {Vogt}, {Wake}, {Wang}, {Weaver}, {Weinberg}, {White},
  {White}, {Yanny}, {Yasuda}, {Yeche}, \& {Zehavi}}]{dr8}
{Aihara}, H. {et~al.} 2011, \apjs, 193, 29

\bibitem[{{Allen}(1995)}]{allen95}
{Allen}, S.~W. 1995, \mnras, 276, 947

\bibitem[{{Annis} {et~al.}(2011){Annis}, {Soares-Santos}, {Strauss}, {Becker},
  {Dodelson}, {Fan}, {Gunn}, {Hao}, {Ivezic}, {Jester}, {Jiang}, {Johnston},
  {Kubo}, {Lampeitl}, {Lin}, {Lupton}, {Miknaitis}, {Seo}, {Simet}, \&
  {Yanny}}]{assbd11}
{Annis}, J. {et~al.} 2011, ArXiv: 1111.6619

\bibitem[{{Annis} {et~al.}(1999)}]{annisetal99}
{Annis}, J., {et~al.} 1999, in Bulletin of the American Astronomical Society,
  Vol.~31, American Astronomical Society Meeting Abstracts, 1391

\bibitem[{{Bahcall} {et~al.}(2003)}]{bahcalletal03b}
{Bahcall}, N.~A., {et~al.} 2003, \apjs, 148, 243

\bibitem[{{Becker} {et~al.}(2007){Becker}, {McKay}, {Koester}, {Wechsler},
  {Rozo}, {Evrard}, {Johnston}, {Sheldon}, {Annis}, {Lau}, {Nichol}, \&
  {Miller}}]{bmkwr07}
{Becker}, M.~R. {et~al.} 2007, \apj, 669, 905

\bibitem[{{Benson} {et~al.}(2013)}]{bensonetal13}
{Benson}, B.~A., {et~al.} 2013, \apj, 763, 147

\bibitem[{{Biviano}(2000)}]{biviano00}
{Biviano}, A. 2000, in Constructing the Universe with Clusters of Galaxies

\bibitem[{{Blackburne} \& {Kochanek}(2012)}]{blackburnekochanek12}
{Blackburne}, J.~A., \& {Kochanek}, C.~S. 2012, \apj, 744, 76

\bibitem[{{Brodwin} {et~al.}(2011){Brodwin}, {Stern}, {Vikhlinin}, {Stanford},
  {Gonzalez}, {Eisenhardt}, {Ashby}, {Bautz}, {Dey}, {Forman}, {Gettings},
  {Hickox}, {Jannuzi}, {Jones}, {Mancone}, {Miller}, {Moustakas}, {Ruel},
  {Snyder}, \& {Zeimann}}]{bsvsg11}
{Brodwin}, M. {et~al.} 2011, \apj, 732, 33

\bibitem[{{Clerc} {et~al.}(2012){Clerc}, {Sadibekova}, {Pierre}, {Pacaud}, {Le
  F{\`e}vre}, {Adami}, {Altieri}, \& {Valtchanov}}]{clercetal12}
{Clerc}, N., {Sadibekova}, T., {Pierre}, M., {Pacaud}, F., {Le F{\`e}vre},
  J.-P., {Adami}, C., {Altieri}, B., \& {Valtchanov}, I. 2012, \mnras, 423,
  3561

\bibitem[{{Csabai} {et~al.}(2007){Csabai}, {Dobos}, {Trencs{\'e}ni},
  {Herczegh}, {J{\'o}zsa}, {Purger}, {Budav{\'a}ri}, \& {Szalay}}]{cdthj07}
{Csabai}, I., {Dobos}, L., {Trencs{\'e}ni}, M., {Herczegh}, G., {J{\'o}zsa},
  P., {Purger}, N., {Budav{\'a}ri}, T., \& {Szalay}, A.~S. 2007, Astronomische
  Nachrichten, 328, 852

\bibitem[{{Dawson} {et~al.}(2013){Dawson}, {Schlegel}, {Ahn}, {Anderson},
  {Aubourg}, {Bailey}, {Barkhouser}, {Bautista}, {Beifiori}, {Berlind},
  {Bhardwaj}, {Bizyaev}, {Blake}, {Blanton}, {Blomqvist}, {Bolton}, {Borde},
  {Bovy}, {Brandt}, {Brewington}, {Brinkmann}, {Brown}, {Brownstein}, {Bundy},
  {Busca}, {Carithers}, {Carnero}, {Carr}, {Chen}, {Comparat}, {Connolly},
  {Cope}, {Croft}, {Cuesta}, {da Costa}, {Davenport}, {Delubac}, {de Putter},
  {Dhital}, {Ealet}, {Ebelke}, {Eisenstein}, {Escoffier}, {Fan}, {Filiz Ak},
  {Finley}, {Font-Ribera}, {G{\'e}nova-Santos}, {Gunn}, {Guo}, {Haggard},
  {Hall}, {Hamilton}, {Harris}, {Harris}, {Ho}, {Hogg}, {Holder}, {Honscheid},
  {Huehnerhoff}, {Jordan}, {Jordan}, {Kauffmann}, {Kazin}, {Kirkby}, {Klaene},
  {Kneib}, {Le Goff}, {Lee}, {Long}, {Loomis}, {Lundgren}, {Lupton}, {Maia},
  {Makler}, {Malanushenko}, {Malanushenko}, {Mandelbaum}, {Manera}, {Maraston},
  {Margala}, {Masters}, {McBride}, {McDonald}, {McGreer}, {McMahon}, {Mena},
  {Miralda-Escud{\'e}}, {Montero-Dorta}, {Montesano}, {Muna}, {Myers},
  {Naugle}, {Nichol}, {Noterdaeme}, {Nuza}, {Olmstead}, {Oravetz}, {Oravetz},
  {Owen}, {Padmanabhan}, {Palanque-Delabrouille}, {Pan}, {Parejko},
  {P{\^a}ris}, {Percival}, {P{\'e}rez-Fournon}, {P{\'e}rez-R{\`a}fols},
  {Petitjean}, {Pfaffenberger}, {Pforr}, {Pieri}, {Prada}, {Price-Whelan},
  {Raddick}, {Rebolo}, {Rich}, {Richards}, {Rockosi}, {Roe}, {Ross}, {Ross},
  {Rossi}, {Rubi{\~n}o-Martin}, {Samushia}, {S{\'a}nchez}, {Sayres}, {Schmidt},
  {Schneider}, {Sc{\'o}ccola}, {Seo}, {Shelden}, {Sheldon}, {Shen}, {Shu},
  {Slosar}, {Smee}, {Snedden}, {Stauffer}, {Steele}, {Strauss}, {Streblyanska},
  {Suzuki}, {Swanson}, {Tal}, {Tanaka}, {Thomas}, {Tinker}, {Tojeiro},
  {Tremonti}, {Vargas Maga{\~n}a}, {Verde}, {Viel}, {Wake}, {Watson}, {Weaver},
  {Weinberg}, {Weiner}, {West}, {White}, {Wood-Vasey}, {Yeche}, {Zehavi},
  {Zhao}, \& {Zheng}}]{dsaaa13}
{Dawson}, K.~S. {et~al.} 2013, \aj, 145, 10

\bibitem[{{Durret} {et~al.}(2011)}]{durretetal11b}
{Durret}, F., {et~al.} 2011, \aap, 535, A65

\bibitem[{{Eisenhardt} {et~al.}(2008){Eisenhardt}, {Brodwin}, {Gonzalez},
  {Stanford}, {Stern}, {Barmby}, {Brown}, {Dawson}, {Dey}, {Doi}, {Galametz},
  {Jannuzi}, {Kochanek}, {Meyers}, {Morokuma}, \& {Moustakas}}]{ebgss08}
{Eisenhardt}, P.~R.~M. {et~al.} 2008, \apj, 684, 905

\bibitem[{{Gal} {et~al.}(2009)}]{galetal09}
{Gal}, R.~R., {et~al.} 2009, \aj, 137, 2981

\bibitem[{{George} {et~al.}(2012){George}, {Leauthaud}, {Bundy}, {Finoguenov},
  {Ma}, {Rykoff}, {Tinker}, {Wechsler}, {Massey}, \& {Mei}}]{glbfm12}
{George}, M.~R. {et~al.} 2012, \apj, 757, 2

\bibitem[{{George} {et~al.}(2011){George}, {Leauthaud}, {Bundy}, {Finoguenov},
  {Tinker}, {Lin}, {Mei}, {Kneib}, {Aussel}, {Behroozi}, {Busha}, {Capak},
  {Coccato}, {Covone}, {Faure}, {Fiorenza}, {Ilbert}, {Le Floc'h}, {Koekemoer},
  {Tanaka}, {Wechsler}, \& {Wolk}}]{glbft11}
---. 2011, \apj, 742, 125

\bibitem[{{Gladders} \& {Yee}(2000)}]{gladderyee00}
{Gladders}, M.~D., \& {Yee}, H.~K.~C. 2000, \aj, 120, 2148

\bibitem[{{Gladders} {et~al.}(2007)}]{gladdersetal07}
{Gladders}, M.~D., {et~al.} 2007, \apj, 655, 128

\bibitem[{{Goto} {et~al.}(2002)}]{gotoetal02}
{Goto}, T., {et~al.} 2002, \aj, 123, 1807

\bibitem[{{Hansen} {et~al.}(2005){Hansen}, {McKay}, {Wechsler}, {Annis},
  {Sheldon}, \& {Kimball}}]{hmwas05}
{Hansen}, S.~M., {McKay}, T.~A., {Wechsler}, R.~H., {Annis}, J., {Sheldon},
  E.~S., \& {Kimball}, A. 2005, \apj, 633, 122

\bibitem[{{Hansen} {et~al.}(2009){Hansen}, {Sheldon}, {Wechsler}, \&
  {Koester}}]{hswk09}
{Hansen}, S.~M., {Sheldon}, E.~S., {Wechsler}, R.~H., \& {Koester}, B.~P. 2009,
  \apj, 699, 1333

\bibitem[{{Hao} {et~al.}(2009){Hao}, {Koester}, {Mckay}, {Rykoff}, {Rozo},
  {Evrard}, {Annis}, {Becker}, {Busha}, {Gerdes}, {Johnston}, {Sheldon}, \&
  {Wechsler}}]{hkmrr09}
{Hao}, J. {et~al.} 2009, \apj, 702, 745

\bibitem[{{Hao} {et~al.}(2010){Hao}, {McKay}, {Koester}, {Rykoff}, {Rozo},
  {Annis}, {Wechsler}, {Evrard}, {Siegel}, {Becker}, {Busha}, {Gerdes},
  {Johnston}, \& {Sheldon}}]{hmkrr10}
---. 2010, \apjs, 191, 254

\bibitem[{{Hasselfield} {et~al.}(2013){Hasselfield}, {Hilton}, {Marriage},
  {Addison}, {Barrientos}, {Battaglia}, {Battistelli}, {Bond}, {Crichton},
  {Das}, {Devlin}, {Dicker}, {Dunkley}, {D{\"u}nner}, {Fowler}, {Gralla},
  {Hajian}, {Halpern}, {Hincks}, {Hlozek}, {Hughes}, {Infante}, {Irwin},
  {Kosowsky}, {Marsden}, {Menanteau}, {Moodley}, {Niemack}, {Nolta}, {Page},
  {Partridge}, {Reese}, {Schmitt}, {Sehgal}, {Sherwin}, {Sievers}, {Sif{\'o}n},
  {Spergel}, {Staggs}, {Swetz}, {Switzer}, {Thornton}, {Trac}, \&
  {Wollack}}]{hasselfieldetal13}
{Hasselfield}, M. {et~al.} 2013, JCAP, 7, 8

\bibitem[{{Henry} {et~al.}(2009)}]{henryetal09}
{Henry}, J.~P., {et~al.} 2009, \apj, 691, 1307

\bibitem[{{Hockney} \& {Eastwood}(1981)}]{hockney81}
{Hockney}, R.~W., \& {Eastwood}, J.~W. 1981, {Computer Simulation Using
  Particles} (Computer Simulation Using Particles, New York: McGraw-Hill, 1981)

\bibitem[{{Hoffleit} \& {Jaschek}(1991)}]{hj91}
{Hoffleit}, D., \& {Jaschek}, C.~. 1991, {The Bright star catalogue} (None)

\bibitem[{{H{\o}g} {et~al.}(2000){H{\o}g}, {Fabricius}, {Makarov}, {Urban},
  {Corbin}, {Wycoff}, {Bastian}, {Schwekendiek}, \& {Wicenec}}]{hfmuc00}
{H{\o}g}, E. {et~al.} 2000, \aap, 355, L27

\bibitem[{{Johnston} {et~al.}(2007){Johnston}, {Sheldon}, {Wechsler}, {Rozo},
  {Koester}, {Frieman}, {McKay}, {Evrard}, {Becker}, \&
  {Annis}}]{johnstonetal07}
{Johnston}, D.~E. {et~al.} 2007, ArXiv e-prints, 709, astro-ph/0709.1159

\bibitem[{{Kaiser} {et~al.}(2002)}]{panstarrs02}
{Kaiser}, N., {et~al.} 2002, in Society of Photo-Optical Instrumentation
  Engineers (SPIE) Conference Series, Vol. 4836, Society of Photo-Optical
  Instrumentation Engineers (SPIE) Conference Series, ed. J.~A. {Tyson} \&
  S.~{Wolff}, 154--164

\bibitem[{{Kepner} {et~al.}(1999){Kepner}, {Fan}, {Bahcall}, {Gunn}, {Lupton},
  \& {Xu}}]{kepneretal99}
{Kepner}, J., {Fan}, X., {Bahcall}, N., {Gunn}, J., {Lupton}, R., \& {Xu}, G.
  1999, \apj, 517, 78

\bibitem[{{Koester} {et~al.}(2007{\natexlab{a}})}]{koesteretal07a}
{Koester}, B., {et~al.} 2007{\natexlab{a}}, \apj, 660, 239

\bibitem[{{Koester} {et~al.}(2007{\natexlab{b}}){Koester}, {McKay}, {Annis},
  {Wechsler}, {Evrard}, {Rozo}, {Bleem}, {Sheldon}, \& {Johnston}}]{kmawe07a}
{Koester}, B.~P. {et~al.} 2007{\natexlab{b}}, \apj, 660, 221

\bibitem[{{Landy} \& {Szalay}(1993)}]{landyszalay93}
{Landy}, S.~D., \& {Szalay}, A.~S. 1993, \apj, 412, 64

\bibitem[{{Lin} \& {Mohr}(2004)}]{linmohr04}
{Lin}, Y.-T., \& {Mohr}, J.~J. 2004, \apj, 617, 879

\bibitem[{{Lopes} {et~al.}(2004)}]{lopesetal04}
{Lopes}, P.~A.~A., {et~al.} 2004, \aj, 128, 1017

\bibitem[{{Lopes} {et~al.}(2006)}]{lccgd06}
---. 2006, \apj, 648, 209

\bibitem[{{LSST Dark Energy Science Collaboration}(2012)}]{lsst12}
{LSST Dark Energy Science Collaboration}. 2012, ArXiv e-prints,
  astro-ph/1211.0310

\bibitem[{{Mahdavi} {et~al.}(2013){Mahdavi}, {Hoekstra}, {Babul}, {Bildfell},
  {Jeltema}, \& {Henry}}]{mahdavietal13}
{Mahdavi}, A., {Hoekstra}, H., {Babul}, A., {Bildfell}, C., {Jeltema}, T., \&
  {Henry}, J.~P. 2013, \apj, 767, 116

\bibitem[{{Mana} {et~al.}(2013){Mana}, {Giannantonio}, {Weller}, {Hoyle},
  {H{\"u}tsi}, \& {Sartoris}}]{manaetal13}
{Mana}, A., {Giannantonio}, T., {Weller}, J., {Hoyle}, B., {H{\"u}tsi}, G., \&
  {Sartoris}, B. 2013, \mnras, 434, 684

\bibitem[{{Mandelbaum} {et~al.}(2008){Mandelbaum}, {Seljak}, \&
  {Hirata}}]{mandelbaumetal08}
{Mandelbaum}, R., {Seljak}, U., \& {Hirata}, C.~M. 2008, JCAP, 8, 6

\bibitem[{{Mantz} {et~al.}(2010)}]{mantzetal10a}
{Mantz}, A., {et~al.} 2010, \mnras, 406, 1759

\bibitem[{{Matthews} {et~al.}(1964){Matthews}, {Morgan}, \& {Schmidt}}]{mms64}
{Matthews}, T.~A., {Morgan}, W.~W., \& {Schmidt}, M. 1964, \apj, 140, 35

\bibitem[{{McNamara} {et~al.}(2006){McNamara}, {Rafferty}, {B{\^i}rzan},
  {Steiner}, {Wise}, {Nulsen}, {Carilli}, {Ryan}, \& {Sharma}}]{mrbsw06}
{McNamara}, B.~R. {et~al.} 2006, \apj, 648, 164

\bibitem[{{Menanteau} {et~al.}(2013)}]{menanteauetal13}
{Menanteau}, F., {et~al.} 2013, \apj, 765, 67

\bibitem[{{Milkeraitis} {et~al.}(2010){Milkeraitis}, {van Waerbeke}, {Heymans},
  {Hildebrandt}, {Dietrich}, \& {Erben}}]{milkeraitisetal10}
{Milkeraitis}, M., {van Waerbeke}, L., {Heymans}, C., {Hildebrandt}, H.,
  {Dietrich}, J.~P., \& {Erben}, T. 2010, \mnras, 406, 673

\bibitem[{{More} {et~al.}(2011){More}, {van den Bosch}, {Cacciato}, {Skibba},
  {Mo}, \& {Yang}}]{mvcsm11}
{More}, S., {van den Bosch}, F.~C., {Cacciato}, M., {Skibba}, R., {Mo}, H.~J.,
  \& {Yang}, X. 2011, \mnras, 410, 210

\bibitem[{{Murphy} {et~al.}(2012){Murphy}, {Geach}, \& {Bower}}]{mgb12}
{Murphy}, D.~N.~A., {Geach}, J.~E., \& {Bower}, R.~G. 2012, \mnras, 420, 1861

\bibitem[{{Navarro} {et~al.}(1995){Navarro}, {Frenk}, \&
  {White}}]{navarroetal95}
{Navarro}, J.~F., {Frenk}, C.~S., \& {White}, S.~D.~M. 1995, \mnras, 275, 56

\bibitem[{Nelder \& Mead(1965)}]{nelder65}
Nelder, J.~A., \& Mead, R. 1965, Computer Journal, 308

\bibitem[{{Oemler}(1976)}]{oemler75}
{Oemler}, Jr., A. 1976, \apj, 209, 693

\bibitem[{{Piffaretti} {et~al.}(2011)}]{piffarettietal11}
{Piffaretti}, R., {et~al.} 2011, \aap, 534, A109

\bibitem[{{Popesso} {et~al.}(2007){Popesso}, {Biviano}, {B{\"o}hringer}, \&
  {Romaniello}}]{popessoetal07}
{Popesso}, P., {Biviano}, A., {B{\"o}hringer}, H., \& {Romaniello}, M. 2007,
  \aap, 464, 451

\bibitem[{{Rozo} {et~al.}(2011){Rozo}, {Rykoff}, {Koester}, {Nord}, {Wu},
  {Evrard}, \& {Wechsler}}]{rrknw11}
{Rozo}, E., {Rykoff}, E., {Koester}, B., {Nord}, B., {Wu}, H.-Y., {Evrard}, A.,
  \& {Wechsler}, R. 2011, \apj, 740, 53

\bibitem[{{Rozo} \& {Rykoff}(2014)}]{rozorykoff14}
{Rozo}, E., \& {Rykoff}, E.~S. 2014, \apj, 783, 80

\bibitem[{{Rozo} {et~al.}(2009){Rozo}, {Rykoff}, {Koester}, {McKay}, {Hao},
  {Evrard}, {Wechsler}, {Hansen}, {Sheldon}, {Johnston}, {Becker}, {Annis},
  {Bleem}, \& {Scranton}}]{rrkmh09}
{Rozo}, E. {et~al.} 2009, \apj, 703, 601

\bibitem[{{Rozo} {et~al.}(2007){Rozo}, {Wechsler}, {Koester}, {McKay},
  {Evrard}, {Johnston}, {Sheldon}, {Annis}, \& {Frieman}}]{rwkme07a}
---. 2007, ArXiv Astrophysics e-prints, astro-ph/0703571

\bibitem[{{Rozo} {et~al.}(2010)}]{rozoetal10a}
{Rozo}, E., {et~al.} 2010, \apj, 708, 645

\bibitem[{{Rykoff} {et~al.}(2012){Rykoff}, {Koester}, {Rozo}, {Annis},
  {Evrard}, {Hansen}, {Hao}, {Johnston}, {McKay}, \& {Wechsler}}]{rkrae12}
{Rykoff}, E.~S. {et~al.} 2012, \apj, 746, 178

\bibitem[{{Schlegel} {et~al.}(1998){Schlegel}, {Finkbeiner}, \&
  {Davis}}]{sfd98}
{Schlegel}, D.~J., {Finkbeiner}, D.~P., \& {Davis}, M. 1998, \apj, 500, 525

\bibitem[{{Schombert}(1986)}]{schombert86}
{Schombert}, J.~M. 1986, \apjs, 60, 603

\bibitem[{{Sheldon} {et~al.}(2012){Sheldon}, {Cunha}, {Mandelbaum},
  {Brinkmann}, \& {Weaver}}]{scmbw12}
{Sheldon}, E.~S., {Cunha}, C.~E., {Mandelbaum}, R., {Brinkmann}, J., \&
  {Weaver}, B.~A. 2012, \apjs, 201, 32

\bibitem[{{Sinnott}(1988)}]{sinnott88}
{Sinnott}, R.~W. 1988, {The complete new general catalogue and index catalogues
  of nebulae and star clusters by J. L. E. Dreyer} (None)

\bibitem[{{Soares-Santos} {et~al.}(2011)}]{soaressantosetal11}
{Soares-Santos}, M., {et~al.} 2011, \apj, 727, 45

\bibitem[{{Song} {et~al.}(2012{\natexlab{a}}){Song}, {Mohr}, {Barkhouse},
  {Warren}, {Dolag}, \& {Rude}}]{smbwd12}
{Song}, J., {Mohr}, J.~J., {Barkhouse}, W.~A., {Warren}, M.~S., {Dolag}, K., \&
  {Rude}, C. 2012{\natexlab{a}}, \apj, 747, 58

\bibitem[{{Song} {et~al.}(2012{\natexlab{b}})}]{songetal12}
{Song}, J., {et~al.} 2012{\natexlab{b}}, \apj, 761, 22

\bibitem[{{Stott} {et~al.}(2012)}]{stottetal12}
{Stott}, J.~P., {et~al.} 2012, \mnras, 422, 2213

\bibitem[{{Swanson} {et~al.}(2008){Swanson}, {Tegmark}, {Hamilton}, \&
  {Hill}}]{sthh08}
{Swanson}, M.~E.~C., {Tegmark}, M., {Hamilton}, A.~J.~S., \& {Hill}, J.~C.
  2008, \mnras, 387, 1391

\bibitem[{{Szabo} {et~al.}(2011){Szabo}, {Pierpaoli}, {Dong}, {Pipino}, \&
  {Gunn}}]{spdpg11}
{Szabo}, T., {Pierpaoli}, E., {Dong}, F., {Pipino}, A., \& {Gunn}, J. 2011,
  \apj, 736, 21

\bibitem[{{Thanjavur} {et~al.}(2009)}]{thanjavuretal09}
{Thanjavur}, K., {et~al.} 2009, \apj, 706, 571

\bibitem[{{The DES Collaboration}(2005)}]{des05}
{The DES Collaboration}. 2005, ArXiv: 0510346

\bibitem[{{Tinker} {et~al.}(2012){Tinker}, {George}, {Leauthaud}, {Bundy},
  {Finoguenov}, {Massey}, {Rhodes}, \& {Wechsler}}]{tglbf12}
{Tinker}, J.~L., {George}, M.~R., {Leauthaud}, A., {Bundy}, K., {Finoguenov},
  A., {Massey}, R., {Rhodes}, J., \& {Wechsler}, R.~H. 2012, \apjl, 755, L5

\bibitem[{{van Breukelen} \& {Clewley}(2009)}]{vanbreukelenetal09}
{van Breukelen}, C., \& {Clewley}, L. 2009, \mnras, 395, 1845

\bibitem[{{Vikhlinin} {et~al.}(2009)}]{vikhlininetal09b}
{Vikhlinin}, A., {et~al.} 2009, \apj, 692, 1060

\bibitem[{{Voges} {et~al.}(1999){Voges}, {Aschenbach}, {Boller},
  {Br{\"a}uninger}, {Briel}, {Burkert}, {Dennerl}, {Englhauser}, {Gruber},
  {Haberl}, {Hartner}, {Hasinger}, {K{\"u}rster}, {Pfeffermann}, {Pietsch},
  {Predehl}, {Rosso}, {Schmitt}, {Tr{\"u}mper}, \& {Zimmermann}}]{vogesetal99}
{Voges}, W. {et~al.} 1999, \aap, 349, 389

\bibitem[{{von der Linden} {et~al.}(2007){von der Linden}, {Best}, {Kauffmann},
  \& {White}}]{vbkw07}
{von der Linden}, A., {Best}, P.~N., {Kauffmann}, G., \& {White}, S.~D.~M.
  2007, \mnras, 379, 867

\bibitem[{{von der Linden} {et~al.}(2012)}]{vonderlindeetal12}
{von der Linden}, A., {et~al.} 2012, ArXiv: 1208.0597

\bibitem[{{Wen} {et~al.}(2012)}]{wenetal12}
{Wen}, Z.~L., {et~al.} 2012, \apjs, 199, 34

\bibitem[{{York} {et~al.}(2000){York}, {Adelman}, {Anderson}, {Anderson},
  {Annis}, \& { the SDSS collaboration}}]{yorketal00}
{York}, D.~G., {Adelman}, J., {Anderson}, J.~E., {Anderson}, S.~F., {Annis},
  J., \& { the SDSS collaboration}. 2000, \aj, 120, 1579

\bibitem[{{Zwicky} {et~al.}(1968){Zwicky}, {Herzog}, \& {Wild}}]{zwickyetal68}
{Zwicky}, F., {Herzog}, E., \& {Wild}, P. 1968, {Catalogue of galaxies and of
  clusters of galaxies} (Pasadena: California Institute of Technology (CIT),
  1961-1968)

\end{thebibliography}
\end{document}